%% file: bidding.GEB.full.tex
\documentclass[authoryear]{elsarticle}
\makeatletter
\def\ps@pprintTitle{%
     \let\@oddhead\@empty
     \let\@evenhead\@empty
     \def\@oddfoot{}%
     \let\@evenfoot\@oddfoot}

\usepackage[ruled,vlined]{algorithm2e}

\usepackage{float}
\usepackage{natbib}
\usepackage{helvet}
\usepackage{courier}
\usepackage{mathrsfs}
\usepackage{graphicx}
\usepackage{amsmath}
\usepackage{amssymb}
\usepackage{amsthm}
\usepackage{enumitem}
\usepackage{multirow} 
\usepackage{xcolor}

\usepackage{wrapfig,framed}
\usepackage{subcaption}
\usepackage{tikz}
\usepackage{url}
\usepackage[T1]{fontenc}
\usepackage[letterpaper,  margin=1.5in]{geometry}
\usepackage{accents}
\newlength{\dhatheight}
\newcommand{\doublehat}[1]{%
    \settoheight{\dhatheight}{\ensuremath{\hat{#1}}}%
    \addtolength{\dhatheight}{-0.35ex}%
    \hat{\vphantom{\rule{1pt}{\dhatheight}}%
    \smash{\hat{#1}}}}
\usetikzlibrary{arrows,decorations,decorations.shapes,decorations.markings,decorations.pathreplacing,backgrounds,shapes,petri,topaths}\usepackage{tkz-berge}

\usetikzlibrary{calc}
\newcount\Comments  % 0 suppresses notes to selves in text
\definecolor{darkgreen}{rgb}{0,0.6,0}
\newcommand{\kibitz}[2]{\ifnum\Comments=1{\color{#1}{#2}}\fi}

\newcommand{\rmr}[1]{\kibitz{blue}{[RESHEF:#1]}}

\newcommand{\eps}{\epsilon}

\DeclareMathOperator*{\argmin}{argmin}
\DeclareMathOperator*{\argmax}{argmax}
\def\maj{\text{maj}}
\def\two{\text{two}}
\def\url{\texttt}
\newcommand{\dul}[1]{{\ul{\ul{#1}}}} 
\newcommand{\dol}[1]{{\ol{\ol{#1}}}} 

\input{defs.tex}

\newtheorem{theorem}{Theorem}
\newtheorem{lemma}[theorem]{Lemma}
\newtheorem{proposition}[theorem]{Proposition}
\newtheorem{corollary}[theorem]{Corollary}
\newtheorem{definition}{Definition}
\newcommand{\xqed}{\mbox{\raggedright $\Diamond$}}

\newcommand{\omittext}[1]{}
\def\cite{\citep}
\def\shortcite{\citeyearpar}

\def\ol{\overline}
\def\ul{\underline}

\begin{document}
\begin{frontmatter}

\title{Bidding Games and Efficient Allocations\tnoteref{thanks}}
\tnotetext[thanks]{A preliminary version of this paper appeared in the 16th ACM Conference on Economics and Computation (EC-2015).
The first author is supported by the Israel Science Foundation (ISF) under Grant \#773/16. The second and third authors are supported by advanced ERC grants. Part of the research was performed when all of the authors were affiliated with MSR Herzlia, Israel. 
}
\author[tech]{Reshef Meir\corref{cor1}}
\ead{reshefm@ie.technion.ac.il}
\author[huji]{Gil Kalai}
\ead{kalai@math.huji.ac.il}
\author[tech]{Moshe Tennenholtz}
\ead{moshet@ie.technion.ac.il}

\cortext[cor1]{Corresponding author}
\address[huji]{Hebrew University, Jerusalem, Israel}
\address[tech]{Technion-Israel Institute of Technology, Haifa, Israel}

\begin{abstract}
Richman games are zero-sum games, where in each turn players bid in order to determine who will play next~\cite{laz99}.
We extend the theory to impartial general-sum two player games called \emph{bidding games}, showing the existence of pure subgame-perfect equilibria (PSPE). In particular, we show that  PSPEs form a semilattice, with a unique and natural \emph{Bottom Equilibrium}.

Our main result shows that if only two actions available to the players in each node, then the Bottom Equilibrium has additional properties: (a) utilities are monotone in  budget; (b) every outcome is Pareto-efficient; and (c)  any Pareto-efficient outcome is attained for some budget.

In the context of combinatorial bargaining, we show that a player with a fraction of $X\%$ of the total budget prefers her allocation to $X\%$ of the possible allocations.
In addition, we provide a polynomial-time algorithm to compute the Bottom Equilibrium of a binary bidding game. 

\medskip
\noindent \textit{JEL classification:} C72; C78
\end{abstract}

\begin{keyword}
  Extensive form games; Richman games; Combinatorial games; Bargaining
	\end{keyword}
\end{frontmatter}

\section{Introduction}
Are game-theorists better at playing games? Game-theory and game-playing are often considered to have distinct purposes and different sets of required skills. Exceptional points of intersection are games that are both fun to play and possess interesting theoretical properties; among those, a particularly exquisite example is \emph{Bidding Chess}.\footnote{For rules and historical background, see \cite{bhat2009bidding}.} Unlike standard Chess where turns are alternating among white and black, in the bidding version players each start with a fixed budget, and \emph{bid} at the beginning of each turn. The higher bidder has the right to play next after paying her bid to the other player. It is thus possible for example that a player will move 3 pieces before the other player makes a single move. Yet one needs to carefully balance the bids for the different moves, or else she could lose too much of her budget, leaving her unable to take advantage of her position in the game.
Indeed, even the bidding variation of the simple Tic-Tac-Toe game is non-trivial to play, and our personal experience shows that often a game where one player seems to lead gets turned over.\footnote{Many hours of play have been dedicated in the sake of promoting science.} 

These bidding variations of popular recreational games are special cases of a wider class of games known as \emph{Richman Games}: A Richman game is very similar to a \emph{combinatorial game}~\cite{albert2007lessons}, in that it consists of a directed graph of states (e.g. board states of the Chess game), where states without outgoing edges  assign victory either to the black player or to the white player. In contrast to standard combinatorial games where turns are alternating, a bidding game starts with an \emph{initial budget allocation}, and in each turn players \emph{bid for the right to play next}. The higher bidder pays her bid to the lower bidder, and chooses a directed edge to follow, until either a white or a black terminal is reached. Richman games have been formally studied by Lazarus et al.~\shortcite{laz99} (and later by Develin and Payne~\shortcite{DP10}), whose main contribution was a surprising connection with random-turn games. See  Section~\ref{sec:related} for more details on Richman games and other variations of bidding games.

However, a zero-sum game with bidding phases is still a zero-sum game, and to the best of our knowledge, bidding games with general utility functions have not been previously studied. In this work, we extend the model to general-sum two player games by adding bidding phases to an arbitrary extensive form game. Whereas zero-sum games have a \emph{value} that each player can guaranty, when extending the definition to general-sum games we can only talk about \emph{equilibrium}.  The standard solution concept studied in extensive-form game is \emph{pure subgame-perfect equilibrium} (PSPE), meaning that each player is playing a best-response to the other player at any subtree of the game. Indeed, the purpose of this work is to study the existence and properties of PSPE in general-sum bidding games. We highlight that this is a natural extension of previous work on zero-sum games, since the value of a zero-sum game is a special case of a PSPE outcome.

 %\footnote{Since the modified game  contains simultaneous bidding moves, a-priori there is no reason to believe a pure equilibrium exists.}

Other than the pure theoretical interest in such results, many social and economic interactions can be formalized as two-player extensive form games. For example, we can consider two agents bargaining over a set of heterogeneous items. Then any sequential bargaining protocol is in fact an extensive form game (with complete information), where the utilities for players are the respective values of the subsets of items they get in the end. Indeed, several bargaining and arbitration protocols have been suggested in the literature, some of them of a sequential nature (see Section~\ref{sec:related}).
We can thus use bidding phases as above to design a new bargaining mechanism called \emph{sequential scrip bargaining} (SSB): The agents each start with some fixed budget (reflecting their relative power or entitlement), and use this budget to bid over items auctioned in some order. As mechanism design often deals with efficiency and fairness of proposed mechanisms, it is natural to ask e.g. whether the equilibrium outcomes of SSB correspond to efficient and/or fair allocations.

\paragraph{Paper structure and contribution}
The  definition of bidding games and some examples are provided in Section~\ref{sec:model}. In Section~\ref{sec:exist}, we  show that under a mild assumption on tie-breaking, every bidding game admits at least one PSPE. Further, the set of all PSPEs forms a meet-semilattice, whose lowest point is termed the \emph{Bottom Equilibrium}. 

Our main result is given in Section~\ref{sec:main}, and it states that in any bidding game over a \emph{binary tree} (when at every stage of the game there are two possible actions), the Bottom Equilibrium can be computed by simulating an ascending auction. Further, it admits three desirable properties: 
 (a) each player's utility is weakly monotone in her budget; (b) a Pareto-efficient outcome is reached for any initial budget; and (c) for any Pareto-efficient outcome there is an initial budget such that this outcome is attained.
We complement this result by showing that none of the above properties is guaranteed in non-binary trees.
%As a complementary result, we show that in bidding games on a non-binary tree, none of the properties above is guaranteed, and all PSPE outcomes may be non-monotone and arbitrarily bad for both players.
A key lemma required for the main theorem shows that the Bottom Equilibrium has a simple structure and is not sensitive to small budget fluctuations. 

In Section~\ref{sec:bar} we analyze the properties of the SSB mechanism informally described above, when the agents have arbitrary combinatorial valuations over items. 
As a direct corollary from our main theorem, we get that the mechanism guarantees a Pareto-efficient allocation for any valuation functions agents have over subsets of items. We further show that the allocation is fair in the sense that an agent with a fraction of $X\%$ of the total budget prefers her allocation to $X\%$ of the possible allocations, a criterion known as \emph{minimal satisfaction test}~\cite{de2012selection}.

%By the observation above on the equivalence of bidding games and SSAs, our results immediately apply to the SSA mechanism, guarantying an efficient allocation of items to two agents.  %, and for equal budgets the attained allocation is \emph{fair} in the sense that it is envy bounded by a single good~\cite{Budish11}. 
%These properties hold for any valuation functions agents have over subsets of items, showing that for two rational agents with public information, the SSA mechanism implements the whole range of efficient allocations.  in 
%
\iffalse
Our result can be applied not just to item allocation problems, but also to other domains where two agents have combinatorial preferences over outcomes, including multi-issue voting~\cite{lang2009sequential} and selection of arbitrators~\cite{de2012selection}. Implications of our main result extend to other games such as the Nash bargaining game and Centipede games. 

\fi

Finally, we prove in Section~\ref{sec:compute} that any binary bidding game can be solved in time that is polynomial in the number of states, and show how such an algorithm can be used to efficiently find the Bottom Equilibrium in SSBs with additive valuations and other succinct representations of valuations.

\section{Model}
\label{sec:model}
For an integer $k$, we denote the set $\{1,2,\ldots,k\}$ by $[k]$. 
Unless explicitly mentioned otherwise, we assume that $N=\{1,2\}$, i.e. that there are only two competing agents (sometimes called \emph{players} or \emph{bidders}). We use the notation $-i$ (instead of $3-i$) to denote the player that is not $i$. We name player~1 the \emph{white} player, and player~2 the \emph{black} player. 
%The notation $a \gti b$ means  $a \geq b$ when $i=1$ and  $a>b$ when $i=2$.

\subsection{Bidding games}
\label{sec:bidding_games}
We first define the game structure that is common to combinatorial games, random-turn games and bidding games. The bidding process which is unique to bidding games is defined below.
\begin{definition}
\label{def:game_structure}
An  \emph{game structure} is a tuple $G=\tup{S,s_0,T,g_1,g_2,u_1,u_2}$, where: 
\begin{itemize}
\item  $S$ is a set of game states; 
\item $s_0\in S$ is the initial state; 
\item $T\subseteq S$ is a set of terminal states; 
\item $g_i:(S\setminus T)\rightarrow 2^S$ defines the optional moves of player $i$ in state $s\in S\setminus T$;
\item $u_i:T \rightarrow \mathbb R$ defines the utility for player $i$ in each terminal $t\in T$. 
\end{itemize}
\end{definition}
We make the following assumptions about the transition functions $g_i$. 
 First, unless specified otherwise, we assume that there are no cycles. Formally, there is no sequence $s_1,s_2,\ldots,s_k$ where $s_k=s_1$ and $s_{j+1}\in g_1(s_j) \cup g_2(s_j)$ for all $j<k$. 
Second, unless specified otherwise, we assume that in every non-terminal state both players can play. Formally, that $g_i(s)\neq \emptyset$ for all $i\in N,s\in S\setminus T$. 
The utility function $u_i$ induces a complete preference order over terminals $T$ (see also Sec.~\ref{sec:gen_uniq}). We denote $t \succeq_i t'$ whenever $u_i(t) \geq u_i(t')$.  
Given a game structure $G$ and $s \in S$, we denote by $G|_s$ the subgame of $G$ rooted in $s$.  The \emph{height} of $s$ is the maximal distance between $s$ and a leaf $t\in T(G|_s)$. In particular, $\height(G) = \height(s_0)$.

The following two properties are structural properties of the underlying game tree/DAG:
\begin{definition}
A  game structure $G$ is \emph{impartial}, if $g_1(s)=g_2(s)$ for all $s\in S$. That is, if the same set of moves is available to both players in every state. 
\end{definition}
Unless specified otherwise, all games in this paper are impartial. In impartial games we only need to specify one transition function $g(s)$.\footnote{Chess for example is not an impartial game since different moves are available to black and white. Partial zero-sum games are discussed in Section~\ref{sec:related}, however for \emph{acyclic} zero-sum games partiality does not matter~\cite{DP10}.} 

\begin{definition}
A  game structure  $G$ is \emph{binary}, if $|g(s)|\leq 2$ for all $s\in S\setminus T$.  
\end{definition}

%\paragraph{Bidding for turns}

\paragraph{Playing a bidding game}
In order to complete the definition of a game, we also need a method to determine who plays at every turn. Traditionally, there is a turn function that assigns the current player for every state (e.g., alternating turns in Nim and other combinatorial games). Another way is to randomly select the current player in each turn, as in \cite{peres07}. In this work we follow the bidding framework of \cite{laz99,DP10}, where each player has an initial budget $B_i$ that is used for the bidding. The higher bidder pays her bid to the lower bidder, so that the total budget $B_1+B_2=1$ is constant throughout the game.

We begin with an informal description of the strategies and an example. %, and in Section~\ref{sec:PSPE} we formally describe the strategy space and the equilibria. 

 %The game has both simultaneous moves, and ``standard'' one-player moves. Informally, in each turn players each simultaneously submit a bid that does not exceed their budget, and the player who submitted the higher bid has the right to choose the next game state. 
In each turn (suppose at state $s$), each player submits a bid $b_i\leq B_i$, and a ``next state'' $s^*_i\in g(s)$, which is realized in case $i$ wins the round. We break ties in favor of player~1. Then the winner (say, $i^*$) pays her bid to the loser $-i^*$ and the game proceeds from state $s^*_{i^*}$ with the modified budgets $B_{i^*}-b_{i^*}, B_{-i^*}+b_{i^*}$.  

%
%More formally, a strategy of $i$ in $G$ is composed of a pair $\tup{s^*_i,b_i}$ for every $s\in S\setminus T$, and every $B_i$. We require that $b_i\leq B_i$, and that $s^*_i\in g(s)$, otherwise this is not a valid strategy.  

We consider both discrete and continuous bids. Finite resolution means that all bids and budgets are multiples of $\eps = \frac{1}{R'}$ for some integer $R'$. We say that $\eps$ is \emph{high resolution} if $R'=R\cdot 2^k$ where $k=\height(G)$ and  $R\geq 4$ is an even integer. In particular it means that $\frac{1}{\eps}$ is at least exponential in the height of the tree. Throughout most of the paper, we will assume a high finite resolution.
Later in Sections~\ref{sec:discrete_mon} and \ref{sec:real} we consider the consequences of small and infinite resolutions, respectively. 

%Unless explicitly mentioned otherwise, bids can be real numbers. %we assume that bids can be any non-negative real number that does not exceed the current budget. %We also assume that in case of a tie ($b_1=b_2$), player~1 gains the turn. 
%It is also possible to think of games where bids (and budgets) are restricted to integers, and we will consider this variation in Section~\ref{sec:discrete_mon}.  
For a given $\eps$, we denote by $\calB_\eps=\{(B_1,B_2) \text{ s.t. } B_1+B_2=1,\text{ and } \frac{B_1}{\eps},\frac{B_2}{\eps}\in \mathbb N\}$  the set of possible budget partitions. We omit the subscript $\eps$ when clear from context. Since in either case the total budget is fixed,  $B_2$ can always be inferred from $B_1$ and vice versa. We therefore identify $\calB_\eps$ with $\{0,\eps,2\eps,\ldots,1\}$ and use either $B_1$ or $B_2$ to denote a particular budget partition. %\rmr{use $\calB_0=[0,1]$ in the continuous case}

\begin{definition}\label{def:bidding_game}A (discrete) \emph{bidding game} is a pair $\tup{G,\eps}$, where $G$ is a game structure as per Def.~\ref{def:game_structure}, and $\eps=\frac{1}{R'}$ (for some integer $R'$) is the resolution of budgets and bids. 
\end{definition}
Note that a bidding game $\tup{G,\eps}$ is in fact a collection of $\frac{1}{\eps}+1$ different games, one for every initial budget partition $B_1=0,\eps,2\eps,\ldots,1$. %The allowed strategies in a bidding game are defined in the next section. Before that, we provide an intuitive example.

\paragraph{Example---Majority} Consider the following  (zero-sum) game structure $G_{\maj}$, depicted in Figure~\ref{fig:Gmaj}. In this game there are at most three turns, and the winner is the player which plays at least twice. Formally, $S=(\{0,1,2\}\times \{0,1,2\})\setminus \{(2,2)\}$, and an element in $S$ is simply the number of times that each player played; $s_0=(0,0)$; $T=\{(s_1,s_2)\in S: s_1=2 \vee s_2=2\}$; $g(s_i,s_{-i})=\{(s_i+1,s_{-i}),(s_i,s_{-i}+1)\}$ for all $s\in S\setminus T$; and $u_i(t)=1$ if $t_i=2$ and $0$ otherwise. 

Note that if $G_{\maj}$ is played with alternating turns and without budgets or bidding, the white player can always win.  Consider $\tup{G_\maj,\eps}$ with $\eps = \frac{1}{800}$, and  $B_1=B_2=0.5$, i.e. where each player has an initial budget of $0.5$.  

%\begin{wrapfigure}{r}{0.45\textwidth}
\begin{figure}
\input{example_maj}
\end{figure}
%\end{wrapfigure}
The following is a possible game play:

\begin{itemize}
%\item $s_0 = (0,0)$.
	\item In turn~1, white bids $0.2$ and black bids $0.15$. White gains the turn, plays the round, and the new budgets are $0.3$ and $0.7$. $s = (1,0)$.
	\item In turn~2, white bids $0.14$ and black bids $0.26$. Black plays and pays $0.26$.  $s = (1,1)$, and budgets are updated to $(0.56,0.44)$.
	\item In the last turn, both players exhaust their budgets, so the player with higher remaining budget (white with $0.56$ vs. black with only $0.44$) plays and wins the game.  $s = (2,1) \in T$.
\end{itemize}

Note that a higher budget makes it easier to win in this game. For example, if $B_i>3(B_{-i}+\eps)$ for some player, then player $i$ can always gain the first two rounds by bidding more than the other player's budget. % and win the game.

%
%Note that if $G_{\maj}$ is played with alternating turns and without budgets or bidding, the white player can always win.  Consider $G_\maj$ with $B_1=B_2=8$, i.e. where each player has an initial budget of $8$ (thus $\ol B=16$).  
%
%\begin{wrapfigure}{r}{0.45\textwidth}
%\input{example_maj}
%\end{wrapfigure}
%The following is a possible game play:
%
%
%\begin{itemize}
%%\item $s_0 = (0,0)$.
	%\item In turn~1, white bids $6$ and black bids $4$. White gains the turn, plays the round, and the new budget partition is $B_1=4$ (which entails $B_2=16-4=12$). White chooses the next state $s_1 = (1,0) \in g(s_0)$.
	%\item In turn~2, white bids $3$ and black bids $5$. Black pays $5$ and chooses  $s_2 = (1,1) \in g(s_1)$. The budget partition is updated to $B_1=4+5=9$.
	%\item In the last turn, both players exhaust their budgets, so the player with higher remaining budget (white with $9$ vs. black with only $7$) plays and wins the game.  $s_3 = (2,1) \in T$.
%\end{itemize}
%
%
%Note that a higher budget makes it easier to win in this game. For example, if $B_i>3(B_{-i}+1)$ for some player, then player $i$ can always gain the first two rounds. % and win the game.

\paragraph{Strategies}
We next define the allowed strategies for players in bidding games, and the main solution concept we apply.
%Suppose players are in some node $s\in S$, with budgets $B_1,B_2$. %Formally, the node $s$ corresponds to three levels in the game tree $G|_s$. In the first level, both players submit their bids; in the second level, the winner decides whether she wants to play or relegate the turn; and in the third level either the winner or the player to whom the turn was relegated selects the next state. 
%Note that the second and third levels are ``standard'' in the sense that only one player acts, whereas the first level is a simultaneous step.

%We will denote by $\gamma$ a strategy profile for the two players Formally,
\begin{definition}\label{def:profile}
A \emph{strategy profile} in a bidding game $\tup{G,\eps}$ is a mapping
$$\gamma:S\times \calB \rightarrow N \times (\calB)^{2}  \times  \calB \times S^2 \times T,$$
where $\calB=\calB_\eps$ and $\gamma(s,B_1)$ specifies the actions taken in state $s$ under budget partition $(B_1,B_2)$ (recall that $B_1+B_2=1$ so the partition is determined by a single parameter).
\end{definition}
More specifically, for all $s\in S, B_1\in \calB$, $\gamma(s,B_1)=(i^*,b_1,b_2,B^*_{i^*},s^*_1,s^*_2,t)$, where: $i^*$ is the winner; $b_1$ and $b_2$ are the bids of both players; $B^*_{i^*}$ is the remaining budget of the winner; $s^*_i$ is the next state chosen by each $i$; and $t\in T$ (also denoted by $\gamma(\cdot)_T$) is the outcome.
 While some of these elements can be inferred from others, the ability to reference them directly will become useful later on.

\begin{definition}\label{def:valid}
$\gamma$ is a \emph{valid strategy profile} if for all $s\in S\setminus T$ and $B_1\in \calB$, $\gamma(s,B_1)=(i^*,b_1,b_2,B^*_{i^*},s^*_1,s^*_2,t)$ holds the following consistency constraints:
\renewcommand{\labelitemi}{$\bullet$}
\begin{itemize}
\item $b_1 \leq B_1$ and $b_{2}\leq B_{2}$. 
	\item $i^*=1$ if $b_1\geq b_2$ and otherwise $i^*=2$. That is, highest bidder wins the round with ties broken in favor of white.
	\item $B^*_{i^*} = B_{i^*}-b_{i^*}$. That is, the winner pays her bid to the other player.
	\item $s^*_1,s^*_2 \in g(s)$.
	\item $t = \gamma(s^*_{i^*},B^*_{i^*})_T$. That is, players reach to the same terminal from  state $s$ and from the next state $s^*_{i^*}$.
	\end{itemize}
For a terminal $t\in T$ we  define $\gamma(t,B_1)=(1,0,0,B_1,t,t,t)$ in every valid strategy. 
\end{definition}
We denote by $\Gamma(G,\eps)$ the set of all valid profiles in the bidding game $\tup{G,\eps}$.

%Since the strategy in each state is composite as explained above, we 

 %Thus the strategy of a player $i$ in a node $s$ must specify (a) her bid $b_i$; (b) conditional on gaining the turn, either a state $s^*\in g_i(s)$, or a $\bot$ action; (c) conditional on getting the turn from the other player and on the other player's bid, a state in $g_i(s)$. 

\subsection{Subgame perfect equilibria}
\label{sec:PSPE}
\begin{definition}\label{def:PSPE}
A strategy profile $\gamma$ is a \emph{pure subgame perfect equilibrium} (PSPE) in a bidding game $\tup{G,\eps}$, if it is a valid strategy profile, and each player plays a best-response strategy to the other player at any subtree, for any budget.
\end{definition}
We denote by $\Gamma^*(G,\eps)\subseteq \Gamma(G,\eps)$ the set of all PSPEs in bidding game $\tup{G,\eps}$.

Since there are many possible responses of a player in a given strategy profile, we use the following lemma to specify all of them explicitly. The proof is immediate, by noticing that the lemma covers all possible selections of bids and states.

%Since there are multiple types of actions, including some actions that are simultaneous, 

%\begin{definition}[PSPE]
\begin{lemma}\label{lemma:PSPE_def}
A valid strategy profile $\gamma$ is a PSPE in the bidding game $\tup{G,\eps}$ if and only if the following incentive constraints hold for all $s\in S\setminus T$ and $B_1\in \calB$ (where player~$i^*$ is the winner in $\gamma(s,B_1)$):
	\begin{itemize}
	%\item $s^*_i \in \ar ,,  dsagmax_{s'\in g(s)}u_i(\gamma(s',B^*_i)_T)$. That is, the next state is optimal for the winner (given her remaining budget).
	%\item $u_i(t) \geq \max_{b'_i\gti b_{-i}}u_{i}(\gamma(s^*_i,B_i-b'_{i})_T)$. That is, the winner $i$ cannot benefit by using a different bid and still win the turn.
	\item For any $s'_{i^*}\in g(s)$ and any $b'_{i^*}\in \calB$ s.t. $b'_{i^*}\leq B_{i^*}$, $b'_{i^*} > b_{-i^*}$ (weak inequality when $i^*=1$), it holds that  $u_{i^*}(t) \geq u_{i^*}(\gamma(s'_{i^*},B_{i^*}-b'_{i^*})_T)$. That is, the winner $i^*$ cannot gain by still winning the turn and use a different bid and/or action.
	\item For any $s'_{-i^*}\in g(s)$ and any $b'_{-i^*}\in \calB$ s.t. $b'_{-i^*}\leq B_{-i^*}$, $b'_{-i^*} > b_{i^*}$ (weak inequality when $i^*=2$), it holds that $u_{-i^*}(t) \geq u_{-i^*}(\gamma(s'_{-i^*},B_{i^*}+b'_{-i^*})_T)$. That is, the loser $-i^*$ cannot gain by increasing his bid and win the turn.
		\item $u_{i^*}(t) \geq  u_{i^*}(\gamma(s^*_{-i^*},B_{i^*}+b_{-i^*})_T)$.  That is, the winner $i^*$ cannot gain by lowering her bid and losing the turn (does not apply if $i^*=1$ and $b_2=0$). Note that that the next state selected by  the loser ($s^*_{-i^*}$) must be part of the description of the strategy.
\end{itemize}
\end{lemma}
\renewcommand{\labelitemi}{$-$}
We emphasize that a strategy profile determines the actions and outcome for any budget partition in any internal node, and in particular for any initial budget partition.
For each $s\in S\setminus T$ denote $\mu_s(B_1) = \gamma(s,B_1)_T$.  
Therefore, every profile $\gamma$ for $\tup{G,\eps}$ induces a mapping $\mu_\gamma$ from $\calB_\eps$ to outcomes $T(G)$. That is, $\mu_\gamma(B_1) = \gamma(s_0,B_1)_T$.
 We say that the profiles $\gamma,\gamma'$ are \emph{outcome-equivalent} in $\tup{G,\eps}$ if $\gamma(s,B_1)_T=\gamma'(s,B_1)_T$ for all $s\in S,B_1\in \calB_\eps$.

\section{Existence and Structure of Equilibria}
\label{sec:exist}

We start by  showing the following basic result.
For $s\in S\setminus T$, denote by $\gamma|_s$ the profile $\gamma$ on the subtree rooted in $s$. Clearly if $\gamma$ is a PSPE then so is $\gamma|_s$. 
\begin{theorem}
\label{th:PSPE_exist}
Any bidding game $\tup{G,\eps}$ has a PSPE, i.e. $\Gamma^*(G,\eps)\neq \emptyset$. 

Further, if for all $s_j\in g(s_0)$ there is a PSPE $\gamma^{(j)}$ in $\tup{G|_{s_j},\eps}$, then there is a PSPE $\gamma'$ in $\tup{G,\eps}$ s.t. for all $s_j\in g(s_0)$, $\gamma'|_{s_j}=\gamma^{(j)}$.
\end{theorem}

Intuitively, we construct such an equilibrium by traversing the game tree bottom-up, where in each node the players simulate an ascending auction (for every possible budget) to determine the equilibrium bids. We emphasize that in the actual bidding game bids are simultaneous, but we use an ascending auction in order to determine the equilibrium bids.  

\subsection{Existence of PSPE}

Let a \emph{first-price game} be a two-player one-shot game of the following form: each player submits a bid $b_i\in \calB$, and the utilities of both players (denoted $u_1$ and $u_2$) are an arbitrary function of the higher bid.

Given a bid profile $(b_1,b_2)$, a \emph{better-reply} of player~$i$ is a bid $b'_i$ that results in a strictly better outcome, i.e., such that $u_i(\max\{b'_i,b_{-i}\}) > u_i(\max\{b_i,b_{-i}\})$.

\begin{lemma}
\label{lemma:auction}
Consider any sequence of alternating better-replies in a first-price game. Then after the first time a bidder raises her bid, no bidder ever lowers her bid.
%A first-price game  where $\calB$ is finite has a pure Nash equilibrium. Moreover, any sequence of better-replies must converge. % (i.e., the game has an ordinal potential).
\end{lemma}

\begin{proof}
Recall that tie-breaking is in favor of bidder~1.
%Assume, toward a contradiction, that there is a cycle of improvements. 
Suppose that at some step bidder~1 increases her bid from $(b_1,b_2)$ to $b'_1$. Note that $b'_1\geq b_2$, as otherwise $b_1 < b'_1 < b_2$ and $(b_1,b_2);(b'_1,b_2)$ lead to the same outcome (if the first increase is by bidder~2 then $b'_2>b_1$). Any further step of bidder~1 must keep $b'_1\geq b_2$, otherwise we just go back to the outcome of $(b_1,b_2)$ (which is weakly worse for bidder~1). 

Consider the next reply by bidder~2---he must also increase his bid to some $b'_2>b'_1\geq b_2$ in order to have any effect. Then bidder~1 must increase again and so on. Thus no bidder ever reduces her bid (after the first increase). % and budgets are bounded, the process must converge to a pure Nash equilibrium.
\end{proof}

Let $\tup{G,\eps}$ be a bidding game, $(s,B_1)$ be some state and budget, and $\boldsymbol\gamma=(\gamma_j)_{s_j\in g(s)}$ be some action profile in all the subgames $G|_{s_j}$.
We observe that $\tup{G,\eps},(s,B_1)$ and $\boldsymbol\gamma$ together induce a first-price game played in state $(s,B_1)$:
\begin{itemize}
	\item In each $s_j$ consider the mapping $\mu_j :\calB \rightarrow T(G)$;
	\item For every $B^*_i\in \calB_\eps$, let $s^*_i(B^*_i) \in \argmax_{s_j\in g(s)}u_i(\mu_j(B^*_i))$, i.e., the best selected state for $i$ under remaining budget $B^*_i$;
  \item  For every $b_1 \geq b_2$, the outcome is $t^*=\mu_{s^*_1(B_1-b_1)}(B_1-b_1)\in T(G)$, and thus utilities $u_1(t^*),u_2(t^*)$ are only determined by the higher bid $b_1$. Similarly when $b_2>b_1$.  
\end{itemize}
We denote this first-price game by $G(\eps,s,B_1,\boldsymbol \gamma)$.

\begin{proof}[Proof of Theorem~\ref{th:PSPE_exist}]
Let $k=\height(G)$.
We prove by induction on the height of the tree. The base case is trivial since in a tree of height $0$, $S\setminus T$ is empty.

We first set the next states players select in $s$.  Let $g(s) = \{s_1,\ldots,s_q\}$. By the induction hypothesis, for each subgame $G|_{s_j}$ there is some PSPE $\gamma^{(j)}$. 
This defines a first-price game $G(\eps,s,B_1,\boldsymbol \gamma)$ played in state $(s,B_1)$, where $\boldsymbol\gamma=(\gamma^{(j)})_{s_j\in g(s)}$.

Suppose we initialize both bids to 0. By Lemma~\ref{lemma:auction}, bids can only increase. % By Lemma~\ref{lemma:intervals}, whenever white increases her bid $b_1$, she can w.l.o.g. set the bid $b'_1$ so that $B_1-b'_1$ is a multiple of $\eps$. 
Thus there is only a finite number of bid increments, from any initial state.\footnote{This uses the fact that there is only a finite number of possible bids. In the continuous case there are some caveats, see Section~\ref{sec:real}.} We then set the bids in $\gamma'(s,B_1)$ to be the equilibrium bids of the first-price game, and action $s^*_i=\argmax_{s_j\in g(s)}u_i(\mu_{s_j}(B_i-b_i))$ for $i\in \{1,2\}$. We complete the definition of $\gamma'$ by setting $\gamma'|_{s_j} = \gamma^{(s_j)}$.
\end{proof}
%We further note that if $\mu_j$ is monotone for all $s_j$ (that is, more budget never hurts the player), then the minimal increase in Algorithm~\ref{alg:PSPE} takes a particularly simple form. If a player gains by bidding $b'_i > b_i$, it is always best to increase by the minimal amount that guarantees winning the turn (and leaves the player with maximal remaining budget). Thus all increments are by one unit of $\eps$: The black player always increases to $b'_2 = b_1+\eps$, and the white player always increases to $b'_1 = b_2$. 

\subsection{Genericity and uniqueness}
\label{sec:gen_uniq}
Generic games are games where agents have strict preferences over all outcomes.
In classical extensive-form games (without bidding), it is known that genericity entails the existence of a unique PSPE. %It can be similarly shown that weak genericity entails that all PSPEs reach the same outcome (possibly via different paths) in a DAG. 
However in our game there are  simultaneous steps, and thus genericity may not be sufficient for uniqueness.

Our next example shows that in the general case, there might be multiple PSPEs that are not outcome-equivalent.

\begin{proposition}
\label{th:two_PSPE}
There is a generic bidding game with PSPEs that lead to different outcomes under the same budget.
\end{proposition}
\input{example_two_PSPEs_b}
\begin{proof}
Consider the game $G_{\two}$ in Fig.~\ref{fig:two_PSPE}, with some high budget resolution.

We first describe a single equilibrium  $\gamma|_x$ in the subtree of node $x$. 
Any player with budget more than $0.5$ (or $\geq 0.5$ for white) after bidding at $x$ clearly goes to $y$ and gets $9$, while the other player gets $1$. The player $i$ with the higher budget at $x$ bids the difference $b_i=B_i(x)-0.5$ if $i=1$ and $b_i=B_i(x)-0.5-\eps$ if $i=2$, and selects $y$. The other player $-i$ bids $\min\{B_{-i},b_i\}$ if $i=2$ and $\min\{B_{-i},b_i+\eps\}$ if $i=1$, and selects $(5,5)$. 
 Thus if the budget is $B_1(x)\in[0.25,0.75-\eps]$, the outcome is $(5,5)$; if $B_1(x)<0.25$ the outcome is $(1,9)$; and if $B_1(x)\geq 0.75$ the outcome is $(9,1)$. Note that this is a PSPE at $x$: if $y$ is selected then the loser at $x$ cannot raise, 
and if $(5,5)$ is selected and the loser raises the bid to win, they will get to $y$ with not enough budget to win again.
 
\medskip
Next, we set the initial budget at $s_0$ to $(0.5,0.5)$, and describe two distinct equilibrium bids at $s_0$.

In the first PSPE $\gamma^*$,  the bids at $s_0$ are $b_1=b_2=0$, and both players select $x$. White wins and remains with budget $0.5$.  According to $\gamma|_x$, the game ends at terminal  $(5,5)$ (marked with $*$). This is clearly a PSPE.

% with bids of $(0,\eps)$ in $x$, since if the winner in $x$ goes to $y$ he will lose the next round and get $1$. We note that black would bid strictly above $0$, since otherwise white would bid $0$ and still reach $(9,1)$. 

\medskip
In the second PSPE $\gamma^{**}$, the bids at $s_0$ are $b_1=b_2=0.5$, and both players select $(2,2)$ (marked with $**$). %consider a PSPE that starts with bids $b_1=b_2=0.5$. %White wins the turn and remains with a budget of $0$. If she selects $x$, then black can play two turns in a row and reach $(1,9)$. Thus white will select terminal $(2,2)$ .  
This is an equilibrium, since white remains with budget $0$ after winning at $s_0$, and selecting $x$ means a utility of $1<2$ for her. Black selects $(2,2)$ for the same reason, and cannot deviate by changing his bid. Finally, if white lowers her bid the outcome does not change.
\end{proof}

\paragraph{Enforcing generic preferences} In the general case, it is possible that a player is indifferent between two outcomes $t,t'\in T$. In such cases, we will assume throughout the paper that the player has a strict preference towards the outcome that is also better for the other player.\footnote{To see why this assumption is necessary, note that without it even the most simple game with a single decision node and two terminals may result in a Pareto-dominated outcome if the player with a higher budget is indifferent between the two terminals.}
 Thus every player has a strict preference order over all distinct outcomes. Recall that $t\succ_i t'$ means that player $i$ strictly prefers $t$ over $t'$. In the remainder of the paper, $t\succ_i t'$  if either $u_i(t)>u_i(t')$, or $u_i(t)=u_i(t')$ and $u_{-i}(t)>u_{-i}(t')$. We highlight that this assumption only increases the number of potential deviations from a given state, and thus cannot add new equilibria. %As the example above shows, even after this reduction there may be more than one PSPE. 

\subsection{The PSPE semilattice}
In general, $\Gamma(G,\eps)$ may contain multiple PSPEs. However, PSPEs can be partially ordered according to the bids of the players. We define a partial order over all valid profiles. 
Intuitively, profiles with lower winning bids will be lower in this order, however since each profile consists of many bids at different states, we need to construct the partial order carefully. In particular, profiles with smaller gaps between bids will be lower, and the order is first decided over nodes closer the bottom. Only if both profiles are equivalent in all nodes at some level or deeper, we check the bids at a higher level for a difference. 

Given strategy profile $\gamma$ and state $s\in S$, the function $\gamma_s$ maps any budget $B_1\in \calB$ to bids $b_1$ and $b_2$, where $\gamma_s(B_1) = (\gamma(s,B_1)_{b_1},\gamma(s,B_1)_{b_2})$ (do not confuse with $\gamma|_s$, which is the restriction of $\gamma$ on the subtree rooted in $s$).

\begin{definition} \label{def:lattice}The partial order $\pi$  over $\Gamma(G,\eps)$ is defined as follows.
\begin{itemize}
	\item For a given budget level $B_1$, $\gamma'(s,B_1) >_\pi \gamma(s,B_1)$ if either:
	\begin{enumerate}
		\item the winning bid in $\gamma'$ is \emph{higher}, i.e. $\max\{b'_1,b'_2\} > \max\{b_1,b_2\}$;
		\item the winning bids are equal, and the losing bid  in $\gamma'$ is \emph{lower};
		\item $b'_1=b_2 > b_1=b'_2$ (i.e. bids are the same), but white wins in $\gamma'(s,B_1)$.
	\end{enumerate}
	Thus for every $(s,B_1)$, $\pi$ is a linear order over profiles, and in particular over PSPEs. 
	\item For $s\in S\setminus T$, we define $\gamma'_s >_\pi \gamma_s$ if $\gamma'(s,B_1) \geq_\pi \gamma(s,B_1)$ for all $B_1\in \calB$, with at least one strict inequality. Thus for each node $s$, $\pi$ is only a partial order over profiles (see Fig.~\ref{fig:lattice_tab} for an example). 
	\item For $s\in S$, we define the partial order $\sigma(s)$ recursively (see Fig.~\ref{fig:lattice_tree} for an example): 
	\begin{itemize}
	    \item $\gamma' =_{\sigma(t)} \gamma$ for all $t\in T$;
		  \item $\gamma' >_{\sigma(s)} \gamma$ if both:
		  \begin{enumerate}
		       \item $\gamma' \geq_{\sigma(s')} \gamma$ for all $s'\in g(s)$, and
					 \item either $\gamma' >_{\sigma(s')} \gamma$ for some $s'\in g(s)$, or $\gamma'_s >_\pi \gamma_s$.
		  \end{enumerate}
	\end{itemize}		
			%\item If $g(s)\subseteq T$ (i.e., $s$ is a leaf of $S\setminus T$) then $\gamma' >_{\pi(s)} \gamma$ iff herwise, $\gamma' >_{\pi(s)} \gamma$ if $\gamma' \geq_{\pi(s')} \gamma$ for all $s'\in g(s)$ with at least one strict 
		%\end{enumerate}
	%$\gamma' >_{\pi(s)} \gamma$ if $
	%\item Define an arbitrary postorder $O$ over nodes of $G$ (i.e., $s' \prec_O s$ for any $s\in S,~s'\in g(s)$). We define $\gamma' >_\pi \gamma$ if there is a node $s^*\in S\setminus T$ such that:
		%\begin{enumerate}
			%\item $\gamma'_s=\gamma_s$ for all $s \prec_O s^*$; and
			%\item $\gamma'_{s^*} >_\pi \gamma_{s^*}$.
		%\end{enumerate}
		\item Finally, $\gamma' >_\pi \gamma$ if $\gamma' >_{\sigma(s_0)} \gamma$.
\end{itemize}
\end{definition}

\begin{figure}
\begin{framed}
$$
\begin{array}{|c|cccccc|}
\hline
 B_1       &  0        & \eps    & 2\eps       & \ldots & 1-\eps & 1 \\
\hline
\gamma^1 &   - &(0,4\eps)&(2\eps,2\eps)& \ldots  & - & - \\
\gamma^2 &   - &(\eps,4\eps)&(\eps,2\eps)& \ldots  & - & - \\
\gamma^3 &   - &(0,5\eps)&(0,5\eps)& \ldots  & - & - \\
\gamma^4 &   - &(0,4\eps)&(2\eps,6\eps)& \ldots  & - & - \\
\hline
\end{array}
$$
\caption{\label{fig:lattice_tab}The table shows the bids under four different profiles at state $s$, for various budget levels. Suppose that the bids for all $B_1\notin \{\eps,2\eps\}$ are the same under all profiles. \\
From the table we can infer the following linear orders over profiles at given budget levels: \\
$\gamma^2(s,\eps) <_\pi \gamma^1(s,\eps) =_\pi \gamma^4(s,\eps) <_\pi \gamma^3(s,\eps)$; and $\gamma^1(s,2\eps) <_\pi \gamma^2(s,2\eps) <_\pi \gamma^3(s,2\eps) <_\pi \gamma^4(s,2\eps)$.\\
Next, we can infer a partial order over profiles at state $s$: $\gamma^1_s,\gamma^2_s <_\pi \gamma_s^3,\gamma_s^4$. However the relation e.g. between $\gamma^1_s$ and $\gamma^2_s$ is not defined. }
\end{framed}
\end{figure}
 \input{PSPE_lattice.tex}

Any partial order on $\Gamma(G,\eps)$ is in particular a partial order on $\Gamma^*(G,\eps)$. Note that $\pi$ ignores the part of the strategy profile that specifies the selection of the next state. We remark that once bids are set, due to the genericity assumption there is only one way to specify state selection \emph{in equilibrium}. Thus this would not pose a problem when comparing two PSPEs from $\Gamma^*(G,\eps)$.

\begin{theorem}\label{th:lattice}
 $(\Gamma^*(G,\eps),\pi)$ is a meet-semilattice.
\end{theorem}
\begin{proof}
Given two PSPEs $\gamma,\gamma'$ for the same game $\tup{G,\eps}$, we define their meet $\hat \gamma$ as the highest PSPE $\hat \gamma$ such that $\hat \gamma\leq _\pi \gamma$ and $\hat \gamma\leq_\pi \gamma'$. The meet $\hat \gamma = \gamma \wedge \gamma'$ is unique, and is computed by Algorithm~\ref{alg:meet}.
Intuitively, the algorithm traverses both PSPEs from the bottom up, and takes the ``lower'' bids from each pair $\gamma(s,B_1),\gamma'(s,B_1)$ according to the order $\pi$. 

%
%
%\begin{small}
%\begin{algorithm}[h]
%\caption{\label{alg:meet}\textsc{Compute-meet}($\gamma^1,\gamma^2$)}
%Initialize all terminal labels in $T$ to $\{1,2\}$\;
%Initialize all node labels in $S\setminus T$ to $\emptyset$\;
%%Denote by $S_d$ all leafs whose depth (distance from the root) is $d$\; 
%\For{every node $s\in S\setminus T$  in post-order} {
	   %\If{all labels in $g(s)$ contain `$1$' and $\gamma_s^1 \leq_\hfill\tcp{Case~1.1}} { 
		   %set $\hat \gamma_s = \gamma^1_s$\;
			%}
			  %\If{all labels in $g(s)$ contain `$2$' \hfill\tcp{Case~1.2}} { 
		   %set $\hat \gamma_s = \gamma^2_s$\;
			%}
	   %\If{all of $g(s)$ are untouched, and $\gamma_s \neq \gamma'_s$\hfill\tcp{Case~2}} { 
		     %\For{all $B_1\in \calB$} {
		         %set $\hat \gamma(s,B_1)=\min_{\pi}\{\gamma(s,B_1), \gamma'(s,B_1)\}$\;
				  %}
					%change $s$ to `touched'\;
		 %}	
%
	   %\If{some  $s_j\in g(s)$ are touched \hfill		\tcp{Case~3}}{
		   %Let $\boldsymbol\gamma=(\hat \gamma|_{s_j})_{s_j\in g(s)}$\;
		   %\If{all of $\hat \gamma|_{s_j}$ are PSPEs} {
			    %\For{all $B_1\in \calB$} {
					      %%Consider the first-price game $G(\eps,s,B_1,\boldsymbol \gamma)$\;
							 %%\If{the game $G(\eps,s,B_1,\boldsymbol \gamma)$ has an equilibrium} {
							 %Set $\hat \gamma(s,B_1)$ to the maximal equilibrium of the first-price game $G(\eps,s,B_1,\boldsymbol \gamma)$ according to $\pi$\;
					     %%}
							%%\Else{FAIL}
					%}
					%change $s$ to `touched'\;
				%} 
				%\Else{FAIL}
		 %}
%
		%}
%\end{algorithm}
%\end{small}
%
%
\begin{small}
\begin{algorithm}[h]
\caption{\label{alg:meet}\textsc{Compute-meet}($\gamma,\gamma'$)}
Initialize all nodes as `untouched'\;
%Denote by $S_d$ all leafs whose depth (distance from the root) is $d$\; 
\For{every node $s$  in post-order} {
	   \If{all of $g(s)$ are untouched, and $\gamma_s=\gamma'_s$\hfill\tcp{Case~1}} { 
		   set $\hat \gamma_s = \gamma_s$\;
			}
	   \If{all of $g(s)$ are untouched, and $\gamma_s \neq \gamma'_s$\hfill\tcp{Case~2}} { 
		     \For{all $B_1\in \calB$} {
		         set $\hat \gamma(s,B_1)=\min_{\pi}\{\gamma(s,B_1), \gamma'(s,B_1)\}$\;
				  }
					change $s$ to `touched'\;
		 }	

	   \Else{ \tcp{some  $s_j\in g(s)$ are touched} 
		  	
		     Let $\boldsymbol\gamma=(\hat \gamma|_{s_j})_{s_j\in g(s)}$\;
			   \uIf{$\hat \gamma|_{s_j} = \gamma|_{s_j}$ for all $s_j\in g(s)$\hfill\tcp{Case~3.1}}{
				      $\hat \gamma_s = \gamma_s$\;
			   }
				 \uElseIf{$\hat \gamma|_{s_j} = \gamma'|_{s_j}$ for all $s_j\in g(s)$\hfill\tcp{Case~3.2}}{
				       $\hat \gamma_s = \gamma'_s$\;
				 }
		     \uElseIf{all of $\hat \gamma|_{s_j}$ are PSPEs\hfill\tcp{Case~4}} {
			        \For{all $B_1\in \calB$} {
					      %Consider the first-price game $G(\eps,s,B_1,\boldsymbol \gamma)$\;
							 %\If{the game $G(\eps,s,B_1,\boldsymbol \gamma)$ has an equilibrium} {
							     set $\hat \gamma(s,B_1)$ to the maximal equilibrium of the first-price game $G(\eps,s,B_1,\boldsymbol \gamma)$ according to $\pi$\;
					     }
					
				 } 
				 \Else{FAIL}
				 change $s$ to `touched'\;
		 }
}
\end{algorithm}
\end{small}

Let $\hat \gamma = \hat \gamma(\gamma,\gamma')$ be the output of Algorithm~\ref{alg:meet} on input $\gamma,\gamma'$.
To prove the theorem, we need to show that:
\begin{itemize}
	\item the meet operation is well defined (i.e., that Algorithm~\ref{alg:meet} never fails);
	\item $\hat \gamma$ is a PSPE;
  \item $\hat \gamma=\gamma \wedge \gamma'$;
	%\item the meet operation holds idempotency ($\gamma\wedge\gamma=\gamma$), commutativity, and associativity.
\end{itemize}
The algorithm never fails since by induction by the time we reach node $s$, we already have that all of $\hat \gamma|_{s'}$ for $s'\in g(s)$ are PSPEs. 

To see why $\hat \gamma$ is a PSPE, we assume by induction that this holds for all nodes below $s$, and go over all cases of the construction in Algorithm~\ref{alg:meet}. In Cases~1 and 3, $\hat \gamma|_s=\gamma|_s$ (or $\gamma'|_s$) and is thus a PSPE.
 In Case~2, for every $B_1\in \calB$ both $\gamma_s(B_1)$ and $\gamma'_s(B_1)$ are equilibrium bids, so it does not matter which one we select (note that for all $s'\in g(s)$, $\hat \gamma_{s'}=\gamma_{s'}=\gamma'_{s'}$). Since $\hat \gamma_s(B_1)$ are equilibrium bids for all $B_1\in \calB$, $\hat \gamma|_s$ is a PSPE. 
%Cases~3.1 and 3.2 occur when $\gamma <_{\sigma(s)} \gamma'$ or vice versa according to children of $s$, setting $\hat \gamma$ to the smaller one.
  In Case~4 (when $\gamma,\gamma'$ are incomparable by their children), by Prop.~\ref{th:PSPE_exist} it is possible to complete the bids for any budget level $B_1$ such that $\hat \gamma(s,B_1)$ is a PSPE. As for every state $(s,B_1)$ the relation $\pi$ is a linear order over all profiles, $\hat \gamma(s,B_1)$ is uniquely defined.  In fact, it is composed of the highest winning bid $b_{i^*}\leq B_{i^*}$ and the lowest losing bid $b_{-i^*}$ such that $(b_{i^*},b_{-i^*})$ is an equilibrium of the first-price game $G(\eps,s,B_1,\boldsymbol \gamma)$.
	
	\medskip
\def\dhat{\doublehat}
To show that $\hat \gamma=\gamma \wedge \gamma'$ we need to show that $\hat \gamma$ is the ``largest'' PSPE that is still weakly smaller than both  $\gamma,\gamma'$. That is, (i) $\hat\gamma\leq_\pi \gamma,\gamma'$; and (ii) for any  $\dhat \gamma \leq_\pi \gamma,\gamma'$, it also holds that $\dhat \gamma \leq_\pi \hat \gamma$. % which is immediate from the construction in Case~1 and Case~2 (regardless of the construction in higher nodes). 

We prove (i) and (ii) together. Suppose that $\dhat \gamma \leq_\pi \gamma,\gamma'$, then in particular $\dhat  \gamma \leq_{\sigma(s)} \gamma,\gamma'$ at all $s\in S$.
Assume by induction that $\dhat  \gamma \leq_{\sigma(s_j)} \hat\gamma \leq_{\sigma(s_j)} \gamma,\gamma'$ for all $s_j\in g(s)$  (this vacuously holds at the leafs). 

%For (ii),  and assume by induction that $\dhat  \gamma \leq_{\sigma(s_j)} \hat  \gamma$ at all $s_j\in g(s)$. 
 Thus,  condition~1 in Def.~\ref{def:lattice} holds, and it remains to show that one of the two alternatives for condition~2 also holds, i.e., either the inequality in some children is strict, or there is (weak) inequality in $s$ as well.

Specifically, if some of the inequalities $\hat  \gamma \leq_{\sigma(s_j)} \gamma$  are strict (the first alternative), then $\hat  \gamma <_{\sigma(s)} \gamma$ (and same for $\gamma'$) and we are done with (i). 
Similarly, if some of the inequalities $\dhat  \gamma \leq_{\sigma(s_j)} \hat\gamma$  are strict, then $\dhat  \gamma <_{\sigma(s)} \hat  \gamma$ and we are done with (ii). 

To complete (i), note that if $\hat \gamma_s$ is constructed in Cases~1 or 2, then $\hat \gamma_s\leq_\pi \gamma_s,\gamma'_s$ and the second alternative of condition~2 holds. If Case~3.1 holds then $\hat\gamma|_{s_j},\gamma'|_{s_j}$ must differ at some $s_j\in g(s)$, which means the first alternative holds for $\gamma'$ and the second alternative (with equality) holds for $\gamma$ (in Case~3.2 we get the opposite). If Case~4 holds, then $\hat \gamma$ differs from both $\gamma,\gamma'$ in at least one child $s_j$ (may be in different children). 
In either case we get that  $\hat  \gamma \leq_{\sigma(s)}  \gamma,\gamma'$. When applied to the root $s_0$ we get  $\hat  \gamma \leq_{\pi}  \gamma,\gamma'$.

To complete (ii), note that  when there are no strict inequalities in the children of $s$, both of $\hat \gamma_s$ and $\dhat\gamma_s$ are equilibrium bids in the same first-price game $G(\eps,s,B_1,\boldsymbol\gamma)$. 
 We  show that $\dhat  \gamma_s \leq_{\pi} \hat   \gamma_s$ in each of the four cases by which $\hat \gamma_s$ was constructed. In Cases~1 and 3 this is holds since $\hat\gamma_s$ equals one of $\gamma_s,\gamma'_s$, and $\dhat\gamma_s\leq_\pi \gamma_s,\gamma'_s$. In Cases~2 and 4, by construction for all $B_1$, and since $\pi$ is a linear order over all bidding profiles in $(s,B_1)$, we have $\dhat \gamma(s,B_1) \leq_\pi \hat \gamma(s,B_1)$, and thus $\dhat  \gamma_s \leq_{\pi} \hat  \gamma_s$ as well. This means that the second alternative for condition~2 holds (with weak inequality), and thus $\dhat  \gamma \leq_{\sigma(s)} \hat  \gamma$. In particular $\dhat  \gamma \leq_{\pi} \hat  \gamma$ as required.
%We get idempotency since if $\gamma=\gamma'$ then all nodes remain untouched and $\hat \gamma=\gamma$. We get commutativity and associativity in Case~2 since the $\min$ operation has these properties.  In Case~3, the equilibrium of the first-price game only depends on $\boldsymbol \gamma$, and by induction $\gamma_{s_j}\wedge\gamma'_{s_j}= \gamma'_{s_j}\wedge\gamma_{s_j}$. and likewise for associativity.
\end{proof}

We demonstrate the computation of the meet through an example. Suppose that we compute the meet of $\gamma^3,\gamma^4$ from Figure~\ref{fig:lattice_tab} at node $s$. Assuming that all nodes below $s$ are untouched, we are in Case~2. Then at each budget level we take the pair of bids that is lower, getting $\hat \gamma(s,\eps) = \gamma^4(s,\eps)=(0,4\eps)$, $\hat \gamma(s,2\eps) = \gamma^3(s,2\eps)=(0,5\eps)$.

Next, suppose that the three profiles in Fig.~\ref{fig:lattice_tree} are PSPEs, and consider the meet $\hat \gamma= \gamma' \wedge \gamma''$. We have that $\hat \gamma|_{x}=\gamma''|_{x}$ due to Case~2.  Similarly, $\hat \gamma|_{y}=\gamma'|_{y}$. We then complete $\hat \gamma$ in the other nodes $x_1,xy,s_0$ as specified. Note that as $\gamma',\gamma'' >_\pi \gamma$, their meet $\hat \gamma \geq_\pi \gamma$ as well (in this example it holds with a strict relation).

\medskip
Since $\Gamma(G,\eps)$ is finite, $\Gamma^*(G,\eps)$ is also finite and thus by Theorem~\ref{th:lattice}, $\pi$ has a unique minimum, which we refer to as the \emph{Bottom Equilibrium}. We denote the Bottom Equilibrium of the bidding game $\tup{G,\eps}$ by $BE(G,\eps)$. 
\def\IB{{\cal{IB}}}

\section{Efficiency, Monotonicity, and Binary Trees}
\label{sec:main}
Having established the existence of PSPEs, and in particular the Bottom Equilibrium, we next turn to study the properties of this outcome, how it depends on the budget, and whether it is desired from the perspective of the players. 
This is the primary section of this work, where we lay out our main positive result.
%As we will later see, in the general case some PSPE solutions may be quite unwelcome. On the other hand, there is a wide class of games where a highly desireable PSPE exists. 

\begin{definition}\label{def:monotone_PSPE}
A profile $\gamma\in\Gamma(G,\eps)$ is \emph{monotone}, if $\mu_\gamma(B'_1) \succeq_1 \mu_\gamma(B_1)$ whenever $B'_1>B_1$, and likewise for player~2.
\end{definition}

Let $T_P(G)\subseteq T(G)$ be the set of Pareto-efficient outcomes in $G$. That is, $t\in T_P(G)$ if there is no $t'\in T(G)$ s.t. $t' \succ_1 t,t'\succ_2 t$.\footnote{Recall that by our assumption $\succ_1$ and $\succ_2$ are strict orders.}

\begin{definition}\label{def:PO}
A profile $\gamma\in\Gamma(G,\eps)$ is \emph{Pareto-optimal}, if $\mu_\gamma(B_1)\in T_P(G)$ for all $B_1\in \calB$.
\end{definition}

\begin{definition}\label{def:PS}
A profile $\gamma\in\Gamma(G,\eps)$ is \emph{Pareto-surjective}, if for all $t\in T_P(G)$, there is some budget $B_1\in \calB$, s.t. $\mu_{\gamma}(B_1) = t$.    
\end{definition}
%
%
%The next two properties are \emph{desirable} properties of bidding games:
%
%\begin{definition}[Pareto]
%\label{def:pareto}
%A game $G$ is \emph{Pareto-possible} if there exists a PSPE in $G$ that yields a Pareto-optimal outcome under any initial budget partition. $G$ is \emph{Pareto-necessary} if this holds w.r.t. any PSPE.
%\end{definition}
%Note that monotonicity of a particular PSPE is straight-forward. Definition~\ref{def:mon} is required since there may be more than one PSPE in a game.

Intuitively, our main result is that \emph{binary} bidding games always have a highly desirable PSPE (the Bottom Equilibrium), which has all the above properties. %We will first show that if symmetry  is relaxed, then the game may violate Pareto-possibility. 
We will first show that if the game is not binary, none of these properties is  guaranteed. Violating Pareto-surjective is trivial: just consider a root with three children, whose utilities are $(1,3),(2,2)$,and $(3,1)$. There is no equilibrium where $(2,2)$ is selected.

%The game $G_{\two}$ already demonstrates that a bidding game may not be Pareto-necessary, even if it is both binary and symmetric (since $**$ is Pareto dominated by $*$). 
%Consider the games $G_a,G_b$, depicted in Figure~\ref{fig:bad_PSPE}, respectively. The game $G_a$ is not binary. In game $G_b$ there are nodes where only one player can play. 
\begin{proposition}
\label{th:bad_PSPE_binary}
There is a (non-binary) bidding game, where every PSPE violates monotonicity and Pareto-optimality.
\end{proposition}

\begin{proof}
%\begin{wrapfigure}{r}{0.36\textwidth}
\begin{figure}[h]
\input{example_bad_PSPE}
\end{figure}
%\end{wrapfigure}
Consider $G_{\text{bad}}$ (Fig.~\ref{fig:bad_PSPE}) with some high resolution, and initial budget $B_2=0$. Clearly white can play at will, and thus the only PSPE leads to the outcome $(10,7)$ (marked). Denote by $B_1(x)$ the budget at node $x$.
Note that if players reach $x$, then the outcome will be $(10,7)$ if $B_1(x)\geq 0.75$ and otherwise $(0,9)$. 
Now suppose that initial budgets are $B_1=0.8,B_2=0.2$. If black takes the first round, he will choose $(1,8)$, since by choosing $x$ we will have $B_1(x)\geq B_1 = 0.8 > 0.75$. In order to get to $(10,7)$, white has to get to $x$ with at least $0.75$, that is to bid $b_1\leq 0.05$. However if $b_1<0.2 = B_2$, then black will overbid and take the turn. Thus in every PSPE white must bid at least $b_1=0.2$ in $s_0$, and play $(2,1)$.  This contradicts both monotonicity and Pareto:  the utility of black dropped from $7$ to $1$ when his budget increased from $0$ to $0.2$, and $(2,1)$ is Pareto-dominated by $(10,7)$. 
\end{proof}

\paragraph{Budget resolution}
Without sufficient budget (high resolution), the bidding process may not be powerful enough to get the efficient outcome (see Section~\ref{sec:discrete_mon}). However we argue that with high resolution, we can think of the budget interval as continuous, and this will be useful (in fact crucial) in our proofs later on. This idea can be demonstrated by looking at small game trees.

Consider a game tree with a single decision node $s$ (height~1). It is clear that regardless of the budget resolution $\eps$, it only matters whether $B_1\geq 0.5$, since this will determine who can win the turn. Similarly, if we consider a game of height~2, then there are four relevant ``budget intervals,'' namely $[0,0.25),~[0.25,0.5),~[0.5,0.75)$ and $[0.75,1]$: in the first one the black player can win both turns, in the second black can choose at least one victory, and so on.
%More generally, we can define this property as follows.
We next provide a general definition of this property. 

Recall that by our high resolution assumption, $\eps = \frac{1}{R \cdot 2^k}$ for some $R\geq 4$. %\rmr{needed?}We define $\IB=\IB(k)=\calB_{2^{-k}}$ to be the set of ``integral budgets''.
 For each $j< 2^k$, we define $B_1^j(k) =\frac{j}{2^k} $ and  
$$\calB^j(k) = \{B_1\in \calB_\eps : B_1^j \leq B_1 < B_1^{j+1}\} = \{\frac{j}{2^k} + r\eps : r=0,1,\ldots,R-1\}.$$
That is, $\calB^j$ is the $j$'th budget subinterval, which is ``closed'' from below (includes $\frac{j}{2^k}$) and ``open'' from above.\footnote{The budget $B_1=1$ is included in the last interval $\calB^{2^k-1}$.} We omit the parameter $k$ when clear from the context.

\begin{definition}\label{def:intervals}
A profile $\gamma\in\Gamma(G,\eps)$ is an \emph{interval profile} if  for all $s\in S\setminus T$, all $j< 2^k$ and all $\hat B_1\in \calB^j$, $\mu_\gamma(s,\hat B_1) = \mu_\gamma(s,B_1^j)$. 
\end{definition}

When an equilibrium is not monotone, the equilibrium bids at both ends of the interval $\calB^j(k)$ may reach different intervals at the $k-1$ level.
\rmr{need example?}
%\begin{proposition}Let $\gamma=BE(G,\eps)$. Then
%\end{proposition} 
Thus as we increase the resolution $\eps$, the structure of equilibria may become more and more fragmented. For both practical and technical reasons we would prefer equilibria that are interval profiles. Unfortunately, in the general case the Bottom Equilibrium may not be an interval profile due to lack of monotonicity. However for \emph{binary trees}, we will see that the Bottom Equilibrium is both monotone and an interval profile.

\subsection{Binary bidding games}

\begin{small}
\begin{algorithm}[!t]
\caption{\label{alg:PSPE}\textsc{Find-PSPE}($G,R$)}
\SetKwFunction{compU}{Get-Outcome}
\SetKwFunction{play}{Set-Strategies}
\SetKwFunction{mini}{Minimal-Increase}
Set $\eps\leftarrow \frac{1}{R}2^{ -\height(G)}$ \tcp*{Global variable}
%Initialize\footnote{if bids are continuous, then $\calB$ is also continuous. However, let $k$ be the height of $G$. For every $B_1$ in the range $[j \cdot 2^{-k},(j+1) 2^{-k})$, we have the same possible outcomes. Thus we can treat $\calB$ as if it contains $2^k$ intervals and only compute $A_i(s,B)$ for some $B$ in every interval.}
$\calB\leftarrow \{c\cdot\eps: c\in\{0,1,\ldots,\ceil{1/\eps}\}\}$\;
Initialize tables $A_1,A_2,T^*,I$ of size $|S| \times |\calB|$\; 
%\tcp{intuitively, $A_i$ is the strategy of $i$, where $A_i(s,B)$ will contain the action of $i$ in state $\tup{s,B}$ (both bid and next move)}
%Initialize a table $T^*$  of size $|S| \times |\calB|$\;
%\tcp{intuitively, $T^*(s,B)\in T$ is the outcome that will be reached from state $\tup{s,B}$}
In every leaf $t$, assign $T^*(t,B_1)=t$ for all $B_1\in \calB$\;
%\tcp{Traverse nodes bottom up:}
\For{every node $s\in S\setminus T$ in post-order} { 
		%Let $s_l,s_r$ be the two children of $s$\;
	  \For{every budget partition $B_1\in \calB$:} 
		{ 
					$b_1\leftarrow 0; b_2\leftarrow 0$ \tcp*{Initialize both bids}
					$i^*\leftarrow 1$ \tcp*{Initially player~1 wins}
					$s_1 \leftarrow \argmax_{s'\in g(s) }u_1(T^*(s',B_1))$\;
					$s_2 \leftarrow \argmax_{s'\in g(s) }u_2(T^*(s',B_1))$\;
										%\tcp{Compute the lowest equilibrium of the current step (assuming strategies in the subgames are fixed):}
			   	 $t^* \leftarrow T^*(s_1,B_1)$\;
					\If{$s_1=s_2$} {
					\tcp{No conflict. White wins with bid $0$ and plays $s_1$}
					Break\;}
					}
					\Repeat{} 
					{
					    \tcp{Try deviation by loser $-i^*$:}
							$b'_{-i^*} \leftarrow b_{-i^*}+\eps$\;
							 $t' \leftarrow T^*(s_{-i^*},B_{-i^*}-b'_{-i^*})$\;
							\If{$b'_{-i^*}>B_{-i^*}$ or $t' \preceq_{-i^*} t^*$}{	   Break\tcp*{Deviation impossible or not successful
							}
							\Else {
							  $i^* \leftarrow -i^*$\tcp*{Loser becomes winner}
							 $b_{i^*} \leftarrow b'_{i^*}$\tcp*{Deviation becomes the new bid}
							 $t^*\leftarrow t'$\;  		
							}
					}
					\tcp{Write down the equilibrium strategies in  $\gamma(s,B_1)$:}
					%\play($G,T^*,A_1,A_2,s,B_1,B_2,b_1,b_2$)\
$I(s,B_1) \leftarrow i^*$\tcp*{winner}
$A_i(s,B_1) \leftarrow (b_i,s_i)$ for $i\in \{1,2\}$\tcp*{actions of players (bid+selection)}
%\tcp{the move of the loser on the off-equilibrium path where $i$ lowers her bid:}
%$A_{-i}(s,B_1) \leftarrow (b_{-i},s_{-i})$\tcp*{action of loser $-i$ (bid+selection)} 
$T^*(s,B_1) \leftarrow t^*$\tcp*{terminal}

}
}
	\Return{$\gamma \leftarrow (I,A_1,A_2,T^*)$}\;
	\end{algorithm}

	%
	%\begin{function}[t]
	%\SetKwFunction{compU}{Get-outcome}
%\caption{Minimal-Increase($G,T^*,s,B_1,B_2,b_1,b_2,i$)}
			%$t^* \leftarrow$ \compU($G,T^*,s,B_1,B_2,b_1,b_2$)\;
     %\For{every bid $b'_i \in\{b_{-i},b_{-i}+\eps,b_{-i}+2\eps,\ldots,B_i\}$:}
		%{
		     %$t' \leftarrow$ \compU($G,T^*,s,B_1,B_2,b'_i,b_{-i}$)\;
					%\If{$t' \succ_{i} t^*$}{
			%\Return{$\tup{b'_i,t'}$}\;
			%}
			%}
			%\Return{$\tup{b_i,t^*}$}\;
	%\end{function}
\end{small}

The remainder of this section is dedicated to the proof of our main theorem: 
\begin{theorem} 
\label{th:Pareto}
Let $G$ be a binary bidding game with high resolution $\eps$, and let $\gamma=BE(G,\eps)$. Then:
\begin{enumerate}[label=(\Alph*)]
	\item \label{e:tight} At any point $\gamma(s,B_1)$, either $b_1=b_2$ (if white wins) or $b_2=b_1+\eps$ (if black wins);
	%\item \label{e:alg} Algorithm~\ref{alg:PSPE} computes $\gamma$;
	\item \label{e:int} $\gamma$ is an interval profile;
	\item \label{e:PS} $\gamma$ is Pareto-surjective;
	\item \label{e:PO} $\gamma$ is Pareto-optimal;
	\item \label{e:mon} $\gamma$ is monotone.
\end{enumerate}
% its Bottom Equilibrium $\gamma$ is (A) monotone, (B) Pareto-optimal, and (C)
%\item $\mu$ is monotone. That is, if $B'_1>B_1$, then $\mu(B'_1) \succeq_1 \mu(B_1)$ (and by Pareto, $\mu(B'_1)\preceq_2 \mu(B_1)$). 
	%\item $\mu:\calB \rightarrow T_P(G)$. That is, players end up in a Pareto-efficient outcome from any initial budget.
	%\item $\mu$ is surjective. That is, for any Pareto-efficient outcome $t$, there is a budget such that players end up in $t$.
%\end{enumerate}
%There is a monotone surjective mapping $\mu:[0,1]\rightarrow T_P(G)$, and a PSPE in $G$ implementing $\mu$. That is, if white has $B_1\in [0,1]$, then the PSPE outcome will be $\mu(B_1)$. 
\end{theorem}
%While symmetry facilitates the proof, it is not a necessary requirement. We will show that any binary game is w.l.o.g. neutral. 
%Consider Algorithm~\ref{alg:PSPE} for computing a PSPE. 
% We first select a resolution level $\eps$, so that every bid will be an integer multiple of $\eps$. Intuitively, the algorithm traverses the game tree/DAG from bottom up, and computes the PSPE induced by the \emph{lowest stable bids}---for every possible budget (rounding both budgets and bids upwards to a multiple of $\eps$). 
%We need to set an appropriate value for $\eps$. Clearly, for player~1 the lowest effective raise is to match the current bid of player~2. For player~2 we can set $0<\eps<2^{-k-2}$, where $k$ is the height of the current node $s$. Changing the bid by smaller quantities cannot change the outcome (see Lemma~\ref{lemma:intervals}). 

Recall that we denote by $\gamma|_s$ the \emph{projection} of $\gamma$  on a subgame of $G$ that is rooted in $s$. Similarly, $\mu|_s = \mu_{\gamma|_s}$. 
%First we need to show that $\gamma$ is indeed a PSPE. Then we need to show several things (rephrasing the statements in the theorem):  (1) that $\mu(B_1)\in T_P(G)$ for all $B_1\in \calB$; (2)  that for any Pareto-efficient outcome $t_j\in T_P(G)$ there is some budget $B_j$ s.t. $\mu(B_j)=t_j$; (3) that this mapping is monotone. That is, for all $j'>j$, $B_{j'}>B_j$.\footnote{We note that we do not need Proposition~\ref{th:monotone} for the proof of the current theorem, as monotonicity will follow from the inductive construction of $\mu$. However, we note that Proposition~\ref{th:monotone} holds without any additional assumptions on the budget, and is therefore not entailed by the current theorem.}
We will sometimes use the budget of player~2 as the input for $\mu$, rather than the budget of player~1. This should be clear from the notation. That is, $\mu(B_1)$ and $\mu(B_2)$ always refer to the same outcome in $T(G)$. 

Denote the two children of $s_0$ by $s_l,s_r$. 
As the proof of all parts is by induction on the height of $G$, we assume that $\gamma=BE(G,\eps)$ has been computed on both subtrees $G_l=G|_{s_l}$ and $G_r=G|_{s_r}$, and denote $\gamma^{(l)} = \gamma|_{s_l}, \gamma^{(r)} = \gamma|_{s_r}$ and similarly for $\mu_l,\mu_r$. By construction, we have that $\gamma^{(l)}=BE(G_l,\eps), \gamma^{(r)} = BE(G_r,\eps)$.  Thus by the induction hypothesis both of $\gamma^{(l)},\gamma^{(r)}$ have all properties (A)-(E).

\begin{proof}[Proof of Theorem~\ref{th:Pareto}\ref{e:tight} (near-tie)]
By induction in  each subtree we have that  $\gamma^{(l)},\gamma^{(r)}$ are monotone. Suppose $(b_1,b_2)$ are the Bottom Equilibrium bids at some state $\gamma_s(B_1)$, and assume towards a contradiction that $b_1>b_2$. Consider the bid profile $\gamma'_s(B_1)=(b_1,b_2+\eps)$. Either $(b_1,b_2+\eps)$ are equilibrium bids, or there is a deviation. However the only deviation that is not also available in $(b_1,b_2)$ is by the white player. W.l.o.g. white can deviate by bidding $b'_1=(b_2+\eps)-\eps=b_2$, since as long as she drops below $b_2+\eps$ the outcome is the same (this is a first-price game). Now either $(b'_1,b_2+\eps)$ are equilibrium bids, or black has a deviation. By monotonicity of $\gamma^{(l)}$ and $\gamma^{(r)}$, black cannot gain by bidding $b'_2 > b_2+\eps$, since $\mu_l(B_2-b'_2)\preceq_2 \mu_l(B_2-(b_2+\eps))$ and likewise for $\mu_r$. Thus the only deviation is to decrease his bid and drop the round, w.l.o.g. to $b'_2=b'_1$. Again either $(b'_1,b'_2)$ is an equilibrium, or white drops to $b''_1=b'_2-\eps$ and so on until no player has a deviation at some $(b^*_1,b^*_2)$. Thus we get a PSPE $\gamma^*$ where $\gamma^*_s(B_1)=(b^*_1,b^*_2)$. If there were at least two deviations, then $b^*_{i^*} < b_1$. If there are no deviations, then $b^*_{i^*}=b_1$ and $b^*_{-i^*}>b_2$. If there is exactly one deviation then $b^*_{i^*}=b^*_2=b_1$ and $b^*_1=b_2$. In either case, we get that $(b^*_1,b^*_2) <_\pi (b_1,b_2)$, and thus $\gamma^* <_\pi \gamma$, which is a contradiction to $\gamma$ being the Bottom Equilibrium. The proof for $b_2>b_1+\eps$ is the same except that there are only two cases for $(b^*_1,b^*_2)$ (no deviations, or at least one). 
\end{proof}

\begin{lemma}\label{lemma:alg}Algorithm~\ref{alg:PSPE} computes $\gamma$.
\end{lemma}
\begin{proof}%[Proof of Theorem~\ref{th:Pareto}\ref{e:alg} (Algorithm)]
Recall that $k=\height(G)$.
Note that the algorithm effectively goes over all states $s\in S\setminus T$ from the bottom up, and for for any possible budget allocation, runs an auction where players iteratively modify their bids by the smallest possible increment $\eps$, until none of them can gain by further modifying her bid.
Note that the players never change their selected state when increasing their bid. Indeed, if player~$2$ has an incentive to raise at the initial bids of $(0,0)$ (and select child $s_2$) only if he strictly prefers $t'=\mu_2(B_2-\eps)$ over $t^*=\mu_1(B_2)$, and $\mu_1(B_2)\succeq_2\mu_1(B_2-\eps)$ by monotonicity of $\gamma^{(2)}$. Similarly, now player~1 has an incentive to raise only if she prefers $t'=\mu_1(B_1-\eps)$ over $t^*=\mu_2(B_2-\eps)$, which means that $t'\succ_1 \mu_2(B_2-\eps)=\mu_2(B_1+\eps)\succeq_1\mu_2(B_1-\eps)$ by monotonicity of $\gamma^{(1)}$. This continues with every raise of bid, so player~1 always prefers $s_1$ and player~2 always prefers $s_2$. 

By property~\ref{e:tight}, the winning and losing bids in $\gamma$ are nearly tied or tied. The algorithm goes over all such pairs of bids in increasing order and thus must reach $(b_1,b_2)$ as in $\gamma(s,B_1)$ unless it stops at some lower bids. Suppose there are lower bids $\gamma'_s(B_1)=(b'_1,b'_2)$ where the loser $-i^*$ cannot deviate by $\eps$ and the algorithm stops. Then by monotonicity of $\gamma^{(1)},\gamma^{(2)}$, neither $i^*$ nor $-i^*$ can deviate by further increasing their bid. Further, by Lemma~\ref{lemma:auction}, bids can only increase so there is no deviation by decreasing a bid either, meaning that $\gamma'$ is also a PSPE which is lower than $\gamma$. A contradiction.
\end{proof}

\begin{proof}[Proof of Theorem~\ref{th:Pareto}\ref{e:int} (Intervals)]Let $s$ be a node at height $k$ and $j<2^k$.
Consider $\gamma_s(B_1^j)=(b_1,b_2)$ and $\hat B_1\in \calB^j(k)$. Note that the size of each level $k$ interval is $R$, and the size of level $k-1$ intervals is $2R$. Thus $\hat B_1 = B_1^j + r\eps$ for some $r<R$.
Suppose first that $b_1 \geq b_2$, i.e. that white wins. By property~\ref{e:tight}, $b_1=b_2$. Let $j',j''<2^{k-1}$ be the level $k-1$ budget intervals reached when white wins and if white drops, respectively. Formally, $B^j_1-b_1 \in \calB^{j'}(k-1)$ and $B^j_1+b_1 = B^j_1+b_2 \in \calB^{j''}(k-1)$. 
We argue that there are 3 cases:
 \input{intervals1L}
 \input{intervals1W}
 \begin{description}
   \item[Case~0]  $b_1=b_2=0$ (no conflict). Both agents prefer the same node $s'\in g(s)$ at the current budget $B_1$. So all of interval $\calB^{j'}(k-1)$ is (weakly) preferred to all of interval $\calB^{j''}(k-1)$.

For any $\hat B_1 \in \calB^j(k)$ with $\hat b_1=\hat b_2=0$ this still holds, as $\calB^j(k)$ is contained in both $\calB^{j'}(k-1)$ and $\calB^{j''}(k-1)$ (which are actually the same interval but in different nodes). Thus $\gamma(s,\hat B_1)=(0,0)$ as well.

	 \item[Case~1L] $B^j_1+b_1$ is the maximal point in $\calB^{j''}(k-1)$, i.e. $B^j_1+b_1 = B_1^{j''+1}-\eps=B_1^{j''}+(2R-1)\eps$. This means that $B_1-b_1 = B_1^{j'}+\eps$. By the progress of Algorithm~\ref{alg:PSPE} we can conclude due to Lemma~\ref{lemma:alg} that white prefers outcome $j'$ (and winning the turn) over outcome $j''$ (and losing the turn), otherwise she would not have made the last increase from to $b_1$. Similarly, black prefers $j''$ over $j'$, but prefers $j'$ over $j''+1$, as otherwise he would deviate to $b_2+\eps$. 

Let $\hat B_1 = B_1^j + r\eps$ (where $r<R$). Denote $\hat b_1=b_1-r\eps, \hat b_2=\hat b_1 = b_2-r\eps$.  Equilibrium bids are presented graphically in Fig.~\ref{fig:intervals1L}.
We argue that 
\begin{itemize}
	\item $\hat B_1 +\hat b_2 \in \calB^{j''}(k-1)$; 
	\item $\hat B_1 - \hat b_1 \in \calB^{j'}(k-1)$; and
	\item $\gamma_s(\hat B_1) = (\hat b_1,\hat b_2)$
\end{itemize}
The first claim is obvious, since $\hat B_1 +\hat b_2  = B_1^j+b_2 \in \calB^{j''}(k-1)$.
To see why the second claim holds, note that since $r<R$, 
$$\hat B_1 -\hat b_1 = B_1^j+r\eps - (b_1 -r\eps)  = B_1^j-b_1+2r\eps <    B_1^j-b_1+(2R-1)\eps,$$
which is still in interval $\calB^{j'}(k-1)$. 

For the third claim, suppose the equilibrium bids are lower than $\hat b_1$. If white wins then the outcome is either in  interval $j'$, or in a higher interval which is even worse for black. This means black would have a deviation to $\hat b_2$. Similarly, if black wins with $b'_2\leq \hat b_2$ then the outcome is wither in interval $j''$ or in a lower interval which is worse for white, and thus white would have a deviation to $\hat b_1$. 
By Lemma~\ref{lemma:alg}, it is left to show that Algorithm~\ref{alg:PSPE} stops at $(\hat b_1,\hat b_2)$. Indeed, the next deviation by black would reach interval $j''+1$, and we know black prefers the current outcome $j'$.
Intuitively what we showed is that $B_1^j+b_2 = B_1^{j''}+(2R-1)\eps$ is the ``critical point'' where the auction stops, as long as $B_1\in \calB^j(k)$ (see rightmost dotted line in Fig.~\ref{fig:intervals1L}).

\item[Case~1W] $B^j_1+b_1$ is not the maximal point in its interval $\calB^{j''}(k-1)$. Then we argue that $B^j_1-b_1$ is the highest point of $\calB^{j'}(k-1)$, i.e. $B^j_1(k)-b_1=B^{j'}_1(k-1)+(2R-1)\eps$. This is since both agents prefer interval $j'$ (where white wins) over $j''$ (where black wins), as otherwise black would raise by $\eps$. Then since $b_1=b_2$, we get that $B_1^j + b_2 = B^{j''}_1(k-1)+\eps$, i.e. that the losing bid reaches the second-lowest point on its interval. 

Now, we continue in a similar way to Case~1L, except that here critical point is $B_1^j-b_1=B^{j'}_1(k-1)+(2R-1)\eps$, i.e. determined by the winner $i=1$ (leftmost dotted line in Fig.~\ref{fig:intervals1W}). Thus define  $\hat b_1=\hat b_2 = b_1+r\eps$, and we will show that
\begin{itemize}
	\item $\hat B_1 - \hat b_1 \in \calB^{j'}(k-1)$; 
	\item $\hat B_1 +\hat b_2 \in \calB^{j''}(k-1)$; and
	\item $\gamma_s(\hat B_1) = (\hat b_1,\hat b_2)$
\end{itemize}
The first claim follows since $\hat B_1 -\hat b_1  = B_1^j-b_1 \in \calB^{j'}(k-1)$, and the second since
$$\hat B_1 + \hat b_2 = B_1^j(k)+r\eps + (b_1+r\eps) = B_1^j(k)+b_1 + 2r\eps = B_1^{j''}(k-1)+\eps + 2r\eps \leq B_1^{j''}(k-1)+(2R-1)\eps.$$
As in Case~1L, the algorithm would not stop at lower bids than $\hat b_1$. Any increase by black would reach interval $j''$ or higher (worse for black), but we know black prefers $j'$ over $j''$ so would not deviate.
 \end{description}

In either case, we get by the induction hypothesis that 
$$\mu_\gamma(s,\hat B_1) = \mu_\gamma(s_i,\hat B_1-\hat b_1)=\mu_\gamma(s_i,B^j_1-b_1)=\mu_\gamma(s,B^j_1).$$ 

 \input{intervals2}
If $b_2>b_1$, then Case~0 is impossible, and we get two cases similar to the above analysis (see Fig.~\ref{fig:intervals2}): Case~2L where the critical point is $B_1^j-b_1=B_1^{j'}(k-1)$ and $B^j_1+b_2 = B_1^{j''}(k-1)+\eps$; and case~2W where the critical point is $B^j_1+b_2 = B_1^{j''}(k-1)$, and $B^j_1-b_1 = B_1^{j'}(k-1)+\eps$. In either case we get that there is some critical player $\hat i$, for which $\hat B_{\hat i}-\hat b_{\hat i} = B_{\hat i}^j-b_{\hat i}$, and that $\hat B_{-\hat i}-\hat b_{-\hat i},B_{-\hat i}^j- b_{-\hat i}$ reach the same level $k-1$ budget interval (details omitted but very similar to cases 1L,1W). Then again,  by the induction hypothesis  
$$\mu_\gamma(s,\hat B_1) = \mu_\gamma(s_i,\hat B_2-\hat b_2)=\mu_\gamma(s_i,B^j_2-b_2)=\mu_\gamma(s,B^j_1).$$ 
\end{proof}
Note that in particular we proved the following corollary (phrased as a lemma since will be used later):
\begin{lemma}\label{lemma:four_cases}
For any $s\in S\setminus T$ and $j<2^k$, one of the following five cases applies for the Bottom Equilibrium $\gamma_s(B_1^j)=(b_1,b_2)$.

\begin{small}
\begin{tabular}{l|llll}
Case & Winner & Critical  & Bid of player~1  &  Bid of player~2 \\
\hline  
%\begin{description}
0    & $i^*=1$  &  -       & 0 & 0 \\
	1L & $i^*=1$ & $\hat i=2$&$B^j_1(k)-b_1 = B_1^{j'}(k\!-\!1)+\eps$       & $B^j_1(k)+b_2 =B_1^{j''}+(2R\!-\!1)\eps$ \\ 
	1W& $i^*=1$ & $\hat i=1$ & $B^j_1(k)-b_1 = B_1^{j'}(k\!-\!1)+(2R\!-\!1)\eps$ & $B^j_1(k)+b_2 =B_1^{j''}+\eps$\\
	2L &$i^*=2$ & $\hat i=1$ & $B^j_1(k)-b_1 = B_1^{j'}(k\!-\!1)$           & $B^j_1(k)+b_2 = B_1^{j''}(k\!-\!1)+\eps$\\
2W  &$i^*=2$ & $\hat i=2$ & $B^j_1(k)-b_1 = B_1^{j'}(k\!-\!1)+\eps$      & $B^j_1(k)+b_2 = B_1^{j''}(k\!-\!1)$
%\end{description}
\end{tabular}
\end{small}
\end{lemma}

\medskip
\begin{proof}[Proof of Theorem~\ref{th:Pareto}\ref{e:PS} (Pareto-surjective)]%\!\!\footnote{Note that the proof of (3) does not depend on (2).}
We prove by induction on the height of the tree $G$. For height $0$ it is obvious. For height~$1$ either one leaf weakly Pareto-dominates the other, in which case this leaf is reached in $\mu$ regardless of budgets, or each player has a favorite leaf in which case both leafs are Pareto-efficient.

 Let $t^*\in T_P(G)$, we need to show that there is some initial budget $B\in \calB$ s.t. that  $\mu(B)=t^*$. W.l.o.g. $t^*\in T_P(G_l)$. 

By induction on property~\ref{e:PS}, let $B^*_1$ s.t. $\mu_l(B^*_1) = t^*$ (must exist),
%By the induction hypothesis there is some budget $B^*_1$ (where $B^*_2 = 1-B^*_1$) and some PSPE $\boldsymbol a_l$ s.t. $t$ is realized in $(\boldsymbol a_l,B^*_1,B^*_2)$. For any $B'_2>B^*_2$, we complete $\vec a_l$ so that the realized outcome $t'$ is at least as good for player~2 (which is possible by monotonocity). Since $t$ is Pareto-optimal, $u_1(t')\leq u_1(t)$. This means that under $\vec a_l$, the utility of white in $G_l$ can only decrease if she has a budget lower than $B^*_1$. For any $B'_1>B^*_1$, we complete $\vec a_l$ so that the outcome is at least as good for player~1---and thus can only be worse for black than $t$ (\#). 
% Let $\vec a_r$ be a PSPE in $G_r$ that is monotone for player~2, 
and let $t^*_r=\mu_r(B^*_1)$. %$(\vec a_r,B^*_1,B^*_2)$. 
As $\gamma$ is an interval profile, w.l.o.g. $B^*_1=B^j_1(k)+2\eps$ for some $j<2^k$ (i.e. it an even multiple of $\eps$, and is not on the edge of its respective interval). This is possible by property~\ref{e:int} and since $R\geq 4$. Thus by slightly increasing or decreasing the budget, the outcome will not change:  $\mu_l(B^*_1-2\eps)=\mu_l(B^*_1+\eps) = t^*$ [also equal to $\mu_l(B^*_2-\eps)$]; and likewise for $\mu_r$. %Since the size of budget intervals is $2^{-\height(G)}$, setting $\eps<2^{-\height(G)-2}$ is sufficient.

We know that $t^*_r$ does not strictly Pareto dominate $t^*$ (as $t^*\in T_P(G)$), thus either $t^*_r \prec_1 t^*$ or $t^*_r\prec_2 t^*$. 

Case~I: Both players prefer $t^*$ over $t^*_r$ at budget $B^*_1$. 
%Case~Ia: Suppose that $t^*_r\preceq_2 t^*$. 
We claim that the equilibrium bids are $b_1=b_2=0$. 

%We have that $\mu_r(B^*_2-1)\preceq_2 \mu_r(B^*_2) = t^*_r \preceq_2 t^*$, and thus black has no reason to raise his bid above the initial bid $0$.  %i.e. that black  $s_l$ in $\gamma (since $t^*\in T_P(G)$ this must be an equality). Then we set $B_1=B^*_1$ and claim that $b_1=b_2=0$ is an equilibrium in $s_0$. Since white takes, then by monotonicity of $\mu_l,\mu_r$ she has no incentive to raise her bid.

%As explained above, white has no reason to increase her bid, as this will only result in playing $\vec a_l$ with a lower budget and thus earning less.%\footnote{We need to be careful here: is it possible that $t^*$ is only attainable at a PSPE that is worse than anything acheivable with lower budget? perhaps we should set $B^*_1$ to be the minimal budget for which $u_1(t)$ is attainable and by induction $t$ is attainable in $G_l$.} 
%Of course she cannot lower her bid below $0$, and even if she could drop the tie on purpose she has no reason to do that as her budget would not increase.
Suppose that black raises to $b'>b_2=0$, takes the round and goes right. %If he goes right, then 
then
$$\mu_r(B^*_2-b') \preceq_2 \mu_r(B^*_2) = t^*_r \prec_2 t^*,$$
and thus black does not gain. 
%Then, $\mu_l(B^*_2-b')\preceq_2 \mu_l(B^*_2) = t^*$, so black also does not gain.
Therefore $b_1=b_2=0$, and $\mu(B^*_1)=\mu_l(B^*_1)=t^*$.

Case~II: Suppose that $t^*_r \succ_2 t^*,~ t^*_r \prec_1 t^*$,  i.e. that black strictly prefers child $s_r$ at budget $B^*_2$, whereas white strictly prefers $s_l$. 
%Recall that with any budget $B'_2<B^*_2$, $\vec a_r$ only yields a worse outcome for black than $t_r$, and $\vec a_l$ can only yield a better outcome for black than $t$. 
Let $\hat B_2\leq B^*_2$ be the lowest budget s.t. for every $\beta_2\in[\hat B_2,B^*_2]$, $\mu_r(\beta_2) \succ_2 t^*$ (it is possible that $\hat B_2=0$). Set $\hat t_r=\mu_r(\hat B_2)$. %\hat t_r = (\vec a_r,\hat B_2)$ over $t$. 
By Pareto-optimality of $t^*$ it follows that for every $\beta_2\in[\hat B_2,B^*_2]$, white prefers $s_l$ (since $\mu_r(\beta_1) \prec_1 t^*=\mu_l(B^*_1) \preceq_1 \mu_l(\beta_1)$). 

% there is some budget lower than $B^*_2$ where black strictly prefers $s_r$ (i.e. the outcome $t_r$ realized in $(\vec a_r,B^*_2-\Delta)$), and let $\hat B_2$ be the minimal budget under this condition.

  We set $B_2 = \frac{B^*_2+\hat B_2}{2}$, $B_1  =1-B_2$, and argue that $\mu(B_1)=t^*$. Let $b^*=\frac{B^*_2-\hat B_2}{2}$. % be the minimal bid s.t. $\mu_l(B_1-b^*)=\mu_l(B_2+b^*)=t^*$. Note that $b^*\leq \frac{B^*_2-\hat B_2}{2}$. 
	We will show that the equilibrium in $s_0$ is $b_1=b_2=b^*$. Note that $\hat B_2$ is the lower bound of a budget interval and thus also an even multiple of $\eps$, so $B_2$ and $b^*$ are integer multiples of $\eps$ . 
	
	As long as bids are strictly below $b^*$, the resulting budget is in the ``conflict zone'' $[\hat B_2,B^*_2]$, and the lowest bidder will raise to take the turn from the other bidder. Suppose now that $b_1=b_2=b^*$. If $\hat B_2 = 0$, then $b_2=b^*=B_2$ and black cannot raise. Otherwise, by definition of $b^*$ if black raises his bid to $b'_2=b^*+\eps$, then %$t^* = \mu_l(B^*_2) = \mu_l(B_2+b^*) \succeq_2 \mu_l(B_2-b'_2)$, so black cannot gain by increasing his bid and select $s_l$. By definition of $\hat B_2$, 
	$$t^* \succeq_2 \mu_r(\hat B_2-\eps) = \mu_r(B_2-(b^*+\eps))= \mu_r(B_2-b'_2),$$
	so black cannot gain by selecting $s_r$. Thus $b_1=b_2=b^*$ is an equilibrium, where $\mu(B_1) = \mu_l(B^*_1)$.
		
	%he gains at most $u_1(t)$ by (\#) above.
	
	%. However, $u_2(\vec a_l,\hat B_2)\leq u_2(\vec a_l,B^*_2)=u_2(t)$ thus black has no reason to raise.

Case~III: $t^*_r \succ_1 t^*$, i.e. white strictly prefers $s_r$ at budget $B^*_1$. %This means that in $(\vec a_r,B^*_1)$ players reach some terminal $t_r$ where $u_1(t_r)>u_1(t)$.
Quite expectantly, the proof of case~III is similar to case~II, where white and black change roles. However, there are some fine issues due to tie-breaking, so we lay out the full proof.

Let $\hat B_1\leq B^*_1$ be the lowest budget s.t. for every $\beta_1\in[\hat B_1,B^*_1]$, $\mu_r(\beta_1) \succ_1 t^*$. Set $\hat t_r=\mu_r(\hat B_1)$. %\hat t_r = (\vec a_r,\hat B_2)$ over $t$.
As in case~II, whenever the budget after the bid is in the range $[\hat B_1,B^*_1]$, white prefers $s_r$ whereas black prefers $s_l$.
We define $B_1 = \frac{\hat B_1+B^*_1}{2}$, $b^*=\frac{B^*_1-\hat B_1}{2}$, and argue that the equilibrium reached in $s_0$ is $b_1=b^*,b_2=b^*+\eps$.

Indeed, if $b_1<b_2\leq b^*$, then black selects $s_l$ under the remaining budget $B_2-b_2\geq B^*_2$, and we have
$\mu_r(B_1-b_2) \succ_1 t^* =\mu_l(B^*_1) =\mu_l(B_2-b^*) \succeq_1 \mu_l(B_2-b_2)$.
I.e., white will raise her bid to $b'_1=b_2$.

Similarly, if $b_1=b_2\leq b^*$, then white selects $s_r$, but black wants to raise to $b'_2=b_2+\eps$: since $B_2-b'_2\geq B_2-b^*-\eps = B^*_2-\eps$, we have\footnote{The equality is by our selection of $B^*_2$ so that it is not on the edge of its respective interval. This is the only place where our assumption of high resolution is applied. However this seemingly innocuous assumption is critical, as we show in the next section.} 
$$\mu_l(B_2-b'_2) \succeq_2 \mu_l(B^*_2-\eps) = \mu_l(B^*_2) = t^*\succ_2 \hat t_r = \mu_r(B_2+b^*) \succeq_2 \mu_r(B_2+b_1).$$

It remains to show that $b_2=b_1+\eps=b^*+\eps$ is an equilibrium, i.e. that white will not raise. This is exactly as in case~II. Either white cannot raise, or $b'_1>b^*$. In the latter case,
$\mu_r(B_1-b'_1) \preceq_1 \mu_r(B_1-b^*-\eps) = \mu_r(\hat B_1 - \eps) \preceq_1 t^*$,
where the last inequality is by definition of $\hat B_1$. 
Thus white cannot gain by selecting $s_r$, and clearly not by keeping $s_l$ at a lower budget.  
\end{proof}

\begin{proof}[Proof of Theorem~\ref{th:Pareto}\ref{e:PO} (Pareto-optimal)]
Let $b_1,b_2$ be equilibrium bids in $s_0$ reached by the above process. We use throughout the proof the fact that by monotonicity of both subgames, it only makes sense for a player to raise her bid if she selects the branch that is not currently played. Note that we  may not use the fact that $\mu$ itself is monotone, since this will be the last thing we prove. 

 Assume w.l.o.g. that $b_1= b_2$ (white takes) and that under budget $B_1-b_1$, white prefers $s_l$.
 We argue that $t^*_l = \mu_l(B_1-b_1)$ is Pareto-efficient in $G$. Assume, toward a contradiction, that there is some $t^*_r\in T_P(G_r)$ that Pareto dominates $t^*_l$ (i.e., $t^*_r \succ_{1,2} t^*_l$).  Since $\gamma^{(r)}$ has property~\ref{e:PS} by induction, $t^*_r=\mu_r(B'_1)$ for some budget $B'_1$. 

%Our proof will show that (under weakly generic preferences), the bidding process in $s_0$ does not stop unless it reaches an outcome that is Pareto-efficient (case~1), and once an outcome was reached, the agents will not continue to an outcome that is Pareto-dominated by it (case~2). 

\medskip
Case 1: Suppose first that $B'_1>B_1$, i.e. that white must drop the turn in order to reach $s_r$ with sufficient budget. Denote $b' = B'_1-B_1$ (note that $b'=B_2-B'_2\leq B_2$). We argue that $b'>b_1$. If $b_1=0$, then clearly $b'>0=b_1$, thus suppose $b_1>0$. Then by construction of the equilibrium $(b_1,b_2)$, $t^*_l$ is strictly preferred by white over the outcome of $(b_1-\eps,b_2)$, which is $\mu_r(B_1+b_1)$ (as $b_2=b_1$). Thus,
$$\mu_r(B_1+b') = t^*_r \succ_1 t^*_l \succ_1 \mu_r(B_1+b_1).$$ 
By monotonicity, we must have $b'>b_1=b_2$. Next, recall that $t^*_r \succ_2 t^*_l$.  Thus black could raise his bid to $b'>b_1=b_2$ and strictly gain. This shows that $(b_1,b_2)$ are not an equilibrium in $s_0$, in contradiction to the construction of $\gamma$. 

\medskip
% for non-generic:
%Thus it can only be that $t^*_r =_2 t^*_l, t^*_r \succ_1 t^*_l$. Note that $t^*_r$ cannot be Pareto-dominated by any outcome $t'_l$ in $G_l$, as this would mean that $t'_l$ Pareto dominates $t^*_l$. Thus $t^*_r$ is Pareto-dominant in $G$. We set new equilibrium bids as $b'_2 = b_1+1$. The black takes and plays $s_r$, reaching $t^*_r$. It is left to show that $(b_1,b'_2)$ is indeed an equilibrium in $s_0$. Clearly white cannot gain by raising, as 
%$$\mu_l(B_1-b'_2) \preceq_1 \mu_l(B_1-b_1) \preceq_1 t^*_r;\ \mu_r(B_1-b'_2) \preceq_1 \mu_r(B_1+b') = t^*_r.$$
%Suppose that black lowers his bid to $b_2=b_1$ so that white takes. Then we already know that inder bids $(b_1,b_1)$ white selects $s_l$ and reaches $t^*_l \preceq_2 t^*_r$. Thus black does not gain either and this is an equilibrium.
%
%if white selects $s_l$, we have
%$\mu_l(B_1-(b'-1)) \succeq_1 \mu_l(B_1-b_1) = t^*_l$, and thus (since every $\mu_l(B)$ is Pareto-efficient in $G_l$) $\mu_l(B_1-b'-1) \preceq_2 t^*_l \preceq_2 t^*_r$. 
%Then we go back to the begining where black and white swap roles, and $t^*_r$ replaces $t^*_l$. Clearly case 1 can only occur a finite number of times. 

Case 2: $B'_1=B_1-b'$ for some $b'\geq 0$. %We will show that this is not possible unless $b_1=b_2=0$.  
If $b_1=0$, then $\mu_r(B_1) \succeq_1 \mu_r(B'_1) = t^*_r \succ_1  t^*_l = \mu_l(B_1).$
That is, white prefers $s_r$ but chose $s_l$. A contradiction.

%Note first that if $b_1=0$, then we have a Pareto-efficient PSPE where white selects $s_r$ in $s_0$ (and reaches $t^*_r$). 
Thus suppose $b_1 > 0$, and we will show that this will lead to a contradiction. 

Since $\mu_r(B_1-b') = t^*_r \succ_1 t^*_l \succeq_1 \mu_r(B_1-b_1)$, we must have by monotonicity of $\mu_r$ that $0\leq b'\leq b_1$. %That is, both agents should lower their bid to $b'$, and then white (who still takes) select $s_r$ instead of $s_l$. 
By the construction of the equilibrium bids, white strictly preferred $t^*_l=\mu_l(B_1-b_1)$ over the previous state $\mu_r(B_1+b_2)=\mu_r(B_1+b_1)$ (where the bid of player~1 was strictly lower). Thus $t^*_r \succeq_1  t^*_l \succ_1 \mu_r(B_1+b_1) \succeq_1 \mu_r(B_1-b') = t^*_r$,
which is a contradiction. 
\end{proof}

\begin{proof}[Proof of Theorem~\ref{th:Pareto}\ref{e:mon} (Monotonicity)]
By the induction hypothesis, both of $\mu_l,\mu_r$ are monotone. %Also, by property~\ref{e:PO} (Pareto optimality), $\mu(B_1)\in T_P(G)$ for all $B_1$. 
Let $B'_1=B_1+\eps$, and let $b_1,b_2$ be the equilibrium bids in $\gamma(s_0,B_1)$. Likewise, $b'_1,b'_2$ are the equilibrium bids in $\gamma(s_0,B'_1)$, where $B'_1=B_1+\eps$. 

%We first argue that in $\gamma(s_0,B'_1)$, it is still true that white prefers $s_l$ and black prefers $s_r$ throughout the bidding process, unless both bids are 0. Note that for any bid $\hat b_1\geq \eps$, we have that $\hat B_1 = B'_1-\hat b_1\leq B_1$, and thus white prefers $s_l$. Similarly, for any $\hat b_2\geq 0$, we have that $\hat B_2 = B'_2-\hat b_2< B_2$, and thus black prefers $s_r$. The only exception is
%
%Case~0: When $b'_1=b'_2=0$ is an equilibrium at budget $B'_1$. In this case it is easy to see that white weakly gains by the increased budget, as if $b_1\geq b_2$,
%$$\mu(B'_1)\succeq_1 \mu_l(B'_1) \mu_l(B_1+\eps) \succeq_1 \mu_l(B_1-b_1) = \mu(B_1),$$
%where the first inequality is since white picks her preferred child, and the second by monotonicity of $\mu_l$;
%and if $b_2>b_1$, 
%$$ \mu(B'_1)

%We split the remainder of the proof into two cases, according to the winner in $\gamma(s_0,B_1)$. Each such case further splits according to the bids in $\gamma(s_0,B'_1)$.

\medskip
Case~1:
Suppose white takes under $\gamma(s_0,B_1)$, thus $b_1=b_2$ by property~\ref{e:tight}. By our assumption, player~1  selects $s_l$ at $\gamma(s_0,B_1)$, and the outcome is $t_l=\mu_l(B_1-b_1) \succeq_1 \mu_r(B_1+b_2)$ (in fact the inequality is strict if $b_1>0$, or else player~1 would never bid $b_1$).

We argue that $b'_2\leq b_1+\eps$. Otherwise, it means that Algorithm~\ref{alg:PSPE} reached bids $(b_1+\eps,b_1+\eps)$ under budget $B'_1$ and then black raised again. However, at bid $b_1+\eps$ white remains with $B'_1-(b_1+\eps)=B_1-b_1$ and thus we know the outcome is $t'_l = \mu_l(B_1-b_1) \succeq_2 \mu_r(B_2-(b_1+\eps))$ (since black does not raise to $b_1+\eps$ under $B_1$). By raising under $B'_1$, the outcome for black will be 
$$\mu_r(B'_2-(b_1+2\eps)) = \mu_r(B_2-b_1-3\eps) \preceq_2 \mu_r(B_2-(b_1+\eps)) \preceq_2 t'_l,$$ 
that is, black does not gain.

We conclude that $b'_1,b'_2\leq b_1+\eps$. Then,  by bidding $b''_1=b_1+\eps$ at budget $B'_1$ and selecting $s_l$, white can guarantee a value of $\mu_l(B'_1-b''_1) = \mu_l((B_1+\eps)-(b_1+\eps))= \mu_l(B_1-b_1)$. Thus its current value $\mu(B'_1)$ is at least the same, and we get $\mu(B'_1)\succeq_1 \mu_l(B_1-b_1) =\mu(B_1)$. 

%If $b'_1=b_1,b'_2=b_2+\eps$ (i.e. black wins in $\gamma(s_0,B'_1)$), then the outcome is 
%$$t'_r=\mu_r(B'_2-b'_2) = \mu_r(B'_1+(b'_1+\eps)) \succeq_1 \mu_l(B'_1-(b'_1+\eps)) = \mu_l(B_1-b_1)=t_l,$$
% where the inequality is since otherwise player~1 would raise her bid to $b'_1+\eps$.  
%
%If $b'_1>b_1$ we argue that $b'_1=b'_2=b_1+\eps$ and that white selects $s_l$, meaning we get the same outcome $\mu(B'_1)=\mu_l(B'_1-b'_1) = \mu_l(B_1-b_1)=\mu(B_1)$. 
%We only need to show that according to Algorithm~\ref{alg:PSPE}, player~2 would not choose to increase his bid to $b''_2=b'_2+\eps$ and choose $s_r$. Indeed, 
%$$\mu_r(B'_2-b''_2) = \mu_r(B_2-\eps-(b_2+2\eps)) = \mu_r(B_2-(b_2+\eps)-2\eps)\preceq_2\mu_r(B_2-(b_2+\eps))\preceq_2 \mu_l(B_1-b_1) = \mu_l(B'_1-b'_1),$$
%where the first inequality is due to monotonicity of $\mu_l$, and the second since in $\gamma(s_0,B_1)$ black chooses not to increase his bid from $b_2$. 
 
\medskip
Case~2:
Next, we prove for the case where black takes under  $\gamma(s_0,B_1)$ (i.e.,  $b_2=b_1+\eps$).  Note that in particular $b_2\geq \eps$. 
W.l.o.g. black selects $s_r$, and the outcome is $t_r=\mu(B_2)=\mu_r(B_2-b_2)=\mu_r(B_1+b_2)$. 

%Note that the new outcome $t'=\mu(B'_1) \succeq_1 \mu_r(B'_1+b'_2)$: if black wins then the righthand side is $t'$, and if white wins then the outcome is strictly better, or otherwise she would not raise.
 
%If $b_1=b_2=0$, $t_r=\mu_r(B_1)$ is preferred by both players to $\mu_l(B_1)$. 

By Lemma~\ref{lemma:alg} and the way Algorithm~\ref{alg:PSPE} works, player~1 selects the same child of $s_0$ (w.l.o.g $s_l$) at any bid at or below the equilibrium bids. Thus, if $b_1>0$ then for every $\hat b_1\leq b_1$, $\mu_l(B_1 -\hat b_1)\succ_1 \mu_r( B_1-\hat b_1)$.  Player~2 always prefers the other child ($s_r$), unless both bid $0$ in equilibrium, thus for every $\hat b_2 < b_2$, $\mu_r(B_2-\hat b_2)\succ_2 \mu_l(B_2-\hat b_2)$.

We claim that $b'_2> b_2-2\eps$. Assume, towards a contradiction, that $b'_2\leq b_2-2\eps$ (in particular $b_1,b_2>0$).
Suppose that black wins, i.e. that $b'_2 = b'_1+\eps$ and $t'=\mu_r(B'_2-b'_2)$. 
Note that $b'_2$ (and reaching $t^*_r = \mu_r(B_2-b'_2)$) is not a winning equilibrium bid under $B_1$, i.e., by bidding $b^*_1=b'_2$ white remains with $B_1-b^*_1$ and gets a strictly better outcome $\mu_l(B_1-b'_2) \succ_1 t^*_r$. 
Therefore (the second inequality is by monotonicity of $\mu_r$), 
$$\mu_l(B'_1 - (b'_2+\eps)) = \mu_l(B_1-b'_2) \succ_1 t^*_r = \mu_r(B_2-b'_2) \succeq_1 \mu_r(B'_2-b'_2) = t',$$
 i.e., white has a deviation from $(b'_1,b'_2)$ in  $\gamma(s_0,B'_1)$. 
%
%Let $b^*_2=b'_2+\eps$ and note that $B_2-b^*_2 = (B'_2+\eps) - (b'_2+\eps) = B'_2-b'_2$, and $b^*_2<b_2$. Then at $\gamma(s_0,B_1)$, bid $b^*_2$ is not en equilibrium (or otherwise the algorithm would stop at $b^*_2$ rather than continue to higher bids), meaning that 
%\begin{align*}
%\mu_l(B'_1-b'_2)&=\mu_l((B_1+\eps)-(b^*_2-\eps)) =\mu_l(B_1-b^*_2+2\eps) \\
%&\succeq_1 \mu_l(B_1-b^*_2) \tag{by monotonocity of $\mu_l$} \\
%&\succ_1 \mu_r(B_1+b^*_2)  \tag{since $b^*_2$ is not an equilibrium bid in $\gamma(s_0,B_1)$}\\
%& = \mu_r(B_2-b^*_2)  = \mu_r(B'_2-b'_2) = t'.
%\end{align*}
%But this means that white can strictly gain at $\gamma(s_0,B'_1)$ by raising her bid from $b'_1$ to $b'_2$.

Now suppose that white wins, i.e. that $b'_2=b'_1$ and  $t'=\mu_l(B'_1-b'_1)$. 
We use the fact that $(b_1,b_1)$ is not an equilibrium in $\gamma(s_0,B_1)$, as black raised to $b_2=b_1+\eps$. Note also that $b'_2 < b_2-\eps = b_1$. 
\begin{align*}
\mu_r(B'_2-b_1) &= \mu_r((B_2-\eps)-b_1) = \mu_r(B_2-(b_1+\eps)) = \mu_r(B_2-b_2)\\
&\succ_2  \mu_l(B_2+b_1) \tag{by the last increase of black in $\gamma(s_0,B_1)$}\\
&\succeq_2 \mu_l(B_2 + b'_1) \succeq_2 \mu_l(B'_2+b'_1) \tag{by monotonocity of $\mu_l$}\\
& = \mu_l(B'_1-b'_1) = t'.
\end{align*}
This means that black can strictly gain at $\gamma(s_0,B'_1)$ by raising his bid from $b'_2$ to $b_1$. In either case (black or white win),  we get a contradiction to $(b'_1,b'_2)$ being an equilibrium, which proves the claim.

\medskip
Thus  $b'_2\geq b_2-\eps$. We next lower-bound the utility of the new outcome $t' =\mu(B'_1)$ to player~1: $t' \succeq_1 \mu_r(B'_1+b'_2)$, since if black wins in $\gamma(s_0,B'_1)$ then the right-hand side is $t'$, and if white wins then the outcome is strictly better for her, or otherwise she would not raise.
Then,
$$t' \succeq_1 \mu_r(B'_1+b'_2) =  \mu_r(B_1+b'_2+\eps) \succeq_1 \mu_r(B_1+b_2) = t_r,$$
where the second inequality is by monotonicity of $\mu_r$.

\medskip
Finally, the above proof only shows monotonicity for the white player. Monotonicity for the black player now follows immediately from property~\ref{e:PO}.  Since $\mu(B_2)=\mu(B_1)$ is  (weakly) better for white than $\mu(B_2+\eps)=\mu(B_1-\eps)$, it cannot be strictly better for black.  Otherwise $\mu(B_2+\eps)$ would be Pareto dominated by $\mu(B_2)$. Thus $\mu(B_2+\eps) \succeq_2 \mu(B_2)$.
\end{proof}

%As noted in Section~\ref{sec:gen_uniq}, the PSPE computed by Algorithm~\ref{alg:PSPE} may not be unique, and there may be other PSPEs that do not have these nice properties. This can be seen in the game $G_{\two}$ in Figure~\ref{fig:two_PSPE}, where there is a PSPE leading to the outcome marked with $**$, which is Pareto-dominated by $*$. However, this equilibrium is the most natural one, as it is the same one that is achieved by the ascending auction rule (as in Cor.~\ref{th:ascending}).

\rmr{move to discussion?, only applies for binary}
\subsection{Why Bottom Equilibrium?}
While there may be multiple equilibria, we argue that the Bottom Equilibrium, when exists, is the most natural one. One justification is that players may want to maintain as much budget as they can in each round, given that their utility in the rest of the game is monotone in the remaining budget. In other words, if we focus on the single bidding round, each player is inclined to bid the minimal amount.

A second informal justification is that in the Bottom Equilibrium each player is nearly indifferent between winning and losing the round: small deviations (i.e. higher or lower bid) of the other player $-i$ cannot substantially hurt $i$. Further, the loser  of the round in the Bottom Equilibrium never regrets if the other player bids too low and becomes the loser. On the other hand in other PSPEs the loser has a ``risk'' of becoming a winner with very low budget. Thus it is safer in some sense to play the Bottom Equilibrium. 

Our third justification is that  we can maintain the Bottom Equilibrium as the \emph{unique} PSPE by slightly changing the rules of the auction.\footnote{We thank David Parkes for this observation.} Indeed, suppose that instead of a sealed-bid auction in every step, we hold an ascending auction (similar to an English auction), where in each step the price rise by $\eps$ and the loser has a chance to accept the new price and become a winner. Since this is essentially equivalent to the auction being simulated in Algorithm~\ref{alg:PSPE}, the unique outcome is the Bottom Equilibrium.
%
%Consider the proof of Theorem~\ref{th:PSPE} and the related lemmas. It essentially shows that an ascending auction from any initial bids must converge, and in particular from the initial bids $(0,0)$. Moreover, we can write the ascending auction version of the game as a standard extensive-form game without simultaneous moves. %\footnote{This is clear in the discrete case. In the continuous case we can set a sufficiently low minimal raise. See also Section~\ref{sec:comp}.}  Then, by enforced genericity, the PSPE is unique.% Note that the $**$ outcome in game $G_{\two}$ above cannot be reached by ascending auctions, since no player would raise her initial bid in $s_0$ above $0$. 
\begin{corollary}
\label{th:ascending}
Any binary bidding game $\tup{G,\eps}$ with the above ascending auction rule has a unique PSPE, which is the Bottom Equilibrium of $\tup{G,\eps}$. %This also holds in non-generic games where players weakly prefer outcomes with higher social welfare.
\end{corollary}
%We will continue our analysis using the sealed-bid auction rule, yet we use last observation justifies the ``lowest-bid'' PSPE (the one described in Algorithm~\ref{alg:PSPE}) as the most natural one.

\subsection{Low budget resolution}
\label{sec:discrete_mon}
%\todo{explain that all proofs go through if we allow for a sufficiently large budget (say, $M\geq 2^{4k}$). If budget is low, then the result on monotonicity still holds, but not the main theorem on the Pareto mapping. Show example of a game that fails with one chip.}
%Consider bidding games where the bids and budgets are integers, and the total budget is some $M\in\mathbb N$.
 %Which results still go through and which results change? When considering existence, 
We showed in Section~\ref{sec:exist} that a PSPE exists regardless of the budget resolution. However for our main theorem we also assumed ``high resolution'' that is at least exponential in the height of the game tree. We next show that high resolution is necessary  for Pareto Efficiency (main part of Theorem~\ref{th:Pareto}) to hold. 

%The proof of Theorem~\ref{th:monotone} (monotonicity) works just the same with discrete bids, for any $M$. % We only need to change $\eps$ (the minimal significant budget change) with $1$. 
%The picture becomes more involved if we want a discrete variation of Theorem~\ref{th:Pareto}, since we explicitly used the fact we can set a particular budget to be between certain numbers. 

%
%How about Pareto-optimality? If $M\geq 2^{\height(G)+2}$, then we effectively simulate the continuous case, since the budget interval $\{0,1,\ldots,M\}$ can be partitioned to $2^{\height(G)}$ sub-intervals whose size is at least $3$, and the exact budget within each sub-interval has no effect on the outcome. 
%Thus Theorem~\ref{th:Pareto} holds in the discrete case when $M$ is sufficiently high (i.e., exponential in the height of the game tree). If $M$ is too low, this is no longer true. %See full version for a proof and more details. %See Appendix~\ref{apx:discrete} for a proof and more details.

\begin{proposition}
\label{th:discrete}
For any $k\geq 2$ there is a binary bidding game structure $G_k$ of height $k$, s.t. for any $1<R'<2^{k-1}$, the game $\tup{G_k,\frac{1}{R'}}$  with $B_1<0.5$ has no Pareto-optimal PSPEs.
\end{proposition} 
\begin{proof} Consider the binary game $G_k$, described in Figure~\ref{fig:discrete_PSPE}. In order to reach one of the Pareto-optimal leaves (marked with $*$), the same player must win $k-1$ times in a row after a branch is selected (as each branch is an ordinal 0-sum game).   Since $2^{k-1}B_{2} >  1$, and the budget of the loser doubles with each loss, player~1 cannot win in all $k-1$ times unless $B_2(y_1)=0$. Similarly, since after  every win of player~2, the budget of player~1 will be at least $\frac{1}{R'}>2^{1-k}$,  player~2 cannot win in all $k-1$ times either (even if initially $B_2(x_1)=1$).

Therefore, if after the first turn the selected child is $x_1$, the outcome will be $(8,1)$; and if the selected child is $y_1$ then the outcome will be $(1,8)$ (again, unless $B_2(y_1)=0$). The only way to reach $B_2(y_1)=0$ is if black bids all of his budget $b_2=B_2$ at $s_0$ and then selects $y_1$. However, this is not an equilibrium: since $b_1\leq B_1 < b_2-\frac{1}{R'}$, black can decrease his bid to $B_1+\frac{1}{R'}$ and gain $8$ instead of $7$. 
%
%Thus the only way to get to a Pareto-optimal outcome is if player~2 ``surrenders'' all of his budget by bidding $b_2=B_2$ at $s_0$, whereas player~1 bids less, and then player~2 selects $y_1$, and player~1 can win any number of times by bidding 0, and reach $(9,7)$. This would work e.g. if the initial partition is $B_2\le B_1$.  However if $B_2> B_1+\eps$, then bidding $b_2=B_2$ is not an equilibrium: Player~2 can decrease his bid by $\eps$ and still select $y_1$, thereby guaranteeing the outcome $(1,8) \succ_2 (9,7)$.  
%Thus the game must end in one of the ``side'' leaves, which are Pareto-dominated.

Note that due to our tie-breaking assumption it is impossible for agent~1 to similarly surrender her budget, since agent~2 must bid at least $\frac{1}{R'}$ to win each turn. 
\end{proof}
\input{example_bad_PSPE_discrete}

The example above does not cover the case of $R'=1$ (``single coin''). In that case, Consider the game $G'_k$ (for some $k\geq 2$) and a root node $s'$ whose children are $s_0$ and $(1,8)$. If player~2 starts with the coin ($B_2=1$) then in the only equilibrium he wins the first turn and selects $(1,8)$, as he cannot get more than $7$ by reaching $s_0$. 
\section{Continuous Bids}
\label{sec:real}
We denote by $\tup{G,-}$ the game $G$ with continuous bids, where the set of allowed budget partitions is $\calB=[0,1]$ and $B_1+B_2=1$.
Note that the strategy space in such games is infinite, and thus it is not a-priori clear that these games have a subgame perfect equilibrium, let alone a PSPE. Indeed, we show that if the tie-breaking method can depend on the node, then there are games, even zero-sum games, where no equilibrium exists.

\begin{proposition}
\label{th:no_PSPE}
There is a zero-sum bidding game with node-specific tie breaking, that has no subgame-perfect equilibrium.
\end{proposition}

\input{example_no_SPE}
\begin{proof}
Consider the game depicted in Fig.~\ref{sfig:H}. This is a zero-sum game where the white player is the maximizer. By the tie-breaking rule, white wins in $x'$ if her budget is strictly above $0.5$, and wins in $y'$ if her budget is at least $0.5$. In state $x$, white needs to take at least one turn to win the game. Thus a budget of  $B_1(x)> 0.25$ is sufficient. Similarly, in state $y$ white has to take twice, and thus must have a budget of $b_1(y)\geq 0.75$ to win the game. 

Clearly, if the initial budget partition in $s_0$ is $B_1\geq 0.5+\delta$ for some $\delta>0$, then white can win by bidding $b_1=0.25+\delta/2$: either she takes the turn and reaches $x$ with $0.25+\delta/2$, or loses the turn and reaches to $y$ with $B_1+b_2\geq 0.5+\delta + (0.25+\delta/2) = 0.75+3\delta/2 > 0.75$. Similarly, if $B_1 < 0.5+\delta$ then black can bid $0.25+\delta/2$ and force white to lose.

Now, suppose $B_1=B_2=0.5$.  We first show that there is no value in \emph{pure} strategies, by showing that for any bid of one player, the other player has a bid that leads to victory. The outcome for any pair of deterministic bids $(b_1,b_2)$ is displayed in Fig.~\ref{sfig:s0}. 

Let $b_1\in[0,B_1]$. If $b_1 < 0.25$ then set $b_2=b_1$. Thus black takes the turn and gets to $y$ with $B_2(y)>0.25$, which is sufficient for victory. 
% then black can bid $b_2=0.25$, get to $y$ with $B_2(y) \geq 0.25$, and thus win. 
If $b_1 \geq 0.25$, then black can bid $b_2<b_1$.  Now white remains with $B_1(x)= 0.5-b_1 \leq 0.25$, which means black wins. 
Thus white cannot guarantee any value above $-1$.
 %= \frac{b_1 + 0.25}{2} > b_1$. Then we will get to $x$ with $B_1(x) < 0.25$. 

We next consider the black player. Let $b_2\in [0,B_2]$. 
If $b_2< 0.25$ then denote $b_2=0.25-\delta$, and set $b_1=b_2+\delta/2<0.25$. Then white takes and gets to $x$ with $B_1(x)=0.5-b_1>0.25$, which is sufficient for victory.
If $b_2 \geq 0.25$, then set $b_1<b_2$.   Now black remains with $B_2(y)= 0.5-b_2 \leq 0.25$, which means white wins. 
Thus black cannot guarantee any value below $1$.
 %and get to $x$ with $B_1(x) \geq 0.25$ (and thus win). If $b_2 > 0.25$, then white can bid $b_1 < b_2$, and get to $y$ with $B_1(y) = B_1+b_2 > 0.75$ (and win the game). Therefore black cannot guarantee a victory (i.e. a value of $-1$) either. 

It is not hard to generalize the above argument to see that there is no value even in mixed strategies. Indeed, let $b_2\in \Delta([0,B_2])$ be any mixed strategy of black. There must be an open interval $A=(0.25-\delta,0.25)$ s.t. $b_2$ assigns a probability of less than $0.01$ to $A$. Then white can bid $b_1 = 0.25 - \delta/2$ and win w.p. of at least $0.99$: by our arguments on pure strategies, the only bids of black that beat $b_1$ are in the range $[b_1,0.25)\subseteq A$.  Therefore the value of the game under equal initial budgets is at least $0.99 (1)+ 0.01 (-1) = 0.98$. 

Similarly, let $b_1 \in \Delta([0,B_1])$ be some mixed strategy of white. Again there is some open interval $A' = (0.25-\delta,0.25)$ that is played by white w.p. less than $0.01$. Then by playing $b_2= 0.25-\delta/2$ black can win w.p. of at least $0.99$, and the value cannot be above $-0.98$. 
\end{proof}
 
Consider a game structure $G$. With continuous bids, the Bottom Equilibrium is not well-defined, as positive equilibrium bids may be arbitrarily small. However, we next show that playing with continuous bids is essentially the same as playing with high budget resolution. Denote by $\ol \calB^j(k)= [B_1^j(k),B_1^{j+1}(k))$ the $j$'th continuous budget interval of level $k$.  
\begin{theorem}\label{th:PSPE_cont}
Let $\gamma=BE(G,\eps)$ for some binary game $G$ and high resolution $\eps$. There is a PSPE $\ol \gamma$ of $\tup{G,-}$ such that for all $j<2^k$ and for all $\hat B_1\in \ol \calB^j(k)$, $\ol \mu(\hat B_1)=\ol \mu(B^j_1)=\mu(B^j_1)$ (where $\ol \mu=\mu_{\ol \gamma}$).
\end{theorem}
In other words, there is a PSPE of $\tup{G,-}$ that for any budget in the interval yields exactly the same outcome we would get in the Bottom Equilibrium of $\tup{G,\eps}$ for the same budget interval. %In particular, we get that as long as the resolution is high, it does not matter what is the exact value of $\eps$. 
\begin{proof}The proof is (as usual) by induction. 
Let $(b_1,b_2)=\gamma_s(B_1^j)$ be the equilibrium bids in the discrete game. We denote by $j'$ and $j''$ the respective level $k-1$ intervals of $B_1^j-b_1$ and $B_1^j+b_2$.
By induction,  $\ol \mu(B_1)=\ol \mu(B^{j'}_1)=\mu(B^{j'}_1)$ for all $B_1\in \ol \calB^{j'}(k-1)$ and likewise for $j''$. 
So that in level $k-1$ it only matters in which budget interval we arrive.

 For each $\hat B_1\in \ol \calB^j(k)$ we need to define equilibrium bids $\hat b_1\leq \hat B_1,\hat b_2\leq \hat B_2$ so that: (a) $\hat B_1 - \hat b_1 \in \ol \calB^{j'}(k-1)$; (b) $\hat B_2 - \hat b_2 \in \ol \calB^{j''}(k-1)$; (c) $\hat b_1 \geq \hat b_2 \iff b_1\geq b_2$. This will guarantee that any deviation to a different interval at $\ol \gamma_s(\hat B_1)$ entails a deviation in $\gamma_s(B_1^j)$. In addition, we have to make sure that (d) no agent has a new deviation to a different point inside the interval (i.e. one that was not previously available due to the discrete budgets).

 	Let $\delta = B^{j+1}_1-\hat B_1 >0$. All five cases of Lemma~\ref{lemma:four_cases} map into only three cases. 
% Recall  the proof of Theorem~\ref{th:PSPE}\ref{e:int}, where we showed that one of the bids in $\gamma_s(B_1^j)=(b_1,b_2)$ is ``critical'', and that at any $\hat B_1\in \{B_1^j,B_1^j+\eps,\ldots,B^j_1+(q-1)\eps\}$ the same critical point $B_{i'}^j-b_{i'}$ is reached (in some cases the critical player $i'$ is the winner in some cases the loser). Further, in all cases the non-critical point lies at $B_1^{j^*}(k-1)+\eps$ for some $j^*<2^{k-1}$. 
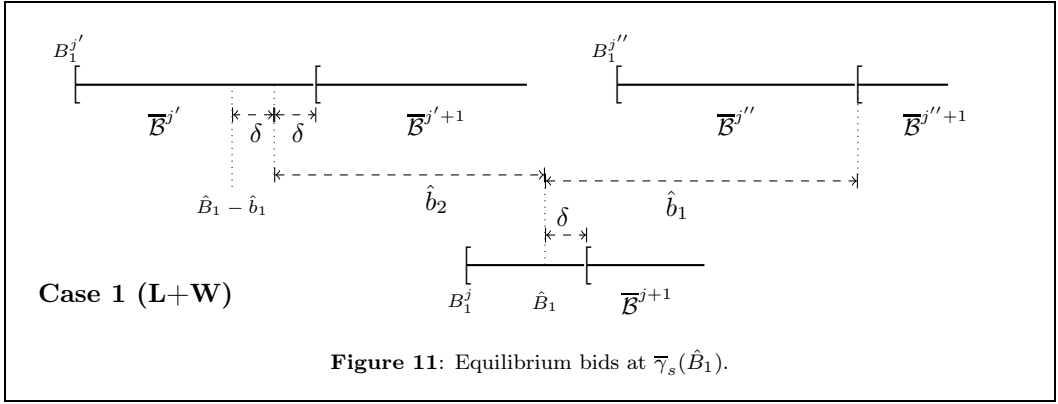
\include{intervals1L_cont}
\begin{itemize}
\item Case~0: $b_1=b_2=0$. We set  $\hat b_1=\hat b_2=0$ as well. (a)+(b)+(c)+(d) trivially hold.
	\item Case~1 (both L and W): $b_1=b_2>0$. %, the loser $-i^*=2$ is critical. %Note that for any $\hat B_1\in \calB^j(k)$, we have $\hat B_1-B^{j
	Note that according to both Cases~1L and 1W of Lemma~\ref{lemma:four_cases}, $B_1^{j'}+B_1^{j''}-2B_1^j+2^{-(k-1)}=0$.
	
	We define $\hat b_1 = B_1^{j''+1}-\hat B_1$, and $\hat b_2 = \max\{0,\hat B_1-B^{j'+1}_1+\delta\}$ (see Fig.~\ref{fig:intervals1_c}).
	%We define $\hat b_1= \hat B_1 -  (B_1^{j'}+2^{-k})$ and $\hat b_2=\hat b_1$. 
	%That is, that $\hat b_2$ is exactly the same size as $\hat b_2$ in the discrete case~1L (in Fig.~\ref{fig:intervals1L}), and $\hat b_2$ is smaller that $b_2$ but still in the same budget interval. 
	Thus (a)+(b) hold by construction.
	(c) holds even when $\hat b_2>0$ since
\begin{align*}
	\hat b_1-\hat b_2 &= B_1^{j''+1}-\hat B_1- (\hat B_1-B^{j'+1}_1+\delta) \\
	&= B_1^{j''+1}-2(B^{j+1}_1-\delta) +B^{j'+1}_1- \delta = B_1^{j''+1}- 2B^{j+1}_1 +B^{j'+1}_1 +\delta\\
	&=   B_1^{j''}+2^{-(k-1)} + B^{j'}_1 +2^{-(k-1)}- 2B^{j}_1 - 2\cdot 2^{-k} +\delta= \delta>0.
	\end{align*}
	As for (d), %we use Case~1L from Lemma~\ref{lemma:four_cases}: 
	%$$B_1^j-(B^{j'}+\eps) = b_1 = b_2 = (B^{j''}_1 + (2R-1)\eps) - B_1^j ~~\Rightarrow~~ 2B_1^j-B_1^{j'} =  B^{j''}_1+2^{-(k-1)}.$$
%Thus 
for any $b'_2>\hat b_1$, we have

$$\hat B_1 + b'_2 > \hat B_1 + \hat b_1  =  B^{j''+1}_1,$$
i.e. we get a point outside the interval $\ol \calB^{j''}$, so black cannot gain by bidding higher. White cannot gain by bidding $b'_1\in[\hat b_2,\hat b_1)$ since $\hat B_1-b'_1\leq \hat B_1-\hat b_2= B^{j'+1}_1-\delta\in \ol \calB^{j'}$, i.e. in the same budget interval. 

 %Note that $\ol \gamma_s(\hat B_1)$ is defined just like $\gamma_s(\hat B_1)$ in Fig.~\ref{fig:intervals1L}, only $\hat B_1$, $\hat b_1$ and $\hat b_2$ are not restricted to integral multiples of $\eps$.
	%\item Case~1W: $b_1=b_2$, the winner $i^*=1$ is critical. We define the bids in the same way as in Case~1L, thus (a)+(b)+(c) hold. (d) is not a problem since the loser $-i^*=2$ prefers $j'$ over $j''$.
	%
	\item Case~2 (both L and W): $b_2=b_1+\eps$. %, the loser $-i^*=1$ is critical.
	%Note that we can write $\hat B_1=B_1^{j}+2^{-k}-\delta$ for some $\delta>0$.
	We set $\hat b_1= \hat B_1 -  B_1^{j'}$ and $\hat b_2 = B_1^{j''+1}-B_1^{j+1}$. Clearly (a)+(b) holds. (d) also holds since (as in Case~1) it is impossible for $-i^*=1$ to deviate and stay in the same interval, and lowering the winning bid of $i^*=2$ still reaches interval $\calB^{j''}(k-1)$. It remains to show (c).

	In both Cases~2L and 2W of Lemma~\ref{lemma:four_cases} we have
	\begin{align*}
	\text{Case~2L:~~~}&B_1^j-B^{j'}_1+\eps = b_1 + \eps = b_2 = (B^{j''}_1 +\eps) - B_1^j ~~\Rightarrow~~ B^{j''}_1+B_1^{j'}- 2B_1^j =0;\\
	\text{Case~2W:~~~}&B_1^j-B^{j'}_1= b_1 + \eps = b_2 = B^{j''}_1  - B_1^j ~~\Rightarrow~~ B^{j''}_1+B_1^{j'}- 2B_1^j =0.
	\end{align*}
	Then we get that similarly to Case~1, 
	\begin{align*}
	\hat b_2-\hat b_1 &= B_1^{j''+1}-B_1^{j+1} -(\hat B_1 -  B_1^{j'}) \\
	&= B_1^{j''+1}-B_1^{j+1} -((B_1^{j+1}-\delta) -  B_1^{j'}) = B_1^{j''+1}+B_1^{j'}-2B_1^{j+1} +\delta \\
	&= B^{j''}_1 + 2^{-(k-1)}+B_1^{j'}- 2B_1^j -2\cdot 2^{-k} +\delta = \delta >0,
	\end{align*}
	%\item Case~2W: $b_2=b_1+\eps$, the winner $i^*=2$ is critical. We set $\hat b_2 = B_1^{j''}-B_1^j$ and $\hat b_1 = \hat b_2-\delta$ ($\delta$ defined as in Case~2L). Thus (b)+(c) hold immediately. (d) holds since the loser $-i^*=1$ prefers $j''$ over $j'$. Showing (a) is similar to Case~2L: 
	%$$B_1^j-B^{j'} = B_1^j-(B^{j'}+\eps)+\eps = b_1 + \eps = b_2 = B^{j''}_1 - B_1^j ~~\Rightarrow~~ 2B_1^j-B_1^{j''} =  B^{j'}_1,$$
	%and then
	%\begin{align*}
	 %\hat B_1 - \hat b_1 &=\hat B_1 -(\hat b_2 - \delta) = \hat B_1 -(B^{j''}_1-\hat B_1)+\delta = 2\hat B_1 - B^{j''}_1 + \delta\\
	%&=2(B^j+2^{-k}-\delta)-B^{j''}_1+\delta = 2B^j_1-B^{j''}_1 + 2^{-(k-1)}-\delta \\
	%&= B^{j'}_1+ 2^{-(k-1)}-\delta\in \ol \calB^{j'}(k-1).
	%\end{align*}
\end{itemize}
	as required.
\end{proof}

One immediate corollary is that as long as the resolution is high, the Bottom Equilibrium induces the same mapping from budget intervals to outcomes (since for any $\eps$ the mapping is the same one as in the continuous case). In the continuous case we may have (infinitely) many PSPEs, all with the mapping $\ol \mu$, but neither of them can be referred to as ``\emph{the} Bottom Equilibrium.''
We thus say that a PSPE $\ol \gamma\in \Gamma^*(G,-)$ is \emph{a Bottom Equilibrium} if $\mu_{\ol \gamma}=\ol \mu$, i.e. if it induces the same outcome as $BE(G,\eps)$ for some high resolution $\eps$.

%%%%%%%%%%%%%%%%%%%%%%%
 
\section{From Bidding Games to Combinatorial Bargaining}
\label{sec:bar}
There are two ways we can think of bidding games. First, we can look for games that inherently have budgets and sequential bidding. While there are some examples of scenarios that arguably fit into this model, typically the incentive structure and/or the bidding structure is quite different (see Section~\ref{sec:other_games} below, and the Related work section).
A more promising way is to think of the sequential bidding process as a \emph{mechanism} that is designed to increase cooperation and welfare among a pair of players in various scenarios. %The scrip auction mechanism is a prime example of such an application, but we do not have to limit our scenario to allocation of items.

\subsection{Sequential scrip bargaining}

\begin{definition}
\label{def:SSA}
A 2-player \emph{sequential scrip bargaining} (SSB) is a tuple $F=\tup{K,v_1,v_2,\tau}$, where:  
	$K$ is the set of items ($k=|K|$); $v_i:2^{K}\rightarrow \mathbb R$ is the value function of agent $i\in N$; and $\tau$ is a permutation over items $K$. 
%\
%\begin{itemize}
	%\item $N=\{1,2\}$ is the set of bidders; 
	%\item $K$ is the set of items. We denote $k=|K|$;
	%\item $v_i:2^{K}\rightarrow \mathbb R$ is the value function of agent $i\in N$;
	%\item $\tau$ is a permutation over items $K$. 
%\end{itemize}
\end{definition}
Intuitively, items $K$ are offered for sale according to order $\tau$, and in each turn agents bid for the current item. To complete the game we need to describe the bidding rules. As expected, we assign a budget $B_i$ to each agent; the highest bidder in each turn pays her bid to the other agent, and decides who gets the item. In case of a tie we treat agent~1 as the winner.

This bargaining mechanism seems like a close reminiscent of \emph{sequential auctions}~\cite{gale2001sequential,bae2007efficiency}.
However, typically in the auction literature items are auctioned in a first- or second-price auction where the highest bidder pays the (first or second) bid to the seller; and utilities are quasi-linear. In particular, an agent's utility depends not just on the allocation, but also on the amount of money the agent paid for her items (see Section~\ref{sec:related}). 

SSBs differ from these standard auctions in two ways. First, while we use the first-price auction rule, the higher bidder pays the other player, rather than paying to the seller, highlighting that this is a bargaining mechanism rather than an auction.  Second and more importantly, the budget $B_i$ is ``scrip money'' and has no value outside the game, thus the utility of $i$ from a bundle $S_i\subseteq K$ is $v_i(S_i)$, regardless of how much of the budget the agent spent.

\subsection{SSBs are Binary Bidding Games}
%It is not hard to verify that binary bidding games and sequential script auctions are in fact equivalent. We describe the formal mapping between them for concreteness and clarity. 

\begin{proposition}
\label{th:SSA_bidding}
Every SSB with $k$ items has an equivalent bidding game structure over a binary tree of height $k$, and vice versa. % Also, any  binary bidding game of height $k$ is equivalent to an SSA with $k$ items.
\end{proposition}

By ``equivalent to'' we mean that for any resolution $\eps$, there is a one-to-one mapping between strategy profiles in the original and induced game, that preserves utilities. Intuitively, we can describe any SSB as a binary tree where any internal node at depth $k'$ corresponds to an allocation of the first $k'$ items. Note that this mapping is independent of the initial budget allocation, the budget resolution, and even of the bidding rule.
\begin{proof}
In the first direction, given $F=\{K,v_1,v_2,\tau\}$, let $K_d\subseteq K$ be the set of first $d$ elements of $K$ (according to $\tau$). We construct a bidding game $G=\tup{S,s_0,T,g,u_1,u_2}$ as follows. Let $S_d$ be the set of all $2^d$ possible bi-partitions  of $K_d$ for $d\in\{0,1,\ldots,k\}$, and $S=\bigcup_{d\leq k}S_d$. Note that these are partial allocations of $K$ to the two players. $s_0$ is the empty allocation (the unique member of $K_0$). Every state can be written as $s=(K^1,K^2) \in S_d$, where $K^i$ is the set of items held by player~$i$ in state $s$. There are two states accessible from $s$: either player~1 or player~2 get item $\tau(d+1)$. Thus $g(s)$ contains exactly two states in $S_{d+1}$. The set of terminals $T$ coincides with $S_k$, i.e. all full partitions of $K$. Finally, for every $t=(K^1,K^2)\in T$, we set $u_i(t) = v_i(K^i)$. Thus $G$ is an impartial binary tree of height $k$, which is clearly equivalent to $F$.

In the other direction, suppose we are given a  binary bidding game $G$ of height $k$. W.l.o.g. $G$ is a balanced tree: if it is in DAG form, we can clone every node with several incoming edges, along with its subtree; if some branches are shorter than $k$, we extend them in the trivial way: replace each terminal $t$ at depth $k'<k$ with a balanced binary subtree of height $k-k'$, all of whose leaves are identical to $t$. 

Now we have a balanced tree of height $k$. For every internal node, arbitrarily label one child as ``left'' and the other as ``right''. We construct an SSB $F=\tup{k,v_1,v_2,\tau}$ as follows.  Let $\tau$ be the identity permutation over $K=[k]$. We identify any path from the root $s_0$ to a terminal $t$ with a partition $(K^1,K^2)$ of $K$ where $K^1$ contains all levels $d$ s.t. the left child was selected, and $K^2$ contains all other levels (i.e., where the right child was selected). Note that any $K'\subseteq K$ appears in exactly one terminal $t'_1$ as $K^1$, and in exactly one terminal $t'_2$ as $K^2$. We set $v_i(K')=u_i(t'_i)$. Thus the set of allocations in $F$ coincides with the set of paths in the original game $G$. 
\end{proof}
%While the proof is in the full version, %Appendix~\ref{apx:SSA_bidding}, 
%it is not hard to see why this is true. Intuitively, we can describe any SSB as a binary tree where any internal node at depth $k'$ corresponds to an allocation of the first $k'$ items. Then taking the left branch means white takes the next item, whereas right branch means the item goes to black. In the other direction, we can assume w.l.o.g. that the binary three is complete. Then we identify each of the $2^k$ leaves of the tree with an allocation of the $k$ items. 
%
For example, the game $G_{\maj}$ (Sec.~\ref{sec:bidding_games}) coincides with a sequential scrip bargaining over three identical items, where each agent assigns a value of $1$ to bundles of size two or more, and $0$ otherwise.

Since SSBs are essentially equivalent to bidding games on binary trees, an immediate corollary of Theorem~\ref{th:Pareto} and Prop.~\ref{th:SSA_bidding} is the following. For a set of items $K$, let $T_P(K)$ be the set of Pareto-efficient allocations of $K$ among two players.
\begin{theorem}
Let $F=\tup{K,v_1,v_2,\tau}$ be a sequential scrip bargaining, and let $\eps$ be some high budget resolution. Then $\tup{F,\eps}$ has a Bottom Equilibrium $\gamma$, s.t. $\mu=\mu_\gamma$ is a monotone and surjective mapping from $\calB_\eps$ to $T_P(K)$.
\end{theorem}

%
%Unfortunately, in some cases there may not be an allocation that meets $m$-maximin share for any constant $m$. In particular, our mechanism (or any other mechanism) cannot find such an allocation. As an example, suppose that items are vertices of an $(m+2)\times (m+2)$ grid. White has a value of $1$ for a full row, whereas black has a value of $1$ for a full column. Since this is a zero-sum game (a ``square'' variation of Hex), in any allocation there is a player $i$ where $v_i(K_i)=0$. However, we can divide $K$ to $m+2$ parts (to rows if $i=1$ and to columns if $i=2$), and the value of each such part to $i$ is $1$. 
%
%Nevertheless, we have the following result.
%\begin{proposition}
%Let $F$ be an SSA, and suppose that initial budgets are equal. Then the allocation attained by Algorithm~1 meets the envy by a single good criterion.
%\end{proposition}
%\begin{proof}
%Since ties are broken in favor of white, and the game is impartial, clearly white does not envy black (even without removing a good). Let $K_1,K_2$ be the final bundles. Let $a’ \in K_1$ be the first  item that white paid for a non-zero amount (i.e., $b’_1>0$). If there is no such item, then black has his optimal bundle, and in particular is not envious.  After $a’$ was allocated, black had more budget, so A2 is preferred over A1\a’.  [what if black won first?]
%\end{proof}
%

%We note that the mapping $\mu$ may depend on the order $\tau$ in which items are sold.
As we mentioned  in the introduction, the set of items $K$ need not be a collection of physical items. We next consider some applications of SSBs. % with  ``virtual items.'' %, where a complete assignment corresponds to an allocation.

%\paragraph{Boolean function games}
%Consider a Boolean function $f:\{0,1\}^k\rightarrow \mathbb R$. Suppose that possible states of the world are vectors $\vec a\in\{0,1\}^k$, and the utility of each agent is defined via some Boolean function $f_i(\vec a)$. Then we can construct a game where the player playing in step $j$ sets the value of $a_j$. An equivalent description of this game is as an SSA, where one player ``gets'' all coordinates that are set to $0$, and the other gets those that are set to $1$. Then $v_i(K')=f_i(\vec a')$, where $a'_l=1$ iff $l\in K'$ (similarly to Prop.~\ref{th:SSA_bidding}).
%

%If turns are set via bidding with budgets, then we get a bidding game over a complete binary tree. By Theorem~\ref{th:Pareto} there is a PSPE where the final assignment is Pareto-efficient.  

\subsection{Applications for SSB mechanisms}
\label{sec:apps}
\paragraph{Multi-issue voting}Consider a set of voters, voting over $k$ binary issues~\cite{winter1997negotiations,lacy2000problem,lang2009sequential}. Here the voters (who are the players) have a complete (weak) preference order over the $2^k$ outcomes, but need not attribute cardinal utilities to them. It is known that when preferences are unrestricted, either truthful or strategic voting may lead to the selection of a Pareto-dominated outcome. Voting on the issues sequentially provides a partial solution, but Pareto-dominated outcomes may still be selected~\cite{lacy2000problem}. While our model only applies for two voters, it provides  strong guaranties of Pareto-efficiency, regardless of the agenda or the structure of preferences. In particular, a Condorcet loser is never selected, and the Condorcet winner is always selected if it exists. 

\paragraph{Arbitration}
For two self-interested parties, a bidding game can be used as an arbitration mechanism that reaches an efficient outcome, and also allows a natural way to take into account parties of different importance, or weight.  As a concrete example, we can think of the Democrats and the GOP bargaining over the various clauses of the Healthcare Act, or some other reform. The initial budget of each party can be set, say, based on its number of seats. Arbitration mechanisms for two parties (not necessarily in a combinatorial setting) have been widely studied in the literature~\cite{sprumont1993intermediate,anbarci2006finite,de2012selection}. 

The \emph{minimal satisfaction test (MST)}~\cite{de2012selection} is an ordinal fairness criterion, that is satisfied if the selected outcome is above the median outcome in every player's preference order. That is, every player weakly prefers the selected outcome to at least half of all possible outcomes. 
De Clippel et al. studied several sequential mechanisms, and showed that three of them---The \emph{Alternate-Strike} mechanism~\cite{anbarci2006finite}, the \emph{Voting by Alternating Voters and Vetoes} mechanism~\cite{sprumont1993intermediate}, and the \emph{Shortlisting} mechanism---each implement some Pareto-efficient outcome that satisfies MST. 

\paragraph{SSB and fair arbitration mechanisms}
%A standard requirement in allocation problems is that the allocation will be \emph{fair}. Of course, if one of the players have the entire budget, then the outcome would very biased in her favor. However we may  still  say that a mechanism is fair if the  utility of a player is in some sense proportional to her budget, or that the envy of a player in the other player is bounded by some function of her initial budget.
%%When dealing with allocation problems, several fairness criteria have been suggested along the lines of proportionality and envy-freeness~\cite{budish}. 
%However, we note that the PSPE outcome in our game only depends on the \emph{ordinal preferences} over outcomes. We can change the cardinal utilities of a game arbitrarily as long as the ordinal relations remain the same, and the equilibrium outcome would not change. There is thus little hope that any notion of fairness that is based on cardinal utilities (such as proportionality) can be guaranteed. 
Consider the following parametrized extension of the MST requirement. 
We say that an outcome $t$ satisfies the $(\alpha_1,\alpha_2)$-satisfaction test, if each player~$i$ weakly prefers $t$ to at least $\ceil{\alpha_i |T|}$ outcomes. 
\begin{proposition}
\label{th:MST}
Let $\tup{G,\eps}$ be a bidding game over a full binary tree, then \emph{any} PSPE outcome satisfies the $(B_1,B_2)$-satisfaction test.
\end{proposition}
In particular, for equal budget we get that the outcome satisfies the minimal satisfaction test of De Clippel at al.~\shortcite{de2012selection}. 
\begin{proof}
We will prove that there is a strategy for player~1 that \emph{guarantees} an outcome weakly preferred to $\ceil{B_1|T|}$ outcomes, and likewise for player~2 and $\ceil{B_2 |T|}$. Since any equilibrium outcome is at least as good as the maximin outcome, this  entails the proposition.

Given $G$ and $B_1$, we define a zero-sum game structure $G'$, that is equivalent to $G$ except for the utility functions. We set $u'_1(t) = 1$ if $t$ is weakly preferred to at least $\ceil{B_1|T|}$ terminals, and $-1$ otherwise; and set $u'_2=-u'_1$. Thus each player can either `win' or `lose' in every terminal. Note that if $B_1 |T|$ is an integer, then there are exactly $|T|-B_1|T| +1 = (1-B_1)|T|+1$ winning nodes, and $B_1|T|-1$ losing nodes. Otherwise there are $\floor{B_1|T|}$ losing nodes, since the next one is weakly preferred to all of them plus itself, i.e. to $\ceil{B_1|T|}$ terminals. Thus the number of losing nodes is exactly $\ceil{B_1|T|}-1$.

According to Lazarus et al.~\shortcite{lazarus1996richman}, the \emph{Richman value} $R(s)$ of a node in a zero-sum win/lose game, is the minimal value s.t. any $B_1>R(s)$ guarantees a victory for player~1. Lazarus et al. show that $R(s)$ equals \emph{exactly} to the probability that player~1 loses in the corresponding ``spinner game'' of $G'$ (the same game only turns are taken at random rather than by bidding). 

This probability is easy to compute. Since $G'$ is a zero-sum game, at each node, white (weakly) prefers one child and black prefers the other. W.l.o.g, white always  prefers the right child and black always prefers the left.
 Thus if players take random turns, they  reach a terminal that is selected uniformly at random from $T$. 
We conclude that 
$$R(s_0)= Pr_{t\sim \text{Uniform}(T)}(u'_1(t) = -1) = \frac{|\{t:u'_1(t) = -1\}|}{|T|}=\frac{\ceil{B_1|T|}-1}{|T|} < B_1,$$
which means that a budget of $B_1$ is sufficient to guarantee a victory to player~1 in $G'$. By definition of $G'$, player~1 can guarantee an outcome in $G$ that is weakly preferred to $\ceil{B_1|T|}$ outcomes. The proof for player~2 is symmetric.
\end{proof}
\iffalse
What if $G$ is not a full tree? Then some terminals may be more likely then others and $(B_1,B_2)$-ST may not hold \emph{ex post}. However if for a given set of outcomes $T$ we construct a game tree at random, it is easy to see that the criterion holds in expectation. Further, We can always construct a tree such that all terminals are at depth $h$ or $h-1$. In such three the probability of selecting a shallower terminal is exactly double. It can be shown that in every such instantiation, the SSB mechanism still guarantees a $(\frac{B_1}{2},\frac{B_2}{2})$-ST outcome.
\fi

%See Appendix~\ref{apx:MST} for the full proof.
%\begin{proof}[ sketch]
%We show that each player has a strategy that \emph{guarantees} her side of the MST. By assuming that the other player is an adversary, our player $i$ can think of the game as a zero-sum game, where she loses iff she gets one of the worst $\ceil{\alpha_i |T|}$ terminals, i.e. a Richman game~\cite{laz99}. Then we use the result from \cite{laz99} that  the budget sufficient to guarantee a victory is exactly the probability of losing in a random turn game --- which is exactly $\ceil{\alpha_i |T|}$. % In a full binary tree random turn is equivalent to choosing a terminal uniformly at random, and thus the probability of losing is exactly $\ceil{\alpha_i |T|}$. 
%Finally, the equilibrium outcome is at least as good as the minimax outcome. %for both players as the outcome of their safety-level strategies. 
%\end{proof}

The SSB mechanism has the additional property of implementing the entire set of Pareto-efficient outcomes. As the proposition shows, every such outcome that is attained under a given budget allocation, satisfies the parametrized MST. 

\paragraph{Communication complexity}
Suppose that Alice and Bob each hold private information $a\in A$ and $b\in B$, respectively. They can exchange arbitrary messages sequentially in some predefined order. 
Given a function $f:A \times B \rightarrow \{0,1\}$, the \emph{communication complexity} of $f$ is the total number of bits Alice and Bob must exchange so that at least one of them knows $f(a,b)$ with certainty~\cite{yao1979some}.

Communication complexity in games was studied mainly in the context of equilibrium computation, where $a$ and $b$ are the utility functions of players in some game, and $f$ is some property of the equilibrium of this game~\cite{conitzer2004communication,hart2010long}. However, for an SSB with $k$ items, the size of the valuation function of each player is already exponential in $k$ (in the general case), so there is no much hope to compute an equilibrium with little communication. 

We have a much more modest requirement.  
Let $H=\tup{\Gamma,\Psi,h}$ be a 2-player game form, where $\Gamma=\Gamma_1 \times \Gamma_2$ is the set of valid pure strategy profiles, and $\Psi$ is the set of possible outcomes, and $h(\gamma_1,\gamma_2)\in \Psi$ is the outcome of playing strategies $\gamma_1,\gamma_2$ in $H$.

\begin{definition} The \emph{playing communication complexity} of $H$ is the number of bits players need to communicate so that at least one of them knows $h(\gamma_1,\gamma_2)$.
\end{definition}
In our case, $\Psi = 2^K$ contains all possible allocations of $K$ among the two players, and $H$ is some mechanism the players use to determine the allocation.  

For example, in the simple mechanism where players each choose sequentially an item from $K$ until all items are selected, there are $k$ rounds, and the communication required in each round is $\log(k)$ bits to specify which item. Thus the playing communication complexity of this mechanism is $O(k\log (k))$ (even though the strategies players use may be complicated).
  
On the other hand, consider the Shortlisting mechanism~\cite{de2012selection}. The mechanism has two rounds where player~1 first marks the $2^{k-1}$ outcomes that she prefers to all others, and then player~2 selects among them. Clearly revealing the full strategy $\gamma_1$ to player~2 takes a number of bits that is exponential in $k$, and there does not seem to be a more efficient way to find $h(\gamma_1,\gamma_2)$. The other mechanisms in \cite{de2012selection} suffer from a similar problem.\footnote{Recall that these mechanisms were not designed for combinatorial valuations so this is not surprising.}

Note that given an order $\tau$ over items, the SSB $F(K,\tau)$ is a valid allocation mechanism that maps any pair of strategies $\gamma=(\gamma_1,\gamma_2)$ to an allocation $\mu_\gamma\in 2^K$. 
\begin{proposition} For any order $\tau$, the playing communication complexity of SSB is $O(k^2)$. 
\end{proposition}
\begin{proof}
An SSB has $k$ rounds. In each round, each player needs to submit a bid, and the winner decides on whether to take or give the current item. For any constant resolution $R$, there are $R2^k$ possible budgets (and thus possible bids). So $\log(R2^k) = O(k)$ bits are sufficient to specify the bid. One additional bit is sufficient to specify the decision on the item. Thus in total each player is required to transfer $O(k^2)$ bits in order to play her strategy. 
\end{proof}
 
%While these properties are also satisfied by some of the mechanisms mentioned above, they require amount of communication that is exponential in $k$. Other auction mechanisms tradeoff between efficiency and communication complexity~\cite{holzman2004bundling}.  

SSB is the first mechanism that guarantees efficiency, fairness, and low playing communication complexity.
Thus SSB may be considered as a feasible and desirable mechanism for combinatorial bargaining/arbitration when the playing parties are asymmetric. 

\paragraph{Zero-sum games}
In a zero-sum game, player~1 gets an outcome that beats $B_1 |T|$ outcomes, if and only if player~2 gets an outcome that beats $B_2 |T|$ outcomes. Thus by Prop.~\ref{th:MST} the set of outcomes and the utility functions define the (unique) PSPE outcome completely, regardless of the tree structure. 

\subsection{Other games}
\label{sec:other_games}
Consider any (impartial) game in tree-form. %\footnote{Of course, many games are not symmetric. E.g., in Chess the white player may not use her turn to move a black pawn.}
If the tree is not binary, we can modify it by recursively breaking decision nodes to a subtree of binary decisions. Then we can divide a large or continuous budget among the players, and let them play the new bidding game. If players adhere to the Bottom Equilibrium, then the initial budget partition will determine the final outcome (from $T_P(G)$).  Thus we can get a good outcome even in games where playing by turns would lead to a poor outcome. In particular, with proper initial budgets we can implement the outcome with maximum social welfare, maximum Egalitarian welfare, etc.

\paragraph{Centipede games}
One class of games that falls under the conditions of Theorem~\ref{th:Pareto}, is that of Centipede games~\cite{rosenthal1981games}. %\footnote{\url{http://en.wikipedia.org/wiki/Centipede_game}.} 
Such games are a notorious example to how rationality leads players to end up in poor outcomes. Under random turn there are still Centipede games where players always choose to finish early, even though staying in the game is eventually significantly better for both.

Nevertheless, in our game setting there is a PSPE where players are guaranteed to continue until they reach one of the Pareto-efficient outcomes (the last two leaves).

\paragraph{Nash bargaining game}
The Nash bargaining game \cite{nash1950bargaining} for two players is typically described by a (convex) set of feasible outcomes $\calF$ in the plane, whose boundary forms the Pareto-efficient frontier, and a status-quo point $q$. Given $\calF$ and $q$, we can think of an extensive-form bargaining game, where players start from $q$ and at each state (point of the grid) they can either ``go right'' (increase the utility of player~1) or ``go up'' (increase the utility of player~2). The terminal states are the outcomes along the Pareto frontier. In this case it is clear that every outcome is Pareto-efficient (as this is how we defined the terminal states), but the properties of monotonicity and Pareto-surjective are still interesting. The (unique) PSPE induces a generalized solution concept for the Nash bargaining game (a point on the Pareto frontier for any budget allocation). %\footnote{Since our model requires a finite number of states, we also need to define a ``step size''. We can then look at the limit of our solution as step size approaches $0$.}

A natural question is how the PSPE solution---in particular for the case of equal budgets---compares with other solution concepts such as the Nash bargaining solution and the Kalai-Smorodinsky bargaining solution~\cite{kalai1975other}. We note that there is much interest in non-cooperative implementations of various bargaining solutions, e.g.~\cite{moulin1984implementing}.

To answer our question, we first modify the game tree described above, by complementing each leaf (a point on the Pareto frontier) to a subtree with identical leafs, such that all leafs are at the same depth. Clearly this does not change the equilibria of the game---it just means the game continues for a few more redundant rounds after the outcome is already determined. 
Since the SSB induced by a Nash bargaining game is ordinally-zero-sum (as all terminals are on the Pareto frontier), binary, and complete, it follows from Proposition~\ref{th:MST} that when the budget is equally partitioned, the realized outcome is reached when players make the same number of moves. In other words, if we normalize $q$ to the origin $(0,0)$, then both players have the same utility (or the closest approximation to same utility), which coincides with the \emph{Egalitarian}, or \emph{proportional}, solution~\cite{kalai1977proportional}. %In fact, different initial budget allocations map directly to different proportional solutions. For example, if $B_1 =\frac34$, 

%We can think of the realized PSPE outcomes for other budget partitions as a generalization of the Egalitarian principle to (two) agents with different weights, or entitlements.  

\section{Computational Issues}
\label{sec:compute}
In this section we elaborate on the algorithmic considerations in computing the Bottom Equilibrium. The key contribution in this section is Algorithm~\ref{alg:PSPE_fast}, presented in Page~\pageref{alg:PSPE_fast}. We also implemented the algorithm in Matlab (along with code to generate simple binary bidding games), and will share the code upon request.

\begin{proposition}\label{th:PSPE_fast}
Let $G$ be a binary bidding game structure. Algorithm~\ref{alg:PSPE_fast} computes a profile $\gamma$ that is outcome-equivalent to a Bottom Equilibrium of $\tup{G,-}$. 
Moreover, Algorithm~\ref{alg:PSPE_fast} runs in time $|S|\cdot \text{poly}(|T|) = poly(|S|)$. 
\end{proposition}
Before showing the full proof, we explain the intuition. How can we even represent the equilibrium efficiently, if at any node there are infinitely many budgets to consider (or exponentially many in $k$, with discrete high resolution)? The answer lies in \emph{monotonicity}. Since we can order all reachable terminals, and the outcome is monotone in the budget of player~1, we only need to find the cutoff points in the interval $[0,1]$ where the next terminal becomes the outcome. There can be at most $|T|-1$ such cutoff points, and the main challenge is to show: (a) that they can be computed efficiently from the cutoff points of both children; and (b) that for every budget between two given cutoff points, we can efficiently compute some bids that reach the correct equilibrium outcome. 
\begin{proof}
We saw that at any level $k$ budget interval $\ol \calB^j(k)$, the equilibrium bids in all $B_1\in  \ol \calB_1^j$ reach the same level $k-1$ intervals (which we previously denoted by $\ol \calB^{j'}(k-1),\ol \calB^{j''}(k-1)$), although they may reach different budget points. Our algorithm only computes this mapping, i.e. from $j$ to $j'$ and $j''$, and then assigns arbitrary bids that reach those intervals from $B_1^j$. These bids may not be stable themselves, but they contain enough information to reconstruct the equilibrium bids from any $B_1\in  \ol \calB_1^j$ according to Theorem~\ref{th:PSPE_cont}.

There are three differences between Algorithm~\ref{alg:PSPE} (see Page~\pageref{alg:PSPE}) and Algorithm~\ref{alg:PSPE_fast}: (a) the fast algorithm only computes the equilibrium in specific budget points ($z_{lr}$), rather than for any possible budget $B\in \calB_\eps$ (or even any $B_1^j(k)$); (b) when black deviates, the fast algorithm increments the bid by a quantity larger than $\eps$; (c) after computing the equilibrium at all critical budget points, the fast algorithm ``filters'' some points out, and does not add them to the description of the PSPE. In contrast to $\eps$ in  Algorithm~\ref{alg:PSPE}, the exact size of $\delta$ is not important, as long as it is sufficiently small (anything below $2^{-\height(G)-1}$ would suffice). 

We need to show that despite those differences we get the same equilibrium outcome, i.e., the same mapping $\mu:[0,1]\rightarrow T(G)$.

The $a$'th budget interval is the union of several consequent subintervals of size $2^{-k}$. Formally, $[F_{s}(a),F_{s}(a+1))=\bigcup_{2^kF_{s}(a)\leq j < 2^kF_{s}(a+1)} \ol \calB^j$.

\rmr{fix this?}
We start by showing (b), i.e. that by
% Note that $B_1$ and $b_1$ are always a multiple of $2^{-\height(G)}$. In the slow algorithm, white raises her bid to $b'_1=b_2$ and remains with a budget of $B_1-b_2$. In the new algorithm,  we have $B_1-b''_1=B_1-b_2-\delta = $ belong to the same budget interval, and thus  by Theorem~\ref{th:PSPE_cont} the outcome is the same. 
 bidding $b'_2 = b_1+\delta$ or \\
$b''_2 = \min\{B_1-F_{s_1}(a_1-1), F_{s_2}(a'_2+1)-B_1-\delta\}$, the same equilibrium is reached. We first argue that $b''_2 > b_1$ (so black indeed takes). 
Since $B_1-b_1\geq F_{s_1}(a_1)$, and $F_{s_1}$ is strictly increasing, clearly $B_1-F_{s_1}(a_1-1)>b_1$. %and thus $B_1-F_{s_1}(a_1-1)-\delta>b_1+\delta = b'_2$ (since $b_1$ and every $F_{s}(a)$ are a multiple of $2^{-\height(G)}>2\delta$). 

Similarly, $B_1+b'_2 \in [F_{s_2}(a'_2),F_{s_2}(a'_2+1))$ by definition of $f_{s_2}$. Thus $F_{s_2}(a'_2+1)-B_1-\delta>b'_2-\delta=b_1$. We conclude that $b''_2 = \min\{B_1-F_{s_1}(a_1-1), F_{s_2}(a'_2+1)-B_1-\delta\}>b_1$.

%	\tcp{$b_2$ yields the same outcome ($t'_2$) as $b_1+\delta$, since $F_{s_2}(a_2+1)$ is the open end of the interval that contains $B+b_1+\delta$.}

Next, note that by bidding $b'_2$ or $b''_2$, black reaches the same budget interval $a'_2$ in $s_2$. The problem with just setting $b''_2 = F_{s_2}(a'_2+1)-B_1-\delta$, is that this may prevent a further deviation by white, that would have been possible after a lower increment. That is, we may reach an equilibrium that is not the lowest-bids equilibrium.
However, if white has a deviation after black, then we only care if she has a deviation that brings her all the way to the next budget interval $a_1-1$ in $s_1$---otherwise black would  deviate  back to $a'_2$ with another $\delta$ increment and so on. Therefore constraining $b''_2 \leq B_1-F_{s_1}(a_1-1)$ would prevent this problem by always allowing white to bid the same. In other words, it guarantees that $a''_1=a'_1$.

As a concrete example, consider such an ascending auction in Fig.~\ref{fig:interval_alg}, for $B_1=0.4$. Whenever black increases his bid $b''_2$, it should be to the next point among $b''_2=F_{s_2}(2)-B_1-\delta = 0.2-\delta,\ b''_2=B_1-F_{s_1}(1)=0.4$, and $b''_2=F_{s_2}(3)-B_1-\delta = 0.6-\delta$, since if there is any deviation, the auction will eventually reach this point. The corresponding remaining budget $B_1+b''_2$ for each of the three options is marked by $d_1,d_2$ and $d_3$, respectively. Note that $d_2$ and $d_3$ both belong to the same budget interval of $s_2$, but skipping $d_2$ may result in looking over an equilibrium where white bids all of $B_1$ and wins. 

\medskip   
We now turn to show (a).	

By induction, the budget between any consecutive critical points $[a_{l}, a_l+1)$ yields the same outcome in $s_l$, and likewise for $s_r$. Fix some $r,l$ and let $B_1\in [z_{lr},z_{lr+})$, where $z_{lr+} = \min \{z_{l'r'} : z_{l'r'} > z_{lr}\}$. We argue that whether white or  black win in $(s,z_{lr})$, the winner reaches the same next state $s'$ and with the same remaining budget as from $(s,B_1)$. The reason is the same as in  the proof of Theorem~\ref{th:Pareto}\ref{e:int} and of Theorem~\ref{th:PSPE_cont} and we do not repeat it. %(there we always take the average between two multiples of $2^{-{k-1}}$, which yields a multiple of $2^{-k}$).

\medskip
It is left to show (c), i.e. that no information is lost when filtering the $(|T|+1)^2$ critical points.
Due to monotonicity, $\ol T$ is sorted, thus $U$ contains the first index of every unique entry in $\ol T$.  The filtering removes redundant parts of the equilibrium, %While in every point in $\ol F$ the bidding strategy may change, 
as there are at most $|T|$ points where the outcome changes. Thus we only consider these points.

\medskip
 Finally, we consider the complexity of the fast algorithm. After filtering, the size of the lists $F_s,T^*_s,A^1_s,A^2_s$ is at most $|T|$ each. With every deviation, either $a_1$  strictly decreases or $a_2$  strictly increases, and thus there can be at most $2|T|$ deviations. The heaviest part in computing a deviation is calculating $f_{s_i}$. Since $|F_{s_i}|\leq |T|$ a na\"ive computation is linear in $|T|$. Therefore $Ascending-Auction()$ runs in $O(|T|^2)$.  It is called at most $|T|^2$ times in every internal node, thus the total complexity is $O(|S| \cdot |T|^4 ) = O(|S|^5)$. 
\end{proof}

\begin{algorithm}[t]
\caption{\label{alg:PSPE_fast}\textsc{Find-PSPE-fast}($G$)}
\SetKwFunction{filter}{Filter}
\SetKwFunction{play}{Set-Strategies}
\SetKwFunction{aaf}{Ascending-Auction}
\SetKwFunction{append}{Append}
\SetKwFunction{init}{Initialize}
		 $\init(\ol F,\ol T^*,\ol A^1, \ol A^2, \ol I)$\;
		\tcp{See description of Initialize() for explanation of variable roles.}
%Initialize an empty list $T^*_s$\;
\For{ every leaf $t$:}
{ 
\tcp{In the leaves there is only one budget interval, which is all the range $[0,1]$.}
    $F_t \leftarrow (0)$ \tcp*{An array of length one}
		$T^*_t \leftarrow (t)$\; % \tcp*{An array of length one}
}
\tcp{Traverse nodes bottom up:}
\For{every node $s$ in post-order} { 
		Let $s_l,s_r$ be the two children of $s$\; 
		For every $a_l= 1,\ldots,|F_{s_l}|+1$ and $a_r= 1,\ldots,|F_{s_r}|+1$, set $z_{lr} = \frac{F_{s_l}(a_l) + F_{s_r}(a_r)}{2}$\;
		Sort $\{z_{lr}\}_{lr}$ in increasing order\;
		\tcp{$z_{lr}$ are the only ``critical points'' of the budget in state $s$. The equilibrium bids for any point in the interval $[z_{lr},z_{lr+})$ reach the same budget interval in the next state (either $s_l$ or $s_r$).} 
	%	Initialize global empty lists $\ol F,\ol T,\ol A_1, \ol A_2$\;
		Initialize global empty lists $ F, T, A_1, A_2$\;
	  \For{every $l,r$ (in increasing order of $z_{lr}$)}
		{ 
		     Set $B_1 = z_{lr}$ \tcp*{Budget of player~1} 
				\tcp{Compute the Bottom equilibrium of the current step for budget $B_1$ (and thus for all its interval)} % budgets between $z_{lr-}$ and $z_{lr}$.}
				$\tup{\tup{s_1,a_1},\tup{s_2,a_2},t}\leftarrow\aaf(G,s,B_1)$\;
				$\append(F,B_1)$;				$\append(T,t)$\;		
						$\append(A_1,\tup{s_1,a_1})$;		    $\append(A_2,\tup{s_2,a_2})$\;

		}
		$U\leftarrow T(1)$\;
		\For{$\ell=2,3,\ldots,|T|$} %also need last interval?
		{
					\If{$ T(\ell) \neq  T(\ell-1)$}
					{
							 $\append(U,\ell)$\;
					}
		}
							
		\tcp{Write down the equilibrium strategies in $s$:}
		%$\ol T^*$ is composed of at most $|T|$ uniform sections $U\subseteq \{1,\ldots,|\ol T^*|\}$\;
					
		$F_s \leftarrow \filter( F,U)$;		$T^*_s \leftarrow \filter( T,U)$\;
		$A^1_s \leftarrow \filter( A_1,U)$;		$A^2_s \leftarrow \filter( A_2,U)$\;
	} 
	\Return{$\gamma \leftarrow \tup{\ol F,\ol T^*,\ol A^1,\ol A^2}$}\;
	\end{algorithm}

\begin{function}[t]
	
\caption{Initialize($\ol F,\ol T^*,\ol A^1, \ol A^2$)}
\For{every node $s\in S$:}
{
    Initialize global empty lists $F_s,T^*_s,A^1_s, A^2_s$\; 
		}
		\tcp{A variable $\ol X$ is a list with one entry per state, i.e. $\ol X = (X_s)_{s\in S}$.}
		\tcp{Intuitively, $F_s$ contains all budget cutoff points. $F_s(1)=0$ always. For every $a\in \{1,2,\ldots,|F_s|\}$, all budgets $B_1\in[F_s(a),F_s(a+1))$ are equivalent, where $F_s(a')=1$ for all $a'>|F_s|$. We use $f_s:[0,1]\rightarrow \{1,2,\ldots,|F_s|\}$ as the inverse of $F_s$, where $f_s(B_1)$ is the index of the budget interval in $s$ that contains $B_1$. Formally, $f_s(B_1)=\max\{a : B_1 \geq F_s(a) \}$.}  
		%\tcp{Intuitively, $F_s$ contains all budget cutoff points. For every $a\in \{1,2,\ldots,|F_s|\}$, all budgets $B_1\in[F_s(a-1),F_s(a))$ lead to the same outcome. We use $f_s:[0,1]\rightarrow \{1,2,\ldots,|F_s|\}$ as the inverse of $F_s$.} 
		
		% last interval is always [1,1)
		\tcp{$T^*_s(a)\in T$ is the outcome that will be reached from state $(s,F_s(a))$.}
		\tcp{$A^i_s$ is the strategy of $i$, where $A^i_s(a)=\tup{s',a'}$.
		 % The equilibrium bid in $s$ for any budget $B_1$ can be derived by subtracting $B_1-F_{s'}(a')$, where $s'\in g(s)$ and $a'$ is the budget index in $s'$.
		}
\end{function}

	\begin{function}[t]
	\SetKwFunction{append}{Append}
\caption{Ascending-Auction($G,s,B_1$)}
	%Assume w.l.o.g. that under current budgets white prefers $s_l$:}
					$b_1,b_2\leftarrow 0$ \tcp*{Initialize bids}
					$a_l \leftarrow f_{s_l}(B_1)$\;
					$a_r \leftarrow f_{s_r}(B_1)$\;
					$s_1 \leftarrow \argmax_{j\in\{l,r\}}u_1(T^*_{s_j}(a_j))$\tcp*{Child preferred by white}
					$s_2 \leftarrow \argmax_{j\in\{l,r\}}u_2(T^*_{s_j}(a_j))$\tcp*{Child preferred by black}
					
					\If{ $s_1=s_2$}
					{ 
					     \tcp{White wins with bid $0$, and plays $s_1$. Reaches budget interval indexed by $a_1$ (either $a_l$ or $a_r$).}
					    %$s' \leftarrow s_1$\;
							 $a_1 \leftarrow f_{s_1}(B_1)$\;
							 $t_1 \leftarrow T^*_{s_1}(a_1)$\;
					     %$\append(\ol F,B)$\;
							 %$\append(\ol A_1 ,\tup{s', a'})$\;
							 %$\append(\ol A_2 ,\tup{s', a'})$\;   
							 %$\append(\ol T^* ,t^*)$\;   
							%Break\;
								\Return $\tup{\tup{s_1, a_1},\tup{s_1, a_1},t_1}$;
					}
					\Repeat{} 
					{
   			      \tcp{Compute outcome under current bids $b_1,b_2$:}
							%$B_w \leftarrow B-b$\; \tcp*{remaining budget if white wins}
							$a_1 \leftarrow f_{s_1}(B_1-b_1)$;
							$t_1 \leftarrow T^*_{s_1}(a_1)$\tcp*{Current outcome}
							%$B_b \leftarrow B+b$\; 
							$a_2 \leftarrow f_{s_2}(B_1+b_2)$\;
							$b'_2 \leftarrow b_1+\delta$\tcp*{Try a deviation of black}
							$a'_2 \leftarrow f_{s_2}(B_1+b'_2)$; 
							$t'_2 \leftarrow T^*_{s_2}(a'_2)$ \tcp*{Outcome after deviation}
							\If{ $b'_2 > 1-B_1$ or $t'_2 \preceq_2 t_1$ }
							{
							   \tcp{Black cannot or would not raise. In this budget interval (at least $B_1=z_{lr}$ and below $z_{lr+}$), white wins. The bid $b_1$ is set so that $B_1-b_1$ falls in the budget interval indexed by $a_1$ in $s_1$.}
					        %$\append(\ol F,B)$\;
							    %$\append(\ol A_1 ,\tup{s_1, a_1})$\;
							    %$\append(\ol A_2 ,\tup{s_2, a_2})$\;   
							    %$\append(\ol T^* ,t_1)$\; 
								 %Break\;
								\Return $\tup{\tup{s_1, a_1},\tup{s_2, a_2},t_1}$;
							}
							\tcp{A higher bid of black with the same outcome as $b'_2$:}
							$b''_2 \leftarrow \min\{B_1-F_{s_1}(a_1-1), F_{s_2}(a'_2+1)-B_1-\delta\}$\;
							% add figure
							$b''_1 \leftarrow b''_2$  \tcp*{Try a deviation of white}
							$a''_1 \leftarrow f_{s_1}(B_1-b''_1)$\; 
							$t''_1 \leftarrow T^*_{s_1}(a''_1)$\;
							\If{$t''_1 \preceq_1 t'_2$}  
							{ 
							   \tcp{White would not raise (Note that $b''_1\leq B_1$ by definition of $b''_2$ so white can always raise). In this budget interval, black wins and plays $s_2$. Reaches budget interval indexed by $a'_2$.}
					        %$\append(\ol F,B)$\;
							    %$\append(\ol A_1 ,\tup{s_1, a_1})$\;
							    %$\append(\ol A_2 ,\tup{s_2, a'_2})$\;   
							    %$\append(\ol T^* ,t'_2)$\; 
								 %Break\;
								\Return $\tup{\tup{s_1, a_1},\tup{s_2, a'_2},t'_2}$;
							 }
							$b_1\leftarrow b''_1$;
							$b_2\leftarrow b''_2$\;
					}
	\end{function}
	
	\begin{function}[t]
		\SetKwFunction{append}{Append}
\caption{Filter($L,U$)}
      Initialize $L^0$ as an empty list\;
			\For{$j=1,\ldots,|U|$}
			{
			    $\ell \leftarrow U(j)$\;
					$\append(L^0, L(\ell))$\;
			}
			\Return{$L^0$}
	\end{function}
	
	\input{intervals_alg}
	
	\subsection{Compiling small DAGs from succinct valuations}
	Suppose we are given a sequential scrip bargaining, with some succinct representation of the value functions $v_i$. While we can apply Proposition~\ref{th:SSA_bidding} to construct an equivalent bidding game, the na\"ive construction would yield a complete binary tree of height $k$. In particular, the number of nodes $|S|$ would be exponential in $k$. 
In the worst case, there is not much we can do. For example, if buyers assign distinct utilities to exponentially many bundles of items, then any bidding game must contain this number of terminal states.  Also, if $|T|$ is fixed, but merely computing an optimal partition of $K$ is NP-complete, then it must also be NP-hard to construct an equivalent succinct bidding game (if such a game even exists). This is since  constructing the tree would in particular require us to efficiently generate all $|T|$ outcomes, and one them is the optimal partition (we just check all of them).

\paragraph{Additive valuations}
We first solve a simple case, where valuations are additive, i.e., $v_i(S) = \sum_{j\in S} v_i(j)$. Suppose that we sold the first $4$ items  out of $k$, whose values to agent~1 are $2,2,4$, and $3$. Then the agent does not care if she currently has the first two items or just the third. If the other agent is also indifferent between these partial partitions, then we can represent them with a single state. 

We can generalize this observation using a dynamic programming technique, similar to the one used for Knapsack problems~\cite{re1962applied}. 

\begin{proposition}\label{th:additive}
Given an additive SSB $F$ with integer valuations, there is an equivalent binary bidding game with $|S|\leq (k+1)\cdot M$ states, where $M=( v_1(K)+1) \times (v_2(K)+1)$. Further, such a bidding game can be constructed efficiently.
\end{proposition}
\begin{proof}
We assign a state $s_{j,m_1,m_2}$ for every $j\in\{0,\ldots,k\}$ and every pair $m_1\in\{0,1,\ldots,v_1(K)\},m_2\in \{0,1,\ldots,v_2(K)\}$. The initial state is $s_{0,0,0}$. For every $j=1,\ldots,k$, we go over all states in level $j-1$. We connect each such state $s_{j-1,m_1,m_2}$ with two children: $s_{j,m_1+v_1(\tau(j)),m_2}$, and $s_{j,m_1,m_2+v_2(\tau(j))}$. That is, either agent~1 or agent~2 gets the $j$'th item (according to order $\tau$). We then add all new children to level $j$.  

Note that any state at level $j$ corresponds to a partition of the first $j$ items. %, thus a state can only belong in one level. 
Finally, we identify the terminals $T$ with the last level $k$ (full partitions), and assign $u_i(s_{k,m_1,m_2}) = m_i$. 
\end{proof}

Using an implementation of Algorithm~\ref{alg:PSPE_fast} in Matlab, we computed the PSPE prices for an SSB with $k$ identical items. When the initial budget is the same ($0.5$ for each player), we witness a clear pattern \emph{increasing prices} over time. This pattern breaks when we start from different initial budget allocations. 

\paragraph{Weight-based games}
A similar technique can be applied in games where the valuations are not additive, but still base on some additive notion of weight. We say that $v_i$ is \emph{weight-based} if every item has a fixed weight $w_j$, and there is a function $f_i:\mathbb N\rightarrow \mathbb R$, s.t. $v_i(S) = f_i(\sum_{j\in S} w_j)$. Note that without further constraints any function is weight-based, as we can make sure that each bundle has a different total weight. However if weights are bounded then we get a succinct representation for a subclass of functions. 

This idea can be further generalized.  We say that $v_i$ is \emph{multi-weight-based} if every item has a vector of fixed weights $(w^1_j,\ldots,w^q_j)$, and there is a function $f_i:\mathbb N\rightarrow \mathbb R$, s.t. $v_i(S) = f_i(\sum_{j\in S} w^1_j, \ldots, \sum_{j\in S} w^q_j)$. 
As a concrete example, suppose that the items are computing machines, each with some properties like storage, memory, bandwidth, etc. The value of a set of machines to a client depends only on the total storage, total memory, and total bandwidth, regardless of how these resources are allocated across among the machines.

Note that additive SSBs are a special case of multi-weight-based SSBs, where there are two dimensions. $f_1(S) = \sum_{j\in S} w^1_j, f_2(S) = \sum_{j\in S} w^2_j$.
Note however that we allow $f_i$ to be a completely arbitrary function, and it does not even need to be monotone.

\begin{proposition}\label{th:weighted}
Given a multi-weight-based SSB $F$, there is an equivalent bidding game with $|S|\leq (k+1)\cdot M^2$ states, where $M= (w^1(K)+1) \times \cdots \times (w^q(K)+1) $. Further, such a bidding game can be constructed efficiently.
\end{proposition}
\begin{proof}
Construction is very similar to Proposition~\ref{th:additive}.
We identify each state at level~$j$ with a tuple $\tup{j,m^1_1,m^1_2,\ldots,m^q_1,m^q_2}$. 
 where $m^{r}_i$ now tracks the total dimension-$r$ weight of the bundle owned by agent $i$. We link each state to its two children by adding all respective weights of item $\tau(j)$ to the bundle owned by one of the agents. 

Finally, in every terminal $t=s_{\tup{k,m^1_1,m^1_2,\ldots,m^q_1,m^q_2}}$ (state that corresponds to a full allocation), we set $u_1(t) = f_1(m^1_1,\ldots,m^q_1), u_2(t) = f_2(m^1_2,\ldots,m^q_2)$.
\end{proof}

As a corollary from Propositions \ref{th:PSPE_fast} and \ref{th:weighted}, a PSPE in multi-weight-based and in additive SSBs can be computed efficiently. That is, in time that is polynomial in the number of items and in the total weight/value.  

\paragraph{Voting with single peaked preferences}
If voters have a strict preference order over all $2^k$ possible outcomes of the game, there cannot be a succinct representation, since we will need $2^k$ terminals. However consider the following scenario. Each voter as an ``ideal'' point $t^*_i \in \{0,1\}^k$. The utility of any other outcome only depends on the Manhattan distance from $t^*_i$, i.e. on the number of coordinates in which $t,t^*_i$ differ. We may also assign a different weight to each coordinate, so that $u_i(t) = -\sum_{j\leq k} w_{ij} |t^*_i(j)-t(j)|$. 

This scenario has a simple mapping to an SSB of additive items: The value to $i$ of every coordinate on which $t^*_1,t^*_2$ differ is $w_{ij}$. If bidder $i$ ``buys'' coordinate $j$ she may set it to $t^*_i(j)$. The value of every coordinate $j$ on which $t^*_1(j) = t^*_2(j)$ can be arbitrary. The bidders will not compete on these coordinates, as it is Pareto-optimal to set $t(j)= t^*_1(j)$.     

\paragraph{Pathfinding}
The items in a pathfinding problem over a graph, are pairs $s=\tup{location,time}$. In a na\"ive representation, all pairs would be allocated, and the value of a bundle of pairs to an agent, would be the optimal legal path from source to target that this bundle contains.  If our game states would each represent a subset of such pairs, we would quickly reach an exponential blowup. We modify the auction so that a state $\tup{location_1,location_2,time}$ means that each agent $i$ is in $location_i$ at $time$, after reaching there from the start. Thus we do not auction states (tuples) in an arbitrary order, but rather construct an alternative bidding game. We present an outline of the construction.

We start from $s_0 = \tup{start_1,start_2,0}$, and link each state $s=\tup{location_1,location_2,time}$ to two other states: the first is just a copy of $s$, and the second is one of the legal moves from $s$, for example $\tup{location'_1,location'_2,time+1}$, where $location'_i$ is either a neighbor of $location_i$ or $location_i$ itself, and it does not hold that $location'_1=location'_2$. We keep enough copies of $s$ to exhaust all legal moves (at most the number of locations, squared).
The terminal states are those where $time=MaxTime$, or both agents reached their targets. After the first agent reaches her target, we record her time. The negative utility to each agent in a terminal is her time.

Thus, in total we have at most $MaxTime^2\times|locations|^4$ states including all copies. 

\rmr{more efficient valuations: * single minded bidders (max over values); XOS with bounded number of quantifiers (single minded/additive have one); Read once MC/XOS of bounded height?}
  
\section{Discussion and Related Work}

We presented a simple and intuitive mechanism---sequential scrip bargaining---that implements the full range of Pareto-efficient allocations of items to two agents with arbitrary valuation functions. Further, our mechanism  can be applied to efficiently and fairly settle other combinatorial bargaining problems that involve two parties. The pure subgame-perfect equilibrium can be computed in time that is polynomial and the size of the full game tree/DAG, and for many classes of valuations, polynomial in the size of the succinct representation. 
%One may ask whether a PSPE can be computed efficiently. While such questions are outside the scope of this paper, we provide in the full version a non-trivial algorithm for computing a PSPE in time that is polynomial in the size of $G$. 

\subsection{Related Work}\label{sec:related}
\paragraph{Richman games}
Lazarus et al.~\shortcite{lazarus1996richman,laz99} were the first to systematically analyze bidding variations of zero-sum games. They coined the term ``Richman games'' in honor of David Ross Richman, the original inventor, and considered games with an infinitely divisible unit of budget. A Richman game is a directed graph (possibly with cycles) with two terminal nodes (say, black and white), and a full play is a path starting from some node and ending in a terminal node. The goal of each player is to end the game in her own terminal. The  main focus of Lazarus et al. was on the following question: ``in every node, what is the minimal fraction of the budget that will guarantee a victory for the white player?'' 
The \emph{Richman function} assigns a unique value to every node $R(s)$, which is the average of the Richman values of its lowest and highest neighbors (normalizing the values of white and black terminals to $0$ and $1$, respectively). %This definition seems to have a recursive nature, it provides a remarkably clean answer to the question above. 
Lazarus et al. show that the Richman function exists on every graph and is also unique if the graph is finite. It turns out that $R(s)$ marks the critical budget: if $B_1>R(s)$ then white can force a victory when starting from node $s$; if $B_1< R(s)$ then black can force a victory. Moreover, $R(s)$ is also  the exact probability that black wins in a game where instead of bidding, the player in each turn is selected by a fair coin toss (see \cite{peres07} for more details on random-turn games). 

Lazarus et al. further study other variations of bidding games by applying different auction rules. In particular, in the \emph{Poorman game}  the highest bidder pays the bank rather than to the other player, so the total budget shrinks with every step. 
%They also suggest various algorithms to compute or approximate $R(s)$, especially in the case of undirected graphs.

Develin and Payne~\shortcite{DP10} extended the theory of zero-sum bidding games in several important aspects.  In most recreational games, such as Tic-Tac-Toe, Chess, and Checkers, white and black can perform different actions when in the same board state $s$, thus games are not necessarily impartial.  Develin and Payne extended some of the previous results of Lazarus et al. to partial games such as Chess and Tic-Tac-Toe. In addition, they considered discrete bids, and the implications of various tie-breaking schemes. To demonstrate their approach, Develin and Payne showed a complete solution of bidding Tic-Tac-Toe for every possible initial budget. Interestingly, in the continuous case the Richman value of the initial state in Tic-Tac-Toe is $133/256\equiv 0.519$, which means that an advantage of $\sim 4\%$ in the initial budget is sufficient to guarantee a victory. In a related work, Payne and Robeva~\shortcite{payne2008artificial} developed a near-optimal algorithm for playing bidding Hex.  In a very recent paper, Larsson and W{\"a}stlund~\cite{larsson2017endgames} explore some intricacies of partiality, and in particular show examples of Chess endgames where no value exists.

\paragraph{Negotiations and bargaining}In typical bargaining problems there is a \emph{conflict deal} that is implemented if agents cannot reach agreement~\cite{nash1950bargaining}, or utility depends on time until agreement~\cite{rubinstein1982perfect,fatima2004agenda}. %This still holds in combinatorial bargaining problems, e.g. allocation of tasks among agents~\cite{zlotkin1996mechanism}.
 In our mechanism there is no conflict deal or discounting and thus it is hard to compare the protocols directly. %\footnote{A detailed comparison with \emph{arbitration mechanisms} (that do not assume a conflict deal) is given in Section~\ref{sec:apps}.} 

Yet, our Bottom Equilibrium can be intuitively thought of as a combinatorial version of the \emph{minimal sufficient concession} principle, that is often employed by agents in the monotonic concession protocol~\cite{zlotkin1996mechanism}: at every node the winner makes the minimal concession (in terms of future bargaining power) that allows her to win the round.

	\paragraph{Sequential combinatorial auctions}	

``Standard'' sequential auctions (with quasi-linear utilities rather than budgets) have been studied by several researchers~\cite{gale2001sequential,rodriguez2009sequential,leme2012sequential}.
These papers focused on particular classes of value functions (unit-demand, submodular, etc.) and generally demonstrated that while pure equilibria exist, they may be substantially inefficient for some of these classes. Inefficient outcomes occur even in 2-buyer, complete information auctions~\cite{bae2007efficiency}. 

%Boutilier et al.~\shortcite{boutilier1999sequential} introduced  a different model of sequential auctions, where the utility model is quite general and may include quasi-linear utilities, as well as utilities that only depend on the acquired items, as in our case. Rather than reasoning about (public or private) valuations of other buyers, each buyer has an explicit assumption over other buyers' bids in each state. Thus computing the optimal bidding policy becomes an online optimization problem, rather than a game.   

%Gale and Setegeman~\cite{gale2001sequential} and Rodriguez~\cite{rodriguez2009sequential} 
 %While for some restricted classes of valuation functions 
\paragraph{All-pay rules and Colonel Blotto}
%The theory of Richman games was recently extended to (zero-sum) games wit all-pay auctions by Menz et al.~\shortcite{menz2015discrete}
Powell~\shortcite{powell2009sequential} models a particular sequential game between an attacker  and a defender (a sequential Colonel Blotto game). This game is essentially a specific general-sum bidding game over a degenerated tree (a path), where the utility in every match---the success probability---is determined by the invested resources of each party.
Our model does not cover this particular game due to several differences. First, in our model the utility is only determined in the leaves rather than accumulated over the entire path;\footnote{This would be a minor difference if the number of outcomes was finite.} More importantly, the utility in every match depends exactly on the ``bids'' (investments) of both parties, and both parties discard their bids (an ``all-pay'' auction). 

Another all-pay variation of bidding games was studied in \cite{menz2015discrete}, who developed a fast algorithm for computing the optimal strategy. Note that the optimal strategy under all-pay rules must be mixed, i.e. use probabilistic bids.
 %However the special case where an attack succeeds if and only if the attacker invensts more than the defender (and then each side gets arbitrary fixed utility), falls under both models. 

\paragraph{Sequential auctions with budgets}
Lastly and closest to our work, Huang et al.~\shortcite{huang2012sequential} study particular bidding game, in which two agents use an initial budget to bid over identical items that are sold sequentially. This is essentially a sequential scrip auction mechanism, yet  their model is not strictly a special case of ours, as they add some additional refinements.\footnote{In \cite{huang2012sequential} players pay to the bank rather to one another (the Poorman variation), and also try to minimize spent budget as a secondary goal.} While we believe that the models are very close, the focus on a particular value function (which makes the game an almost-zero-sum game) allows Huang et al. to provide an accurate characterization of the allocations under PSPE as a function of the budgets. Our results are qualitative in nature, but  apply to arbitrary utility functions. 

Huang et al. prove that item prices \emph{decrease} over time. Curiously, we observe (empirically) that prices \emph{increase}.\footnote{Using a fast implementation of our algorithm %~\ref{alg:PSPE_fast} 
in Matlab, we computed the PSPE prices for an SSB with $k$ identical items. When the initial budget is the same ($0.5$ for each player), we witness a clear pattern of \emph{increasing prices} over time. This pattern breaks when we start from unequal initial budget allocations.}
 One interpretation is that the result of Huang et al. is an artifact of the auction rule, as in the Poorman version the total amount of money is also decreasing in each round. % (pay one another vs. pay the bank).   

\subsection{Variations}
\label{sec:variations}
By changing the auction rules, different types of bidding games arise. For example, an all-pay auction may better describe various real world scenarios like security games, sport matches, and R\&D competitions. However  these games typically do not have a pure equilibrium point, and are more difficult to analyze (see Related work section for specific games that have been studied).
Using a second-price rather than a first-price auction to determine the winner should not have a large effect, since in equilibrium the bids will always be very close. 

An important question is whether our results still hold when the ``Poorman game'' version is played, i.e. when the highest bidder pays the bank rather than the other player. Since in the continuous case the strategy at a given node only depends on the fraction $B_1/B_2$, there is a one-to-one mapping between strategies in the continuous Richman and Poorman variations of a game. We would thus expect our main results (e.g. Pareto-optimality of the Bottom Equilibrium) to hold, but this would require a separate proof. Also computational properties may be far less convenient. The Poorman version has one important advantage over the Richman version---the game definition naturally extends to any number of players. 
It is an open question whether a Pareto-optimal PSPE exists in a binary  game for more than two players, either when playing together, or when connected in a network of bilateral bargaining connections~\cite{Chakraborty:2009}. We believe that the answer to this question is negative, since there are simple Poorman games with three players that are non-monotone.  

\subsection{Future Work}
\label{sec:future}
Other than studying the variations mentioned above, there are several questions that remain open. One natural direction is how changing the order in which items are offered (or the agenda in sequential voting) affects the induced equilibrium. It is known that in general bargaining problems the agenda can strongly affect the outcome~\cite{fershtman1990importance,inderst2000multi}. While the agenda may also affect the outcome in our model, it is possible that there are classes of valuations where all orders yield a similar outcome in PSPE. For example, from Proposition~\ref{th:MST} the order of issues in zero-sum bargaining games does not matter.

More important questions arise when non-determinism enters. For example, we conjecture that our results (both existence and optimality) also hold under random tie-breaking, and that such a randomized mechanism in fact leads to better fairness guarantees. On the other hand random tie-breaking can exponentially increase the number of possible outcomes and thus efficient computation may not be possible.\footnote{We thank Michael Simkin for verifying this assertion.}
In addition, agents' valuations themselves may be private and sampled from some distribution, and we are interested whether similar results regarding efficiency and monotonicity apply in games without complete information. 

Finally, experiments with bargaining games in the spirit of \cite{weg1990two,de2012selection} may reveal whether the game-theoretic analysis is a good proxy to the actual behavior, and whether our mechanism can improve welfare and satisfaction in practice.

\section*{Acknowledgments}
The authors thank David Parkes and Urban Larsson for valuable feedback on the manuscript. 
\bibliography{bidding_games.full}
%\end{small}
%
%\newpage
%\appendix
%%%\bibliographystyle{named}
%\input{bidding_games.appendix.tex}

\end{document}

%% file: defs.tex
\newcommand{\coursename}{(67686) Mathematical Foundations of AI}

\newcommand{\handout}[5]{
   \renewcommand{\thepage}{#1-\arabic{page}}
   \noindent
   \begin{center}
   \framebox{
      \vbox{
    \hbox to 5.78in { {\bf \coursename}
         \hfill #2 }
       \vspace{4mm}
       \hbox to 5.78in { {\Large \hfill #5  \hfill} }
       \vspace{2mm}
       \hbox to 5.78in { {\it #3 \hfill #4} }
      }
   }
   \end{center}
   \vspace*{4mm}
}

\newcommand{\floor}[1]{\left\lfloor #1\right\rfloor}
\newcommand{\ceil}[1]{\left\lceil #1\right\rceil}

\newenvironment{proof-sketch}{\noindent{\bf Sketch of Proof}\hspace*{1em}}{\qed\bigskip}
\newenvironment{proof-idea}{\noindent{\bf Proof Idea}\hspace*{1em}}{\qed\bigskip}
\newenvironment{proof-of-lemma}[1]{\noindent{\bf Proof of Lemma #1}\hspace*{1em}}{\qed\bigskip}
\newenvironment{proof-attempt}{\noindent{\bf Proof Attempt}\hspace*{1em}}{\qed\bigskip}

\makeatletter
\def\fnum@figure{{\bf Figure \thefigure}}
\def\fnum@table{{\bf Table \thetable}}
\long\def\@mycaption#1[#2]#3{\addcontentsline{\csname
  ext@#1\endcsname}{#1}{\protect\numberline{\csname
  the#1\endcsname}{\ignorespaces #2}}\par
  \begingroup
    \@parboxrestore
    \small
    \@makecaption{\csname fnum@#1\endcsname}{\ignorespaces #3}\par
  \endgroup}
\def\mycaption{\refstepcounter\@captype \@dblarg{\@mycaption\@captype}}
\makeatother

\newcommand{\mathify}[1]{\ifmmode{#1}\else\mbox{$#1$}\fi}
\newcommand{\bigO}O

\newcommand\tup[1]{\left\langle #1 \right\rangle}

\newcommand{\height}{\text{height}}

\newcommand{\calB}{{\cal B}}
\newcommand{\calF}{{\cal F}}

\newcommand{\remove}[1]{{}}

%% file: example_maj.tex
%\documentclass[]{article}
%\usepackage{tikz}
%\usepackage{subcaption}
%%\usepackage{tikz}%,fullpage}
%\usetikzlibrary{arrows,%
                %petri,%
                %topaths}%
%\usepackage{tkz-berge}
%
%\begin{document}
%

%
%\begin{figure}
%%%%%%%%%%%%%%%%%%
\centering
\begin{tikzpicture}[scale=1,transform shape]

  \Vertex[x=0,y=0,L=${0,0}$]{s00}
	\Vertex[x=1.5,y=0.6,L=${1,0}$]{s10}
	\Vertex[x=1.5,y=-0.6,L=${0,1}$]{s01}
	\Vertex[x=3,y=0,L=${1,1}$]{s11}

	\tikzstyle{VertexStyle}=[black,draw]
	\Vertex[x=3,y=1.2,L=${2,0}$]{t20}
	\Vertex[x=4.5,y=0.6,L=${2,1}$]{t21}

	\tikzstyle{VertexStyle}=[fill=black!70!white,text=white]
	\Vertex[x=3,y=-1.2,L=${0,2}$]{t02}
		\Vertex[x=4.5,y=-0.6,L=${1,2}$]{t12}
  %\Vertex[x=6,y=0,L=${2,2}$]{t22}
	
	%\tikzstyle{VertexStyle}=[fill=black!10!white]
	%\Vertex[x=1.5,y=3,L=${(1,8)}$]{t1}
	%\Vertex[x=1.5,y=1,L=${(2,1)}$]{t2}
	%\Vertex[x=3,y=3,L=${(0,9)}$]{tx}
	%\Vertex[x=4.5,y=2,L=${(0,9)}$]{ty1}
	%\Vertex[x=4.5,y=0,L=${(10,7)~*}$]{ty2}
	%
  \tikzstyle{LabelStyle}=[fill=white,sloped]
  \tikzstyle{EdgeStyle}=[->]
  
	\Edge[](s00)(s01)
	\Edge[](s10)(t20)
	\Edge[](s01)(s11)
	\Edge[](s01)(t02)
	\Edge[](s11)(t12)
	\tikzstyle{EdgeStyle}=[double,->]
  
\Edge[](s00)(s10)
\Edge[](s10)(s11)
\Edge[](s11)(t21)
	
\end{tikzpicture}
\caption{\label{fig:Gmaj}The game $G_{\maj}$.  White wins in white terminals, black wins in black. Double arrows mark the game play in the example.\vspace{-3mm}}
%\caption{\label{sfig:s0}The value for every bidding profile $(b_1,b_2)$ at state $s_0$. White wins in white areas, black wins in grey areas. The dashed arrows mark the winner in case of a bidding profile that is on the boundary between segments.}

%
%\caption{The game $G_{maj}$. Double arrows mark the game play in the example.}
%%\caption{\label{sfig:s0}The value for every bidding profile $(b_1,b_2)$ at state $s_0$. White wins in white areas, black wins in grey areas. The dashed arrows mark the winner in case of a bidding profile that is on the boundary between segments.}
%\end{figure}
%\end{document}

%% file: example_two_PSPEs_b.tex
%\documentclass[]{article}
%\usepackage{tikz}
%\usepackage{subcaption}
%\usepackage{tikz}%,fullpage}
%\usetikzlibrary{arrows,%
                %petri,%
                %topaths}%
%\usepackage{tkz-berge}
%
%\begin{document}

%%%%%%%%%%%%%%%%%%

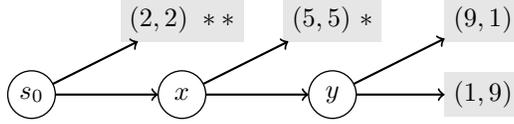
\begin{figure}
\centering
\begin{tikzpicture}[scale=1,transform shape]

  \Vertex[x=0,y=2,L=$s_0$]{s}
	\Vertex[x=2,y=2,L=$x$]{x}
	\Vertex[x=4,y=2,L=$y$]{y}
	
  \tikzstyle{VertexStyle}=[fill=black!10!white]
	\Vertex[x=2,y=3,L=${(2,2)~**}$]{t1}
	\Vertex[x=4,y=3,L=${(5,5)~*}$]{tx}
	\Vertex[x=6,y=2,L=${(1,9)}$]{ty1}
	\Vertex[x=6,y=3,L=${(9,1)}$]{ty2}
	
  \tikzstyle{LabelStyle}=[fill=white,sloped]
  \tikzstyle{EdgeStyle}=[->]
  \Edge[](s)(t1)
	\Edge[](s)(x)
	\Edge[](x)(y)
	\Edge[](x)(tx)
	\Edge[](y)(ty1)
	\Edge[](y)(ty2)

\end{tikzpicture}
\caption{\label{fig:two_PSPE} The game $G_{\two}$. Note that utilities are generic.}
\end{figure}
%\end{document}

%% file: PSPE_lattice.tex
%\documentclass[]{article}
%\usepackage{tikz}
%\usepackage{subcaption}
%%\usepackage{tikz}%,fullpage}
%\usetikzlibrary{arrows,%
                %petri,%
                %topaths}%
%\usepackage{tkz-berge}
%
%\begin{document}
%
%%%%%%%%%%%%%%%%%%

\begin{figure}
%\centering
\begin{framed}
\begin{center}
\begin{subfigure}[b]{0.3\textwidth}
\begin{tikzpicture}[scale=0.7,transform shape]
\tikzstyle{VertexStyle}=[circle,draw,minimum size=7mm]
  \Vertex[x=0,y=0,L=$x$]{x}
	\Vertex[x=1,y=1,L=${x_1}$]{x1}
	\Vertex[x=2,y=2,L=$xy$]{xy}
	\Vertex[x=3,y=1,L=$y$]{y}
	\Vertex[x=1,y=3,L=${s_0}$]{s}
 %gamma%
\tikzstyle{VertexStyle}=[]
\Vertex[x=-0.5,y=1.4,L=bids at ${(x,8)}$:]{xtext0}
\Vertex[x=-0.5,y=1,L=${b_1=3}$]{xtext1}
	\Vertex[x=-0.5,y=0.6,L=${b_2=2}$]{xtext2}

\Vertex[x=3,y=-0.4,L=bids at ${(y,12)}$:]{ytext0}	
	\Vertex[x=3,y=-0.8,L=${b_1=8}$]{ytext}
	\Vertex[x=3,y=-1.2,L=${b_2=6}$]{ytext2}

	\tikzstyle{VertexStyle}=[shape=coordinate]
	\Vertex[x=-1,y=-1,L=$$]{tx1}
	\Vertex[x=0,y=-1,L=$$]{tx2}
	\Vertex[x=1,y=-1,L=$$]{tx3}
	\Vertex[x=1.5,y=0,L=$$]{tx11}
		\Vertex[x=0,y=2,L=$$]{s1}
	\Vertex[x=1,y=2,L=$$]{s2}
		\Vertex[x=3.5,y=0,L=$$]{ty1}
	\Vertex[x=2.5,y=0,L=$$]{ty2}
	%add leafs
	
  \tikzstyle{LabelStyle}=[fill=white,sloped]
  \tikzstyle{EdgeStyle}=[->]

	\Edge[](s)(xy)
	\Edge[](xy)(y)
	\Edge[](xy)(x1)
	\Edge[](x1)(x)
	 \tikzstyle{EdgeStyle}=[dashed]
	  \Edge[](s)(s1)
	\Edge[](s)(s2)
	\Edge[](x)(tx1)
	\Edge[](x)(tx2)
	\Edge[](x)(tx3)
	\Edge[](x1)(tx11)
	\Edge[](y)(ty1)
	\Edge[](y)(ty2)	
\end{tikzpicture}
\caption{$\gamma$}
\end{subfigure}
~ 
\begin{subfigure}[b]{0.3\textwidth}
\begin{tikzpicture}[scale=0.7,transform shape]
\tikzstyle{VertexStyle}=[circle,fill=black!15!white,draw,minimum size=7mm]
  \Vertex[x=0,y=0,L=$x$]{x}
	\Vertex[x=1,y=1,L=${x_1}$]{x1}
	
	\Vertex[x=3,y=1,L=$y$]{y}
	
	\tikzstyle{VertexStyle}=[circle,double,fill=black!15!white,draw,minimum size=7mm]
	\Vertex[x=2,y=2,L=$xy$]{xy}
	\Vertex[x=1,y=3,L=${s_0}$]{s}
%gamma'%
\tikzstyle{VertexStyle}=[]
\Vertex[x=-0.5,y=1.4,L=bids at ${(x,8)}$:]{xtext0}
\Vertex[x=-0.5,y=1,L=${b_1=5}$]{xtext1}
	\Vertex[x=-0.5,y=0.6,L=${b_2=2}$]{xtext2}

\Vertex[x=3,y=-0.4,L=bids at ${(y,12)}$:]{ytext0}	
	\Vertex[x=3,y=-0.8,L=${b_1=8}$]{ytext}
	\Vertex[x=3,y=-1.2,L=${b_2=4}$]{ytext2}

	\tikzstyle{VertexStyle}=[shape=coordinate]
	\Vertex[x=-1,y=-1,L=$$]{tx1}
	\Vertex[x=0,y=-1,L=$$]{tx2}
	\Vertex[x=1,y=-1,L=$$]{tx3}
	\Vertex[x=1.5,y=0,L=$$]{tx11}
		\Vertex[x=0,y=2,L=$$]{s1}
	\Vertex[x=1,y=2,L=$$]{s2}
		\Vertex[x=3.5,y=0,L=$$]{ty1}
	\Vertex[x=2.5,y=0,L=$$]{ty2}
	%add leafs
	
  \tikzstyle{LabelStyle}=[fill=white,sloped]
  \tikzstyle{EdgeStyle}=[->]

	\Edge[](s)(xy)
	\Edge[](xy)(y)
	\Edge[](xy)(x1)
	\Edge[](x1)(x)
	 \tikzstyle{EdgeStyle}=[dashed]
	  \Edge[](s)(s1)
	\Edge[](s)(s2)
	\Edge[](x)(tx1)
	\Edge[](x)(tx2)
	\Edge[](x)(tx3)
	\Edge[](x1)(tx11)
	\Edge[](y)(ty1)
	\Edge[](y)(ty2)	
\end{tikzpicture}
\caption{$\gamma'$}
\end{subfigure}
~ 
\begin{subfigure}[b]{0.3\textwidth}

\begin{tikzpicture}[scale=0.7,transform shape]
\tikzstyle{VertexStyle}=[circle,draw,minimum size=7mm]
  \Vertex[x=0,y=0,L=$x$]{x}
	\Vertex[x=1,y=1,L=${x_1}$]{x1}
	\tikzstyle{VertexStyle}=[circle,fill=black!40!white,draw,minimum size=7mm]

	\Vertex[x=3,y=1,L=$y$]{y}
	
	\tikzstyle{VertexStyle}=[circle,fill=black!40!white,double,draw,minimum size=7mm]

	\Vertex[x=2,y=2,L=$xy$]{xy}
	\Vertex[x=1,y=3,L=${s_0}$]{s}
%gamma''%
\tikzstyle{VertexStyle}=[]
\Vertex[x=-0.5,y=1.4,L=bids at ${(x,8)}$:]{xtext0}
\Vertex[x=-0.5,y=1,L=${b_1=3}$]{xtext1}
	\Vertex[x=-0.5,y=0.6,L=${b_2=2}$]{xtext2}

\Vertex[x=3,y=-0.4,L=bids at ${(y,12)}$:]{ytext0}	
	\Vertex[x=3,y=-0.8,L=${b_1=9}$]{ytext}
	\Vertex[x=3,y=-1.2,L=${b_2=8}$]{ytext2}

	\tikzstyle{VertexStyle}=[shape=coordinate]
	\Vertex[x=-1,y=-1,L=$$]{tx1}
	\Vertex[x=0,y=-1,L=$$]{tx2}
	\Vertex[x=1,y=-1,L=$$]{tx3}
	\Vertex[x=1.5,y=0,L=$$]{tx11}
		\Vertex[x=0,y=2,L=$$]{s1}
	\Vertex[x=1,y=2,L=$$]{s2}
		\Vertex[x=3.5,y=0,L=$$]{ty1}
	\Vertex[x=2.5,y=0,L=$$]{ty2}
	%add leafs
	
  \tikzstyle{LabelStyle}=[fill=white,sloped]
  \tikzstyle{EdgeStyle}=[->]

	\Edge[](s)(xy)
	\Edge[](xy)(y)
	\Edge[](xy)(x1)
	\Edge[](x1)(x)
	 \tikzstyle{EdgeStyle}=[dashed]
	  \Edge[](s)(s1)
	\Edge[](s)(s2)
	\Edge[](x)(tx1)
	\Edge[](x)(tx2)
	\Edge[](x)(tx3)
	\Edge[](x1)(tx11)
	\Edge[](y)(ty1)
	\Edge[](y)(ty2)	
\end{tikzpicture}
\caption{$\gamma''$}
\end{subfigure}

%
%\caption{\label{sfig:s0}The value for every bidding profile $(b_1,b_2)$ at state $s_0$. White wins in white areas, black wins in grey areas. The dashed arrows mark the winner in case of a bidding profile that is on the boundary between segments.}
\caption{\label{fig:lattice_tree}The figure shows a part of three different profiles of the same game. The bids in nodes $x$ and $y$ under particular budgets are shown for each profile. For either $x,y$ we assume that the bids are the same across all profiles for any other budget. We also assume that in all nodes that are not shown the bids are the same across all three profiles under any budget. For the other nodes $\{x_1,xy,s_0\}$ we do not assume anything about the  bids, and they may differ between profiles.\\
In each of $s\in\{x,y\}$ we get a different order $\pi$ over the three profiles: \\
$\gamma'_x >_\pi \gamma_x =_\pi \gamma''_x$ and $\gamma''_y >_\pi \gamma'_y >_\pi \gamma_y$. \\
The shading shows the order $\sigma(s)$, where for any $s$, darker color means the respective equilibrium is higher in the order (unless comparing two nodes with double lines). For example: $\gamma' >_{\sigma(x_1)} \gamma =_{\sigma(x_1)} \gamma''$.  To see why, note that by our construction $\gamma' \geq _{\sigma(s)} \gamma$ in all $s\in g(x_1)$, and the relation is strict for at least one child ($x$). Thus we can ignore the bids in $x_1$ and still determine the relation $\sigma(x_1)$. 
Another example is $\gamma'' >_{\sigma(xy)} \gamma$. \\
 Observe however that $\gamma',\gamma''$ cannot be compared according to $\sigma(xy)$ and $\sigma(s_0)$ (nodes with double line). For $xy$ this is because there are conflicting relations in the children of $xy$, and for $s_0$ this is since the relation in at least one child ($xy$) is undefined. We thus get that $\gamma' >_\pi \gamma$ and $\gamma'' >_\pi \gamma$, whereas $\gamma',\gamma''$ are incomparable. 
}
\end{center}
\end{framed}
\end{figure}
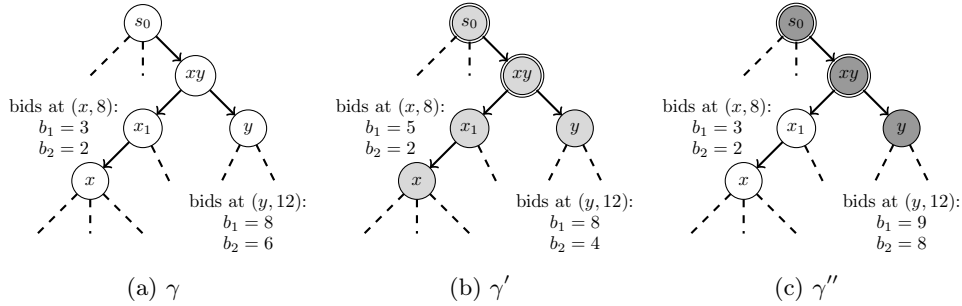
%\end{document}

%% file: example_bad_PSPE.tex
%\documentclass[]{article}
%\usepackage{tikz}
%\usepackage{subcaption}
%%\usepackage{tikz}%,fullpage}
%\usetikzlibrary{arrows,%
                %petri,%
                %topaths}%
%\usepackage{tkz-berge}
%
%\begin{document}

%%%%%%%%%%%%%%%%%%

%\begin{figure}
%\centering
%\begin{subfigure}[b]{0.48\textwidth}
\begin{center}
\begin{tikzpicture}[scale=0.9,transform shape]

  \Vertex[x=0,y=2,L=$s_0$]{s}
	\Vertex[x=1.5,y=2,L=$x$]{x}
	\Vertex[x=3,y=1,L=$y$]{y}
	
  \tikzstyle{VertexStyle}=[fill=black!10!white]
	\Vertex[x=1.5,y=3,L=${(1,8)}$]{t1}
	\Vertex[x=1.5,y=1,L=${(2,1)}$]{t2}
	\Vertex[x=3,y=3,L=${(0,9)}$]{tx}
	\Vertex[x=4.5,y=2,L=${(0,9)}$]{ty1}
	\Vertex[x=4.5,y=0,L=${(10,7)~*}$]{ty2}
	
  \tikzstyle{LabelStyle}=[fill=white,sloped]
  \tikzstyle{EdgeStyle}=[->]
  \Edge[](s)(t1)
	\Edge[](s)(t2)
	\Edge[](s)(x)
	\Edge[](x)(y)
	\Edge[](x)(tx)
	\Edge[](y)(ty1)
	\Edge[](y)(ty2)

\end{tikzpicture}
%\caption{\label{sfig:Ga}A neutral game that is not binary.\\~ \\ ~}
%\caption{\label{sfig:s0}The value for every bidding profile $(b_1,b_2)$ at state $s_0$. White wins in white areas, black wins in grey areas. The dashed arrows mark the winner in case of a bidding profile that is on the boundary between segments.}
\caption{\label{fig:bad_PSPE} The game $G_{\text{bad}}$.}
\end{center}
%\end{figure}
%\end{document}

%% file: intervals1L.tex
%\documentclass[11pt]{article}
%\usepackage{tikz}
%\input{defs}
%\usepackage{wrapfig,framed}
%\begin{document}
%\usetikzlibrary{decorations.pathreplacing}
%\usetikzlibrary{calc}
%

\begin{figure}
\begin{framed}
\begin{tikzpicture}[scale=0.85]

\tikzstyle{dot}=[rectangle,draw=black,fill=white,inner sep=0pt,minimum size=4mm]
\def\tgap{0.6}
\def\egap{1.5}

\node at (0,0) {\Large $[$};
\node at (-0.1,\tgap) {\scriptsize $B^{j'}_1$};
\node at (2,-\tgap) {\small $\calB^{j'}(k-1)$};
  
\coordinate (p) at (4,2pt);
  \foreach \myprop/\mytext [count=\n] in {0.5/$\epsilon$,0.5/$\epsilon$,0.5/$\epsilon$}
  \draw [decorate,decoration={brace,amplitude=2}] (p)  edge [draw] +(0,-4pt) 	-- ++(\myprop,0) coordinate (p) node [midway, above=2pt, anchor=south] {\mytext} ;
\foreach \x in {1,2,3,4,5,6,7,9,10,11,12}
    {        
      \coordinate (A\x) at ($(0,0)+(\x*0.5,0)$) {};
      \draw ($(A\x)+(0,2pt)$) -- ($(A\x)-(0,2pt)$);
      %\node at ($(A\x)+(0,3ex)$) {\x};
    }
  \node at (4,0) {\Large $[$};
\node at (6,-\tgap) {\small $\calB^{j'+1}(k-1)$};

%%%%%

\node at (9,0) {\Large $[$};
\node at (9-0.1,\tgap) {\scriptsize $B^{j''}_1$};
\node at (11,-\tgap) {\small $\calB^{j''}(k-1)$};

\foreach \x in {1,2,3,4,5,6,7,9,10,11,12}
    {        
      \coordinate (A\x) at ($(9,0)+(\x*0.5,0)$) {};
      \draw ($(A\x)+(0,2pt)$) -- ($(A\x)-(0,2pt)$);
      %\node at ($(A\x)+(0,3ex)$) {\x};
    }
  \node at (13,0) {\Large $[$};
\node at (14.3,-\tgap) {\small $\calB^{j''+1}(k-1)$};

%%%%%  Case 1L B_1
\node at (1,2*\egap+0.5) {\textbf{Case~1L}};
\node at (3,2*\egap) {$\gamma_s(B^j_1)$};
\node at (6.5,2*\egap) {\Large $[$};
\node at (6.5-0.1,2*\egap+\tgap) {\scriptsize $B^{j}_1$};
\node at (7.5,2*\egap+\tgap) {\small $\calB^{j}(k)$};

\foreach \x in {1,2,3,4,5,6}
    {        
      \coordinate (A\x) at ($(6.5,2*\egap)+(\x*0.5,0)$) {};
      \draw ($(A\x)+(0,2pt)$) -- ($(A\x)-(0,2pt)$);
      %\node at ($(A\x)+(0,3ex)$) {\x};
    }
  \node at (8.5,2*\egap) {\Large $[$};
\node at (9.5,2*\egap+\tgap) {\small $\calB^{j+1}(k)$};

\draw[|<->|,dashed] (0.5,\egap) -- (6.5,\egap);
\node at (3.5,\egap + 0.3) {$b_1$};
\draw[|<->|,dashed] (6.5,\egap+0.1) -- (12.5,\egap+0.1);
\node at (9.5,\egap + 0.4) {$b_2$};
\draw[dotted] (6.5,\egap) -- (6.5,2*\egap);
\draw[dotted] (0.5,\egap) -- (0.5,0);
\draw[dotted] (12.5,\egap) -- (12.5,0);
%%%%%  Case 1L \hat B
\node at (3,-2*\egap) {$\gamma_s(\hat B_1)$};
\node at (6.5,-2*\egap) {\Large $[$};
\node at (6.5-0.1,-2*\egap-\tgap) {\scriptsize $B^{j}_1$};
%\node at (7.5,-2*\egap-\tgap) {\small $\calB^{j}_1(k)$};  
\node at (8,-2*\egap-\tgap) {\scriptsize $\hat B_1$};

\foreach \x in {1,2,3,4,5,6}
    {        
      \coordinate (A\x) at ($(6.5,-2*\egap)+(\x*0.5,0)$) {};
      \draw ($(A\x)+(0,2pt)$) -- ($(A\x)-(0,2pt)$);
      %\node at ($(A\x)+(0,3ex)$) {\x};
    }
  \node at (8.5,-2*\egap) {\Large $[$};
\node at (9.5,-2*\egap-\tgap) {\small $\calB^{j+1}(k)$};

\draw[|<->|,dashed] (6.5,-2*\egap+0.5) -- (8,-2*\egap+0.5);
\node at (7.25,-2*\egap+0.8) {$r\epsilon$};

\draw[|<->|,dashed] (3.5,-\egap) -- (8,-\egap);
\node at (6,-\egap - 0.4) {$\hat b_1$};
\draw[|<->|,dashed] (8,-\egap-0.1) -- (12.5,-\egap-0.1);
\node at (10,-\egap - 0.5) {$\hat b_2$};
\draw[dotted] (8,-\egap) -- (8,-2*\egap);
\draw[dotted] (3.5,-\egap) -- (3.5,0);
\draw[dotted] (12.5,-\egap) -- (12.5,0);

\end{tikzpicture}
%%%%%%%%%%%%%%%%%%%%%%%%%%%%%%%%%%%%%%%%%%%%%%%%%%%%
\caption{\label{fig:intervals1L}The middle row shows intervals $j'$ and $j''$ of level $k-1$ (each of size $2R\cdot \eps$). The top row shows an interval $j$ of level $k$ (of size $R\cdot \epsilon$) and the equilibrium bids in Case~1L at $(s,B^j_1)$. The bottom row shows the equilibrium bids at $\hat B_1=B^j_1+r\eps$. Note that $b_1=b_2$ and $\hat b_1=\hat b_2$. }
\end{framed}
\end{figure}
%\end{document}

%% file: intervals1W.tex
%\documentclass[11pt]{article}
%\usepackage{tikz}
%\input{defs}
%\usepackage{wrapfig,framed}
%\begin{document}
%\usetikzlibrary{decorations.pathreplacing}
%\usetikzlibrary{calc}

\begin{figure}
\begin{framed}
\begin{tikzpicture}[scale=0.85]

\tikzstyle{dot}=[rectangle,draw=black,fill=white,inner sep=0pt,minimum size=4mm]
\def\tgap{0.6}
\def\egap{1.5}

\node at (0,0) {\Large $[$};
\node at (-0.1,\tgap) {\scriptsize $B^{j'}_1$};
\node at (2,-\tgap) {\small $\calB^{j'}(k-1)$};
  
%\coordinate (p) at (4,2pt);
  %\foreach \myprop/\mytext [count=\n] in {0.5/$\epsilon$,0.5/$\epsilon$,0.5/$\epsilon$}
  %\draw [decorate,decoration={brace,amplitude=2}] (p)  edge [draw] +(0,-4pt) 	-- ++(\myprop,0) coordinate (p) node [midway, above=2pt, anchor=south] {\mytext} ;
\foreach \x in {1,2,3,4,5,6,7,9,10,11,12}
    {        
      \coordinate (A\x) at ($(0,0)+(\x*0.5,0)$) {};
      \draw ($(A\x)+(0,2pt)$) -- ($(A\x)-(0,2pt)$);
      %\node at ($(A\x)+(0,3ex)$) {\x};
    }
  \node at (4,0) {\Large $[$};
\node at (6,-\tgap) {\small $\calB^{j'+1}(k-1)$};

%%%%%

\node at (9,0) {\Large $[$};
\node at (9-0.1,\tgap) {\scriptsize $B^{j''}_1$};
\node at (11,+\tgap) {\small $\calB^{j''}(k-1)$};

\foreach \x in {1,2,3,4,5,6,7,9,10,11,12}
    {        
      \coordinate (A\x) at ($(9,0)+(\x*0.5,0)$) {};
      \draw ($(A\x)+(0,2pt)$) -- ($(A\x)-(0,2pt)$);
      %\node at ($(A\x)+(0,3ex)$) {\x};
    }
  \node at (13,0) {\Large $[$};
\node at (14.3,-\tgap) {\small $\calB^{j''+1}(k-1)$};

%%%%%  Case 1W
\node at (1,2*\egap+0.5) {\textbf{Case~1W}};
\node at (3,2*\egap) {$\gamma_s(B^j_1)$};

\node at (6.5,2*\egap) {\Large $[$};
\node at (6.5-0.1,2*\egap+\tgap) {\scriptsize $B^{j}_1$};
\node at (7.5,2*\egap+\tgap) {\small $\calB^{j}(k)$};  

\foreach \x in {1,2,3,4,5,6}
    {        
      \coordinate (A\x) at ($(6.5,2*\egap)+(\x*0.5,0)$) {};
      \draw ($(A\x)+(0,2pt)$) -- ($(A\x)-(0,2pt)$);
      %\node at ($(A\x)+(0,3ex)$) {\x};
    }
  \node at (8.5,2*\egap) {\Large $[$};
\node at (9.5,2*\egap+\tgap) {\small $\calB^{j+1}(k)$};

\draw[|<->|,dashed] (3.5,\egap) -- (6.5,\egap);
\node at (5,\egap + 0.3) {$b_1$};
\draw[|<->|,dashed] (6.5,\egap+0.1) -- (9.5,\egap+0.1);
\node at (8,\egap + 0.4) {$b_2$};
\draw[dotted] (6.5,\egap) -- (6.5,2*\egap);
\draw[dotted] (3.5,\egap) -- (3.5,0);
\draw[dotted] (9.5,\egap) -- (9.5,0);

%%%%%  Case 1W  \hat B_1
\node at (3,-2*\egap) {$\gamma_s(\hat B_1)$};
\node at (6.5,-2*\egap) {\Large $[$};
\node at (6.5-0.1,-2*\egap-\tgap) {\scriptsize $B^{j}_1$};
%\node at (7.5,-2*\egap-\tgap) {\small $\calB^{j}_1(k)$};  
\node at (7.5,-2*\egap-\tgap) {\scriptsize $\hat B_1$};

\foreach \x in {1,2,3,4,5,6}
    {        
      \coordinate (A\x) at ($(6.5,-2*\egap)+(\x*0.5,0)$) {};
      \draw ($(A\x)+(0,2pt)$) -- ($(A\x)-(0,2pt)$);
      %\node at ($(A\x)+(0,3ex)$) {\x};
    }
  \node at (8.5,-2*\egap) {\Large $[$};
\node at (9.5,-2*\egap-\tgap) {\small $\calB^{j+1}(k)$};

\draw[|<->|,dashed] (3.5,-\egap) -- (7.5,-\egap);
\node at (5.5,-\egap - 0.4) {$\hat b_1$};
\draw[|<->|,dashed] (7.5,-\egap-0.1) -- (11.5,-\egap-0.1);
\node at (9.5,-\egap - 0.5) {$\hat b_2$};
\draw[dotted] (7.5,-\egap) -- (7.5,-2*\egap);
\draw[dotted] (3.5,-\egap) -- (3.5,0);
\draw[dotted] (11.5,-\egap) -- (11.5,0);

\end{tikzpicture}
\\
%%%%%%%%%%%%%%%%%%%%%%%%%%%%%%%%%%%%%%%%%%%%%%%%%%%%
\caption{\label{fig:intervals1W}Equilibrium bids at $(s,B^j_1)$ and $(s,\hat B_1)$ in Case~1W.}
\end{framed}
\end{figure}
%\end{document}

%% file: intervals2.tex
%\documentclass[11pt]{article}
%\usepackage{tikz}
%\input{defs}
%
%\begin{document}
%\usetikzlibrary{decorations.pathreplacing}
%\usetikzlibrary{calc}

\begin{figure}
\begin{framed}
\begin{tikzpicture}[scale=0.85]

\tikzstyle{dot}=[rectangle,draw=black,fill=white,inner sep=0pt,minimum size=4mm]
\def\tgap{0.6}
\def\egap{1.5}

\node at (0,0) {\Large $[$};
\node at (-0.1,-\tgap) {\scriptsize $B^{j'}_1$};
\node at (2,-\tgap) {\small $\calB^{j'}(k-1)$};
  
\coordinate (p) at (4,2pt);
%  \foreach \myprop/\mytext [count=\n] in {0.5/$\epsilon$,0.5/$\epsilon$,0.5/$\epsilon$}
 % \draw [decorate,decoration={brace,amplitude=2}] (p)  edge [draw] +(0,-4pt) 	-- ++(\myprop,0) coordinate (p) node [midway, above=2pt, anchor=south] {\mytext} ;
\foreach \x in {1,2,3,4,5,6,7,9,10,11,12}
    {        
      \coordinate (A\x) at ($(0,0)+(\x*0.5,0)$) {};
      \draw ($(A\x)+(0,2pt)$) -- ($(A\x)-(0,2pt)$);
      %\node at ($(A\x)+(0,3ex)$) {\x};
    }
  \node at (4,0) {\Large $[$};
\node at (6,-\tgap) {\small $\calB^{j'+1}(k-1)$};

%%%%%

\node at (9,0) {\Large $[$};
\node at (9-0.1,\tgap) {\scriptsize $B^{j''}_1$};
\node at (11,-\tgap) {\small $\calB^{j''}(k-1)$};

\foreach \x in {1,2,3,4,5,6,7,9,10}
    {        
      \coordinate (A\x) at ($(9,0)+(\x*0.5,0)$) {};
      \draw ($(A\x)+(0,2pt)$) -- ($(A\x)-(0,2pt)$);
      %\node at ($(A\x)+(0,3ex)$) {\x};
    }
  \node at (13,0) {\Large $[$};
\node at (14,-\tgap) {\small $\calB^{j''+1}(k-1)$};

%%%%%  Case 2L
\node at (2,2*\egap) {\textbf{Case~2L}};
\node at (4.5,2*\egap) {\Large $[$};
\node at (4.5-0.1,2*\egap+\tgap) {\scriptsize $B^{j}_1$};
\node at (5.5,2*\egap+\tgap) {\small $\calB^{j}(k)$};

\foreach \x in {1,2,3,4,5,6}
    {        
      \coordinate (A\x) at ($(4.5,2*\egap)+(\x*0.5,0)$) {};
      \draw ($(A\x)+(0,2pt)$) -- ($(A\x)-(0,2pt)$);
      %\node at ($(A\x)+(0,3ex)$) {\x};
    }
  \node at (6.5,2*\egap) {\Large $[$};
\node at (7.5,2*\egap+\tgap) {\small $\calB^{j+1}(k)$};

\draw[|<->|,dashed] (0,\egap) -- (4.5,\egap);
\node at (2.25,\egap + 0.3) {$b_1$};
\draw[|<->|,dashed] (4.5,\egap+0.1) -- (9.5,\egap+0.1);
\node at (7,\egap + 0.4) {$b_2$};
\draw[dotted] (4.5,\egap) -- (4.5,2*\egap);
\draw[dotted,double] (0,\egap) -- (0,0);
\draw[dotted] (9.5,\egap) -- (9.5,0);
%%%%%  Case 2W
\node at (2,-2*\egap) {\textbf{Case~2W}};
\node at (4.5,-2*\egap) {\Large $[$};
\node at (4.5-0.1,-2*\egap-\tgap) {\scriptsize $B^{j}_1$};
\node at (5.5,-2*\egap-\tgap) {\small $\calB^{j}(k)$};  

\foreach \x in {1,2,3,4,5,6}
    {        
      \coordinate (A\x) at ($(4.5,-2*\egap)+(\x*0.5,0)$) {};
      \draw ($(A\x)+(0,2pt)$) -- ($(A\x)-(0,2pt)$);
      %\node at ($(A\x)+(0,3ex)$) {\x};
    }
  \node at (6.5,-2*\egap) {\Large $[$};
\node at (7.5,-2*\egap-\tgap) {\small $\calB^{j+1}(k)$};

\draw[|<->|,dashed] (0.5,-\egap) -- (4.5,-\egap);
\node at (2.5,-\egap - 0.3) {$b_1$};
\draw[|<->|,dashed] (4.5,-\egap-0.1) -- (9,-\egap-0.1);
\node at (7.75,-\egap - 0.4) {$b_2$};
\draw[dotted] (4.5,-\egap) -- (4.5,-2*\egap);
\draw[dotted] (0.5,-\egap) -- (0.5,0);
\draw[dotted,double] (9,-\egap) -- (9,0);

\end{tikzpicture}
\\
%%%%%%%%%%%%%%%%%%%%%%%%%%%%%%%%%%%%%%%%%%%%%%%%%%%%
\caption{\label{fig:intervals2}Equilibrium bids at $(s,B^j_1)$ in Case~2L and 2W. Note that in both cases $b_2=b_1+\eps$. The critical point in each case is marked with a double dotted line.}
\end{framed}
\end{figure}
%\end{document}

%% file: example_bad_PSPE_discrete.tex
%\documentclass[]{article}
%\usepackage{tikz}
%\usepackage{subcaption}
%\usepackage{tikz}%,fullpage}
%\usetikzlibrary{arrows,%
                %petri,%
                %topaths}%
%\usepackage{tkz-berge}
%
%\begin{document}

%%%%%%%%%%%%%%%%%%

\begin{figure}
\centering
\begin{tikzpicture}[scale=1,transform shape]

  \Vertex[x=0,y=2,L=$s_0$]{s}
	\Vertex[x=1.5,y=1.4,L=$x_1$]{x1}
	\Vertex[x=3,y=1.4,L=$x_2$]{x2}
	\Vertex[x=6,y=1.4,L=$x_{k-2}$]{x3}
	\Vertex[x=7.5,y=1.4,L=$x_{k-1}$]{x4}
	\Vertex[x=1.5,y=2.6,L=$y_1$]{y1}
	\Vertex[x=3,y=2.6,L=$y_2$]{y2}
	\Vertex[x=6,y=2.6,L=$y_{k-2}$]{y3}
	\Vertex[x=7.5,y=2.6,L=$y_{k-1}$]{y4}

  \tikzstyle{VertexStyle}=[fill=black!10!white]
	\Vertex[x=9,y=1.4,L=${(7,9)~*}$]{tx}
	\Vertex[x=9,y=2.6,L=${(9,7)~*}$]{ty}
	\Vertex[x=6,y=0.4,L=${(8,1)}$]{tbx}
	\Vertex[x=6,y=3.6,L=${(1,8)}$]{tby}
	
  \tikzstyle{LabelStyle}=[fill=white,sloped]
  \tikzstyle{EdgeStyle}=[->]
  \Edge[](s)(x1)
	\Edge[](s)(y1)
	\Edge[](x1)(x2)
	\Edge[](x3)(x4)
	\Edge[](x4)(tx)
	\Edge[](y1)(y2)
	\Edge[](y3)(y4)
	\Edge[](y4)(ty)

	\Edge[](x2)(tbx)
	\Edge[](x3)(tbx)
	\Edge[](x4)(tbx)

	\Edge[](y2)(tby)
	\Edge[](y3)(tby)
	\Edge[](y4)(tby)
	
\tikzstyle{LabelStyle}=[fill=white,sloped]
  \tikzstyle{EdgeStyle}=[dashed,->]
  \Edge[](x2)(x3)
	\Edge[](y2)(y3)
	
	\tikzstyle{EdgeStyle}=[bend left=15,->]
		\Edge[](y1)(tby)
			\tikzstyle{EdgeStyle}=[bend left,->]
  %\Edge[](s)(tby)
	\tikzstyle{EdgeStyle}=[bend right,->]
   %\Edge[](s)(tbx)
		\tikzstyle{EdgeStyle}=[bend right=15,->]
		\Edge[](x1)(tbx)

\end{tikzpicture}
\caption{\label{fig:discrete_PSPE} The game $G_{k}$. Pareto-optimal terminals are marked with $*$.}
\end{figure}
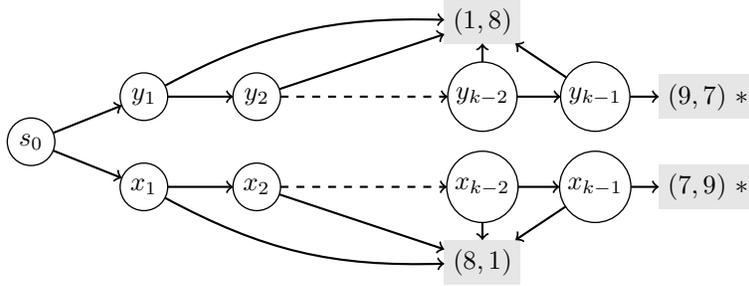
%\end{document}

%% file: example_no_SPE.tex
%%\documentclass[]{article}
%\documentclass[authoryear]{elsarticle}
%\usepackage{tikz}
%\usepackage{subcaption}
%%\usepackage{subfigure}
%%\usepackage{wrapfig}
%%\usepackage{tikz}%,fullpage}
%\usetikzlibrary{arrows,%
                %petri,%
                %topaths}%
%\usepackage{tkz-berge}
%
%\begin{document}
%
%%%%%%%%%%%%%%%%%%%

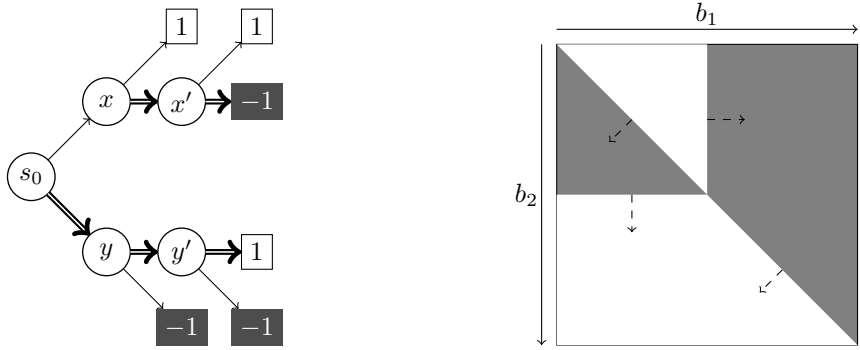
\begin{figure}
\centering

\begin{subfigure}[b]{0.45\textwidth}

\begin{tikzpicture}[scale=1,transform shape]

  \Vertex[x=0,y=3,L=$s_0$]{s}
	\Vertex[x=1,y=4,L=$x$]{x1}
	\Vertex[x=1,y=2,L=$y$]{y1}
	\Vertex[x=2,y=4,L=$x'$]{x2}
	\Vertex[x=2,y=2,L=$y'$]{y2}
	
  \tikzstyle{VertexStyle}=[draw,black]
	\Vertex[x=2,y=5,L=$1$]{tx1}
	\Vertex[x=3,y=5,L=$1$]{tx2}
	\Vertex[x=3,y=2,L=$1$]{ty3}
	\tikzstyle{VertexStyle}=[fill=black!70!white,text=white]
	\Vertex[x=2,y=1,L=$-1$]{ty1}
  \Vertex[x=3,y=1,L=$-1$]{ty2}
  \Vertex[x=3,y=4,L=$-1$]{tx3}

  \tikzstyle{LabelStyle}=[fill=white,sloped]
  \tikzstyle{EdgeStyle}=[thin,->]
  \Edge[](s)(x1)

\Edge[](x1)(tx1)
	\Edge[](x2)(tx2)
	
	\Edge[](y1)(ty1)
	\Edge[](y2)(ty2)

 \tikzstyle{EdgeStyle}=[double,->]
\Edge[](s)(y1)
	\Edge[](x1)(x2)
	\Edge[](x2)(tx3)

	\Edge[](y1)(y2)
	\Edge[](y2)(ty3)
%  \Edge[label=$630$]{S}{B}
%  \Edge[label=$210$]{S}{N}
%  \Edge[label=$230$]{S}{M}

\end{tikzpicture}
\caption{\label{sfig:H}The value of each terminal is marked with a square. White always prefers the upper state in each step, black always prefers the lower state. The double arrows indicate which step is played in case of a tie in the auction.
%caption{\label{fig:no_SPE} The zero-sum game from Proposition~\ref{th:no_PSPE}.}
}
%\end{figure}

%\begin{figure}
\end{subfigure}
~
\begin{subfigure}[b]{0.45\textwidth}
\begin{tikzpicture}[scale=1,transform shape]
\draw[fill=gray] (0,0) rectangle (4,4);
\draw[->] (0,4.2) -- (4,4.2) node at (2,4.2) [black,midway,yshift=0.2cm] {$b_1$};
\draw[->] (-.2,4) -- (-.2,0) node at (-.2,2) [black,midway,xshift=-0.2cm] {$b_2$};
\fill[white] (0,4) -- (2,2) -- (2,4);
\fill[white] (0,0) -- (4,0) -- (2,2) -- (0,2);
\draw[dashed,->] (1,3) -- (1-0.3,3-0.3);
\draw[dashed,->] (3,1) -- (3-0.3,1-0.3);
\draw[dashed,->] (2,3) -- (2+0.5,3);
\draw[dashed,->] (1,2) -- (1,2-0.5);
\end{tikzpicture}
\caption{\label{sfig:s0}The value for every bidding profile $(b_1,b_2)\in [0,\frac12]^2$ at state $s_0$. White wins in white areas, black wins in grey areas. The dashed arrows mark the winner in case of a bidding profile that is on the boundary between segments.}
\end{subfigure}
\caption{\label{fig:no_SPE}A game with continuous budgets, inconsistent tie-breaking,  and no SPE.}
\end{figure}
%\end{document}

%% file: intervals1L_cont.tex
%\documentclass[11pt]{article}
%\usepackage{tikz}
%\input{defs}
%\usepackage{wrapfig,framed}
%\begin{document}
%\usetikzlibrary{decorations.pathreplacing}
%\usetikzlibrary{calc}
%\def\eps{\epsilon}

\begin{figure}[h]
\begin{framed}
\begin{tikzpicture}[scale=0.8]

\tikzstyle{dot}=[rectangle,draw=black,fill=white,inner sep=0pt,minimum size=4mm]
\def\tgap{0.6}
\def\egap{1.5}

\node at (0,0) {\Large $[$};
\node at (-0.1,\tgap) {\scriptsize $B^{j'}_1$};
\node at (1.5,-\tgap) {\small $\ol \calB^{j'}$};
   \draw[thick] (0,0) -- (3.95,0); 
	\draw[thick] (4,0) -- (7.5,0); 
\coordinate (p) at (4,2pt);
%  \foreach \myprop/\mytext [count=\n] in {0.5/$\epsilon$,0.5/$\epsilon$,0.5/$\epsilon$}
 % \draw [decorate,decoration={brace,amplitude=2}] (p)  edge [draw] +(0,-4pt) 	-- ++(\myprop,0) coordinate (p) node [midway, above=2pt, anchor=south] {\mytext} ;
%\foreach \x in {1,2,3,4,5,6,7,9,10,11,12}
%    {        
 %     \coordinate (A\x) at ($(0,0)+(\x*0.5,0)$) {};
 %     \draw ($(A\x)+(0,2pt)$) -- ($(A\x)-(0,2pt)$);
      %\node at ($(A\x)+(0,3ex)$) {\x};
 %   }
  \node at (4,0) {\Large $[$};
\node at (6,-\tgap) {\small $\ol \calB^{j'+1}$};

%%%%%

\node at (9,0) {\Large $[$};
\node at (9-0.1,\tgap) {\scriptsize $B^{j''}_1$};
\node at (11,-\tgap) {\small $\ol \calB^{j''}$};
 \draw[thick] (9,0) -- (12.95,0);
\draw[thick] (13,0) -- (14.5,0); 
%
%\foreach \x in {1,2,3,4,5,6,7,9,10,11,12}
    %{        
      %\coordinate (A\x) at ($(9,0)+(\x*0.5,0)$) {};
      %\draw ($(A\x)+(0,2pt)$) -- ($(A\x)-(0,2pt)$);
      %%\node at ($(A\x)+(0,3ex)$) {\x};
    %}
  \node at (13,0) {\Large $[$};
\node at (14.3,-\tgap) {\small $\ol \calB^{j''+1}$};

%%%%%  Case 1L B_1
%\node at (1,2*\egap-0.5) {\textbf{Case~2L}};
%%\node at (3,2*\egap) {$\gamma_s(B^j_1)$};
%\node at (6.5,2*\egap) {\Large $[$};
%\node at (6.5-0.1,2*\egap+\tgap) {\scriptsize $B^{j}_1$};
  %\node at (8.5,2*\egap) {\Large $[$};
%\node at (9.5,2*\egap-\tgap) {\small $\calB^{j+1}_1$};
%\draw[thick] (6.5,2*\egap) -- (8.45,2*\egap);
%\draw[thick] (8.5,2*\egap) -- (10.45,2*\egap);  
%
%\draw[dotted] (7.8,\egap) -- (7.8,2*\egap);
%\node at (7.8,2*\egap+\tgap) {\scriptsize $\hat B_1$};
%\draw[|<->|,dashed] (7.8,2*\egap-0.5) -- (8.5,2*\egap-0.5);
%\node at (8.1,2*\egap-0.8) {$\delta$};
%\draw[|<->|,dashed] (0,\egap) -- (7.8,\egap);
%\node at (3,\egap + 0.4) {$\hat b_1$};
%\draw[dotted] (0,\egap) -- (0,2*\egap);
%\draw[|<->|,dashed] (7.8,\egap) -- (16.3,\egap);
%\node at (3,\egap + 0.4) {$\hat b_2$};
%\draw[dotted] (0,\egap) -- (0,2*\egap);
%
%
%\foreach \x in {1,2,3,4,5,6}
    %{        
      %\coordinate (A\x) at ($(6.5,2*\egap)+(\x*0.5,0)$) {};
      %\draw ($(A\x)+(0,2pt)$) -- ($(A\x)-(0,2pt)$);
      %%\node at ($(A\x)+(0,3ex)$) {\x};
    %}
  %\node at (8.5,2*\egap) {\Large $[$};
%\node at (9.5,2*\egap+\tgap) {\small $\calB^{j+1}_1(k)$};
%
%\draw[|<->|,dashed] (0.5,\egap) -- (6.5,\egap);
%\node at (3.5,\egap + 0.3) {$b_1$};
%\draw[|<->|,dashed] (6.5,\egap+0.1) -- (12.5,\egap+0.1);
%\node at (9.5,\egap + 0.4) {$b_2$};
%\draw[dotted] (6.5,\egap) -- (6.5,2*\egap
%\draw[dotted] (12.5,\egap) -- (12.5,0););
%\draw[dotted] (0.5,\egap) -- (0.5,0);
%%%%%  Case 1L \hat B
%\node at (3,-2*\egap) {$\overline \gamma_s(\hat B_1)$};
\node at (1,-2*\egap-0.5) {\textbf{Case~1 (L+W)}};

\node at (6.5,-2*\egap) {\Large $[$};
\draw[thick] (6.5,-2*\egap) -- (8.45,-2*\egap);
\draw[thick] (8.5,-2*\egap) -- (10.45,-2*\egap);  
\node at (6.5-0.1,-2*\egap-\tgap) {\scriptsize $B^{j}_1$};
%\node at (7.5,-2*\egap-\tgap) {\small $\calB^{j}_1(k)$};  
\node at (7.8,-2*\egap-\tgap) {\scriptsize $\hat B_1$};

%
%\foreach \x in {1,2,3,4,5,6}
    %{        
      %\coordinate (A\x) at ($(6.5,-2*\egap)+(\x*0.5,0)$) {};
      %\draw ($(A\x)+(0,2pt)$) -- ($(A\x)-(0,2pt)$);
      %%\node at ($(A\x)+(0,3ex)$) {\x};
    %}
  \node at (8.5,-2*\egap) {\Large $[$};
\node at (9.5,-2*\egap-\tgap) {\small $\ol \calB^{j+1}$};

\draw[|<->|,dashed] (7.8,-2*\egap+0.5) -- (8.5,-2*\egap+0.5);
\node at (8.1,-2*\egap+0.8) {$\delta$};

\draw[|<->|,dashed] (3.3,-\egap) -- (7.8,-\egap);
\node at (6,-\egap - 0.4) {$\hat b_2$};
\draw[|<->|,dashed] (7.8,-\egap-0.1) -- (13,-\egap-0.1);
\node at (10,-\egap - 0.5) {$\hat b_1$};

\draw[|<->|,dashed] (3.3,-0.5) -- (4,-0.5);
\node at (3.7,-0.8) {$\delta$};

\draw[|<->|,dashed] (3.3,-0.5) -- (2.6,-0.5);
\node at (3,-0.8) {$\delta$};
\draw[dotted] (7.8,-\egap) -- (7.8,-2*\egap);
\draw[dotted] (3.3,-\egap) -- (3.3,0);
\draw[dotted] (13,-\egap) -- (13,0);
\draw[dotted] (2.6,-\egap-0.2) -- (2.6,0);
\node at (2.6,-\egap - 0.5) {\scriptsize{$\hat B_1-\hat b_1$}};

\end{tikzpicture}
%%%%%%%%%%%%%%%%%%%%%%%%%%%%%%%%%%%%%%%%%%%%%%%%%%%%
\caption{\label{fig:intervals1_c}Equilibrium bids at $\overline \gamma_s(\hat B_1)$. }
\end{framed}
\end{figure}
%\end{document}

%% file: intervals_alg.tex
%\documentclass[11pt]{article}
%\usepackage{tikz}
%\input{defs}
%\usepackage{wrapfig,framed}
%\begin{document}
%\usetikzlibrary{decorations.pathreplacing}
%\usetikzlibrary{calc}

\begin{figure}
\begin{framed}
\centering
\begin{tikzpicture}[scale=0.95]

\tikzstyle{dot}=[rectangle,draw=black,fill=white,inner sep=0pt,minimum size=4mm]
\def\tgap{0.6}
\def\egap{1.5}

\node at (0,0) {\Large $[$};
\node at (2,0) {\Large $)[$};
\node at (7,0) {\Large $)[$};
\node at (10,0) {\Large $]$};
\node at (0,0.7) {$F_{s_1}(1)$};
\node at (2,0.7) {$F_{s_1}(2)$};
\node at (7,0.7) {$F_{s_1}(3)$};

\node at (0,-3) {\Large $[$};
\node at (6,-3) {\Large $)[$};
\node at (10,-3) {\Large $]$};
\node at (0,-3.7) {$F_{s_2}(1)$};
\node at (6,-3.7) {$F_{s_2}(2)$};

\node at (0,-1.5) {$|$};
\node at (1,-1.5) {$|$};
\node at (3,-1.5) {$|$};
\node at (3.5,-1.5) {$|$};
\draw[|<->|,dashed] (2,-1) -- (4,-1);
\draw[|<->|,dashed] (4,-2) -- (6,-2);
\draw[dotted] (2,-1) -- (2,-0.3);
\draw[dotted] (6,-2) -- (6,-2.7);
\node at (0,-1) {$*$};
\node at (5.8,-2.3) {$d_1$};
\node at (8,-2.3) {$d_2$};
\node at (9.8,-2.3) {$d_3$};
\node at (4,-1.5) {$|$};
\node at (5,-1.5) {$|$};
\node at (6,-1.5) {$|$};
\node at (6.5,-1.5) {$|$};
\node at (8,-1.5) {$|$};
\node at (8.5,-1.5) {$|$};
\node at (10,-1.5) {$|$};

\end{tikzpicture}
%%%%%%%%%%%%%%%%%%%%%%%%%%%%%%%%%%%%%%%%%%%%%%%%%%%%
\caption{\label{fig:interval_alg}The top and bottom lines show the budget intervals in both children, where $F_{s_1}=(0,0.2,0.7)$ (say, for the left child), and $F_{s_2}=(0,0.6)$. Any budget maps to a unique interval in each child. E.g., $f_{s_1}(0.3)=2$ since $0.3\in [F_{s_1}(2),F_{s_1}(3))$. Note that the rightmost interval always ends at $1$. 
The middle line shows all potential cutoff points (before filtering) in node $s$: in increasing order, $(z_{lr})_{lr} = (0,0.1,0.3,0.35,0.4,0.5,0.6,0.65,0.8,0.85,1)$. For example, $z_{lr}= 0.4 = \frac{0.2+0.6}{2}$ is obtained for $l=2,r=2$, as shown in the figure.
The points marked by $d_j$ are the only deviations that black needs to consider, where $d_1,d_3$ correspond to the the right edge of $s_2$ intervals (minus $\delta$), and $d_2$ corresponds to the left edge of $s_1$ interval (marked with $*$).}
\end{framed}
\end{figure}
%\end{document}